\documentclass[12pt]{article}
\usepackage{jheppub}
\pdfoutput=1
\usepackage{epstopdf}
\usepackage{amsmath,mathtools}
\usepackage{jheppub}
\usepackage{mathrsfs}
\usepackage{psfrag}
\usepackage{color}
\usepackage{hyperref}
\usepackage[dvipsnames]{xcolor}
\usepackage{slashed}
\usepackage{feynmp-auto}
\usepackage{simplewick}
\usepackage{cancel}

\usepackage{amsmath,bbm,array,amsfonts,graphicx,wrapfig,arydshln,lscape,float,multirow,longtable,rotating,makecell}
\usepackage{url,xcolor}

\newcommand{\la}{\langle}
\newcommand{\ra}{\rangle}

\title{Towards the Gravituhedron: New Expressions for NMHV Gravity Amplitudes}

\author{Jaroslav Trnka}

\affiliation{Department of Physics $\&$ Center for Quantum Mathematics and Physics (QMAP),\\ 
University of California, Davis, CA 95616, USA}

\emailAdd{trnka@ucdavis.edu}

\preprint{}
\abstract{In this paper, we present new expressions for $n$-point NMHV tree-level gravity amplitudes. We introduce a method of factorization diagrams which is a simple graphical representation of $R$-invariants in Yang-Mills theory. We define the gravity analogues which we call ${\cal G}$-invariants, and expand the NMHV gravity amplitudes in terms of these objects. We provide explicit formulas of NMHV gravity amplitudes up to eight points in terms of ${\cal G}$-invariants, and give the general definition for any number of points. We discuss the connection to BCFW representation, special behavior under large momentum shift, the role of momentum twistors and the intricate web of spurious poles cancelation. Because of the close connection between $R$-invariants and the (tree-level) Amplituhedron for Yang-Mills amplitudes, we speculate that the new expansion for gravity amplitudes should correspond to the triangulation of the putative Gravituhedron geometry.}

\begin{document}
\maketitle

\section{Introduction}

In recent years we have seen some remarkable progress in the study of perturbative scattering amplitudes in quantum field theory. One of the most surprising developments is the connection between the particle scattering and the positive geometry \cite{Arkani-Hamed:2013jha,Arkani-Hamed:2017tmz}. The primary example is the Amplituhedron \cite{Arkani-Hamed:2013jha,Arkani-Hamed:2017vfh}, which reproduces tree-level amplitudes and loop integrands in planar ${\cal N}=4$ maximal supersymmetric Yang-Mills theory. The perturbative amplitudes correspond to differential forms with logarithmic singularities on the boundaries of the Amplituhedron. This provides a completely new formulation of the perturbative S-matrix where the mathematical principles are primary and the physical properties, such as locality and unitarity, follow from the intriguing geometry. The discovery of the Amplituhedron was based on earlier polytope picture for NMHV tree-level amplitudes \cite{Hodges:2009hk,ArkaniHamed:2010gg}, loop level recursion relations \cite{ArkaniHamed:2010kv} and the positive Grassmannian \cite{ArkaniHamed:2009dn,ArkaniHamed:2009vw,Mason:2009qx,ArkaniHamed:2009sx,ArkaniHamed:2009dg,Arkani-Hamed:2016byb}. The geometry of Amplituhedron has been studied extensively in last years, including the boundary structure \cite{Franco:2014csa, Lukowski:2019kqi, Lukowski:2020bya, Ferro:2020lgp, Prlina:2017tvx, Dennen:2016mdk,Galloni:2016iuj}, triangulations \cite{Kojima:2020tjf,Kojima:2020gxs}, Yangian invariance \cite{Ferro:2016zmx,Ferro:2016ptt} and more rigorous mathematical understanding of the geometry \cite{Lam:2014jda,Karp:2016uax,Karp:2017ouj,Galashin:2018fri,Lukowski:2020dpn,Mohammadi:2020plf}. The geometric picture has been used to provide some all-loop results which could not be obtained using standard methods \cite{Arkani-Hamed:2018rsk,Langer:2019iuo,An:2017tbf}. The important open question is the existence of the dual Amplituhedron
geometry \cite{Arkani-Hamed:2014dca,Ferro:2015grk,Herrmann:2020qlt}; in this picture the scattering amplitude should be given by a volume rather than a differential form. 

The crucial question is how general the positive geometry picture for scattering amplitudes is and if such mathematical structures underlie other quantum field theories, possibly leading to a complete reformulation of perturbative quantum field theory. In last several years, there have been a huge progress in searching for positive geometries for scalar theories. It was discovered that Associahedron underlies tree-level amplitudes in bi-adjoint $\phi^3$ theory \cite{Arkani-Hamed:2017mur} (see also \cite{He:2018pue,delaCruz:2017zqr,Frost:2018djd,Bazier-Matte:2018rat}) and one-loop generalization \cite{Salvatori:2018aha,Arkani-Hamed:2019vag}. Some progress has been made for more general $\phi^p$ scalar theories
\cite{Banerjee:2018tun,Aneesh:2019cvt,Jagadale:2020qfa,John:2020jww,Kojima:2020tox,
Kalyanapuram:2020vil,Aneesh:2019ddi,Herderschee:2019wtl}. Positive geometry was tied to certain one-loop integrands \cite{Herrmann:2020oud}, string amplitudes \cite{Arkani-Hamed:2019mrd,Huang:2020nqy,He:2020onr,He:2020ray,Herderschee:2020lgb} as well as cosmological \cite{Arkani-Hamed:2017fdk,Benincasa:2019vqr} or CFT correlators \cite{Arkani-Hamed:2018ign}. 

The Amplituhedron was originally defined as a certain map from the positive Grassmannian, but it was later reformulated directly in the kinematical space using topological sign flip conditions. In this picture scattering amplitudes are differential forms living directly in kinematical (momentum twistor) space \cite{Arkani-Hamed:2017tmz}. This ideas have been later used to study more generally the kinematical differential forms \cite{He:2018okq,He:2018svj} and it also lead to the discovery of the Momentum Amplituhedron \cite{Damgaard:2019ztj,Damgaard:2020eox} in the spinor helicity space. The immediate big question is if the non-planar ${\cal N}=4$ SYM amplitudes can be also associated with canonical differential forms over geometric space. We know that these amplitudes enjoy many special properties similar to their planar counterparts \cite{Arkani-Hamed:2014via,Arkani-Hamed:2014bca,Bern:2014kca,Bern:2015ple,Bourjaily:2016mnp,Bourjaily:2018omh,
Bourjaily:2019iqr,Bourjaily:2019gqu,Bern:2018oao,Bern:2017gdk} but an explicit geometric construction is still to be found.

In this paper, we focus on four-dimensional on-shell graviton scattering amplitudes, and initiate the search for the positive geometry which we tentatively call {\it Gravituhedron}. The geometric space should capture all poles and singularities of tree-level gravity scattering amplitudes while associated differential forms should reproduce the explicit expressions. Before going into specific objectives we review some known (and surprising) aspects of tree-level graviton amplitudes, and how they fit in our new picture.

\subsection*{Mysteries of gravity amplitudes}

Graviton scattering amplitudes have been studied extensively in last several decades and these efforts uncovered many interesting properties which are yet to be completely explained. There are two main themes: double copy, and large momentum behavior.  

\subsection*{$\clubsuit$ Double copy}

The KLT relations between open string and closed string amplitudes \cite{Kawai:1985xq} imply the relations between gluon and graviton amplitudes: the $n$-pt tree-level graviton amplitude ${\cal M}_n$ can be expressed as a ``square" of gluon amplitudes ${\cal A}_n$. The square stands for a particular linear combination of products of color-ordered gluon amplitudes ${\cal A}_n$ with different cyclic orderings multiplied by kinematical factor ${\cal S}(\alpha,\beta)$ called KLT kernel,
\begin{equation}
 {\cal M}_n = \sum_{\alpha,\beta} {\cal A}_n(\alpha) {\cal S}(\alpha,\beta) {\cal A}_n(\beta) 
\end{equation}
where the sum is over certain permutations $\alpha,\beta$ of external labels. This concept has been further explored directly in the context of Feynman diagram expansion known as famous BCJ relations or color-kinematics duality \cite{Bern:2008qj,Bern:2010ue,Bern:2019prr}. The full color-dressed gluon amplitude can be expressed as a sum of cubic graphs,
\begin{equation}
    {\cal A}_n = \sum_{\Gamma} \frac{n_kc_k}{D_k}
\end{equation}
where $D_k$ stands for the set of Feynman propagators associated with a single cubic graph, $c_k$ is the color factor and $n_k$ is a kinematical numerator. The numerators $n_k$ satisfy a 3-term identity if the corresponding color factors satisfy the Jacobi identity,
\begin{equation}
    c_i+c_j=c_k\,\,\,\rightarrow n_i+n_j=n_k
\end{equation}
where $i,j,k$ labels the trio of cubic graphs. Once we find such numerators $n_i$, the $n$-pt graviton tree-level amplitude can be then written as 
\begin{equation}
    {\cal M}_n = \sum_{\Gamma} \frac{n_k^2}{D_k}
\end{equation}
This ``squaring" rule is not limited just to gluon and graviton amplitudes but it has a vast generalization to supersymmetric theories (take two different gauge theory numerators $n_k$ and $\widetilde{n_k}$) and the web of relations between various other quantum field theories (NLSM, Born-Infeld theory, Galileons..). The double copy relations are manifest in the CHY formalism \cite{Cachazo:2013iea,Cachazo:2013hca,Cachazo:2014xea} and also appear in the context of celestial amplitudes \cite{Casali:2020uvr,Casali:2020vuy}.

The color-kinematics duality has been successfully used to construct multi-loop gravity integrands \cite{Bern:2012uf,Bern:2017yxu,Bern:2018jmv}, and there many intriguing results when using the double copy idea for the classical solutions in Yang-Mills theory and gravity \cite{Monteiro:2014cda,Luna:2016hge,Kosower:2018adc}, as well as powerful amplitudes techniques used for gravitational wave physics \cite{Bern:2019crd,Bern:2019nnu,Bern:2020uwk}.

\subsection*{$\clubsuit$ Large momentum scaling} 

The four-dimensinal gravity tree-level amplitudes can be constructed using BCFW on-shell recursion relations \cite{Britto:2004ap,Britto:2005fq}. In the spinor helicity formalism each momentum is represented by two spinors as $p_\mu = \sigma_{ab}^\mu\lambda_a\widetilde{\lambda}_b$. We shift two of the spinors using the complex parameter $z$ using $\{i,j\}$ shift,
\begin{equation}
    \{i,j\}\equiv \{\widetilde{\lambda}_i - z\widetilde{\lambda}_j,\lambda_j+z\lambda_i\} \label{shift}
\end{equation}
The Cauchy formula for the shifted amplitude $A_n(z)$ allows us to reconstruct the original amplitude $A_n=A_n(0)$ in terms of residues of $A_n(z)$ which are products of lower point amplitudes,
\begin{equation}
    \oint dz\frac{A_n(z)}{z} = 0 \quad \rightarrow\quad
    A_n = -\sum_k \frac{A_L(z^\ast)A_R(z^\ast)}{P_k^2} 
\end{equation}
where $P_k$ is a sum of external momenta and $P_k(z^\ast)^2=0$. The sub-amplitudes $A_L$ and $A_R$ are evaluated at shifted kinematics. This works if the shifted amplitude $A_n(z)$ vanishes at infinity $z\rightarrow\infty$, otherwise there are boundary terms. For some specific choices of shifts (aligned with helicities) gluon amplitudes scale as $A_n(z)\sim {\cal O}\left(\frac{1}{z}\right)$ and the condition is exactly satisfied. Gravity amplitudes vanish even faster \cite{Cachazo:2005ca,Bedford:2005yy,ArkaniHamed:2008yf},
\begin{equation}
    {\cal M}_n(z) = {\cal O}\left(\frac{1}{z^2}\right)\quad\mbox{for}\quad z\rightarrow\infty\label{largeZ}
\end{equation}
This is especially surprising given the high powercounting of gravity amplitudes. From the Lagrangian point of view the Yang-Mills cubic vertex has one derivative $A^2(\partial A)$ while the gravity vertex has two derivatives $h(\partial h)^2$. This would naively suggest that the dependence on the shifted momenta in the numerator of gravity amplitudes is much stronger and ${\cal M}_n(z)$ would grow much faster for $z\rightarrow\infty$ than for Yang-Mills amplitudes. However, the opposite is the case. While the scaling (\ref{largeZ}) has been proven in \cite{Cachazo:2005ca,Bedford:2005yy,ArkaniHamed:2008yf}, the deeper understanding is still missing. It is worth mentioning that the large-$z$ behavior of gluon amplitudes is closely related to the dual conformal symmetry \cite{Drummond:2008vq,Drummond:2009fd}, while for gravity we do not have any symmetry explanation.
Surprising large momentum behavior has been also observed in the structure of cuts of gravity loop integrands \cite{Herrmann:2016qea,Herrmann:2018dja,Bourjaily:2018omh,Edison:2019ovj}. While in ${\cal N}=4$ SYM there are no poles at infinity for $\ell\rightarrow\infty$, the gravity integrand generally do have these poles due to higher powercounting. Because of the lack of non-planar labels and the unique non-planar integrands, the poles at infinity can only be tested on cuts where all loop momenta are cut. The cut loop integrand is then given by products of tree-level amplitudes, and it is an on-shell gauge-invariant object. On the cut we can scale one or more remaining loop parameters $z$ to infinity, which sends the cut loop momenta to $\ell_k\rightarrow\infty$. 
\begin{equation}
    {\cal M}_n = {\cal O}(z^{p-2}),\qquad z\rightarrow\infty\quad \mbox{for which}\quad \ell_k\rightarrow\infty.
\end{equation}
${\cal N}=4$ SYM amplitudes have no such poles, $p\leq0$, while in gravity these poles are present and are naively controlled by the ``worst" behaved diagram contributing on a given cut. However, in many instances there are surprising cancelations between individual diagrams when evaluated at $\ell_k\rightarrow\infty$, and the amplitude exhibits an improved behavior. This first happens at two-loops when the scaling at infinity on the multi-unitarity cut has an improved behavior with respect to the double-box diagrams,
\vspace{-0.3cm}
$$
\includegraphics[scale=0.8]{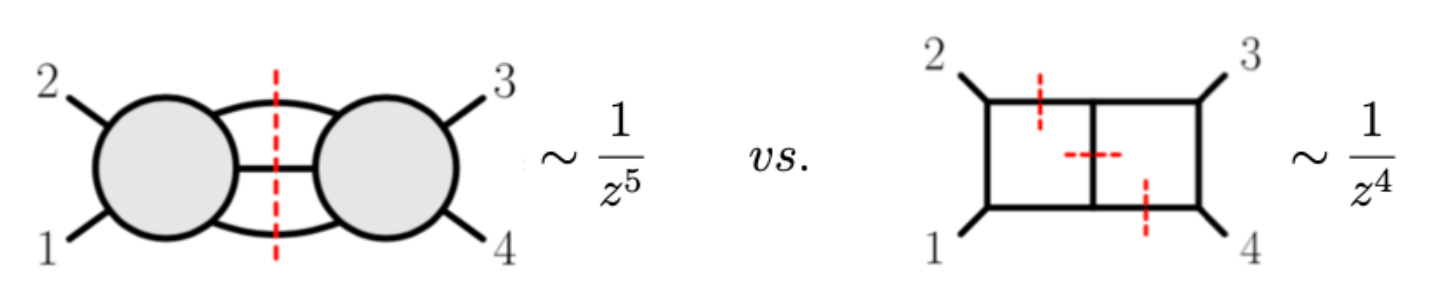}
$$
At higher-loops the improved behavior forces intricate cancelations between many diagrams. This property seems similar in nature to the improved large-$z$ behavior of tree-level amplitudes under the BCFW shift. In fact, it is equivalent to certain large-$z$ scaling of tree-level amplitudes under a particular multi-line shift (which then feeds in the cuts). This suggests that the BCFW scaling is not an isolated property but rather a special case of more general story of how the graviton amplitudes scale at infinity. Interestingly, the improved scaling at infinity for loop integrands is generally only present for $D=4$, and crucially depends on spinor-helicity formalism and the permutational symmetry of graviton amplitudes. In a different open problem it was suggested that ${\cal N}=8$ supergravity amplitudes might be UV finite for $D=4$ (as critical dimension) which would require cancelations of UV divergencies between various diagrams \cite{Bjornsson:2010wm,Kallosh:2008rr,Beisert:2010jx,Bern:2018jmv,Bern:2007xj,Bern:2017lpv,Bern:2006kd} known as enhanced cancelations. We have seen similar cancelations on cuts, but it is not clear yet how these two problems are directly related. 

\subsection*{$\clubsuit$ Towards the Gravituhedron} 

In this paper we focus on four-dimensional tree-level graviton amplitudes using the spinor helicity formalism, and will closely follow the lessons we learnt from the tree-level Amplituhedron for gluon amplitudes. The ultimate goal is to construct the Gravituhedron space and find the appropriate differential form to extract the scattering amplitudes from geometric data. This is a difficult task as the graviton amplitudes are notoriously complicated, and there are many new singularities and kinematical structures that need to be captured by the geometry. Furthermore, it is clear that the logarithmic forms which are attached to Amplituhedron and Associahedron geometries to reproduce amplitudes, must be replaced by something else. Graviton amplitudes have double poles in soft limits, which is obvious e.g. from gravity soft-factors (\ref{GRsoft}), when $\lambda_n\rightarrow0$, and the singularities are not logarithmic. Rather than attacking these hard problems directly -- searching for geometry and differential form -- we take a somewhat different route in this paper. We focus on NMHV tree-level amplitudes, which played a crucial role in the discovery of Amplituhedron, and find a new formula for which suggests analogous geometric interpretation.
$$
\includegraphics[scale=0.27]{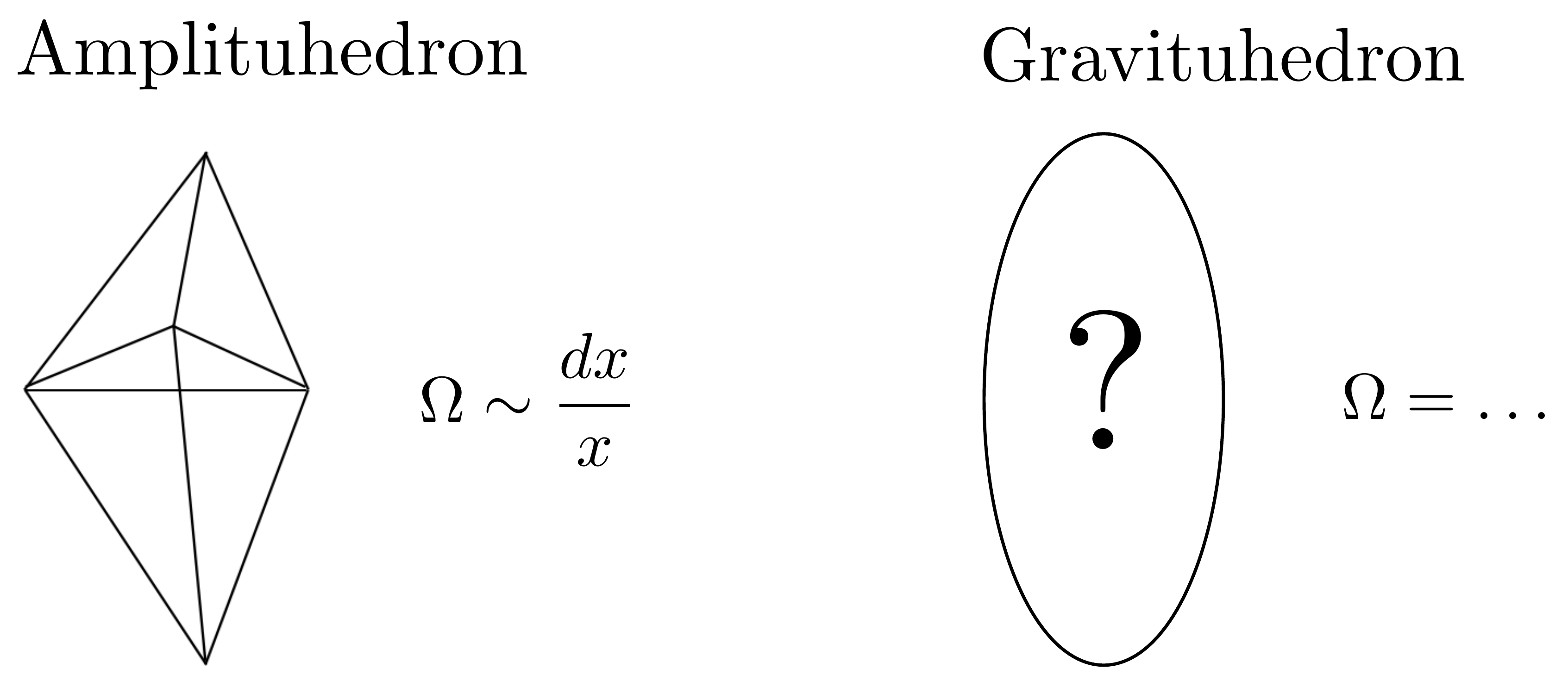}
$$

\vspace{-0.3cm}

First, we reverse-engineer the gluon amplitudes calculation in terms of $R$-invariants, and interpret individual terms graphically as new building blocks which we call {\it factorization diagrams}. The same diagrams can be defined in the context of gravity, and we use them to write down the formula for a general $n$-pt NMHV graviton amplitudes. This formula exhibits some remarkable simplicity and makes manifest the permutational symmetry $S_3\times S_{n{-}3}$ in external legs (in 3 positive and $(n{-}3)$ negative gravitons). Most importantly, the formulas for factorization diagrams are very reminiscent of $R$-invariants for gluons, and we call these objects {\it ${\cal G}$-invariants} suggesting an existence of the hidden symmetry analogous to the dual conformal symmetry. We conjecture that our formula corresponds to a natural triangulation of yet-to-be found Gravituhedron space, similar to the $R$-invariant triangulation of the Amplituhedron. We leave the direct attempts to search for new symmetries and the Gravituhedron geometry using our new formula for future work.

\vspace{0.2cm}

\noindent {\bf Organization of this paper:} In section 2, we review known expressions and facts about tree-level graviton amplitudes. In section 3, we take the representation of the gluon amplitudes in terms of $R$-invariants and interpret it using factorization diagrams. In section 4, we conjecture the new formula for $n$-pt NMHV graviton formula, and provide explicit expressions for six, seven and eight-point amplitudes. In section 5, we discuss various properties of our result such as large-$z$ behavior or spurious poles cancelation. In section 6, we provide a super-symmetric formula for the six-point NMHV amplitude and discuss some new features. In the Outlook, we outline next steps in the search for the Gravituhedron geometry.

\section{Review of gravity tree-level amplitudes}

The central object of our interest is the $n$-pt tree-level  N$^{k-2}$MHV graviton amplitude with the following helicity configuration,
\begin{equation}
    {\cal M}_n(1^-2^-\dots k^-(k{+}1)^+\dots n^+).
\end{equation}
Under the little group scaling $\lambda_i\rightarrow t\lambda_i$, $\widetilde{\lambda}_i\rightarrow \frac{1}{t}\widetilde{\lambda}_i$ the amplitude scales as ${\cal M}_n \rightarrow t^4{\cal M}_n$ if $i$ corresponds to the negative helicity graviton, and ${\cal M}_n \rightarrow \frac{1}{t^4}{\cal M}_n$ for positive helicity gravitons. The amplitudes are invariant under $S_k \otimes S_{n-k}$ symmetry, which is permutational symmetry in all negative, resp. positive helicity labels. At 3pt amplitudes are completely fixed by little group weights, and we get two independent objects,
\begin{equation}
    {\cal M}_3(1^-2^-3^+) = \frac{\la12\ra^6}{\la23\ra^2\la13\ra^2},\qquad {\cal M}_3(1^-2^+3^+) = \frac{[23]^6}{[12]^2[13]^2},
\end{equation}
The all-plus and all-minus amplitudes have higher powercounting and are generated by $R^3$ term. There is one independent 4pt amplitude,
\begin{equation}
    {\cal M}_4(1^-2^-3^+4^+) = \frac{\la 12\ra^4[34]^4}{stu} = \frac{\la 12\ra^6[34]}{\la 13\ra\la14\ra\la23\ra\la24\ra\la34\ra}
\end{equation}
where $[34]/\la12\ra$ has permutation symmetry in all four labels (due to restricted 4pt kinematics) which allows to write the 4pt amplitude in many equivalent forms. At 5pt we have two helicity configurations, MHV and NMHV, but they are related by parity conjugation $h^\pm\leftrightarrow h^\mp$ and $\la ij\ra \leftrightarrow [ij]$, and there is only one independent expression,
\begin{equation}
    {\cal M}_5(1^-2^-3^+4^+5^+) = \frac{\la 12\ra^7\,{\rm Tr}(1234)}{\la13\ra\la14\ra\la15\ra\la23\ra\la24\ra\la25\ra\la34\ra\la35\ra\la45\ra} \label{GR5}
\end{equation}
where the trace is a shortcut for
\begin{equation}
    {\rm Tr}(1234) \equiv {\rm Tr}(\slashed{p_1}\slashed{p_2}\slashed{p_3}\slashed{p_4}) =  \la12\ra[23]\la34\ra[41]-[12]\la23\ra[34]\la41\ra \label{GR5b}
\end{equation}
Note that the trace is completely antisymmetric in all five labels, e.g. ${\rm Tr}(1234) = -{\rm Tr}(1235)$ etc. At six point and higher there are more independent helicity amplitudes.

\medskip

\noindent The simplest helicity sector corresponds to MHV amplitudes with two negative and $(n{-}2)$ positive helicity gravitons. MHV amplitudes have been studied very extensively in the past leading to a number of interesting expressions. This was motivated by the discovery of the famous Parke-Taylor factor for the $n$-pt MHV gluon amplitudes \cite{Parke:1986gb},
\begin{equation}
    {\cal A}_n(1^-2^-3^+\dots n^+) = \frac{\la12\ra^3}{\la23\ra\la34\ra\dots \la n1\ra}
\end{equation}
which is stunningly simple and was the first evidence for hidden structures in the perturbative S-matrix. For the MHV amplitude we can always factor out the helicity factor $\la ij\ra^4$ where $i,j$ are negative helicity gluons,
\begin{equation}
    {\cal A}_n(1^-2^-3^+\dots n^+) = \la12\ra^4\times A_n \qquad\mbox{where}\qquad A_n = \frac{1}{\la12\ra\la23\ra\dots \la n1\ra}\label{MHV2}
\end{equation}
where $A_n$ is cyclically invariant. In maximally supersymmetric (${\cal N}=4$) Yang-Mills theory, the full superamplitude is then given by \cite{Nair:1988bq},
\begin{equation}
    {\cal A}_n = \frac{\delta^{8}(Q)}{\la12\ra\la23\ra\dots \la n1\ra} \label{MHVsuper}
\end{equation}
where the super-momentum delta function is 
\begin{equation}
    \delta^{8}(Q)\equiv \delta^{8}(Q_{aI}) = \delta^{8}\left(\sum_{k=1}^n \lambda_{ak}\eta_{Ik}\right)\label{delta}
\end{equation}
with $a=1,2$ and $I=1,2,3,4$. At tree-level, the gluon amplitudes in pure Yang-Mills theory are identical to ${\cal N}=4$ SYM gluon amplitudes. The super-amplitudes enjoy a complete cyclic symmetry which is absent in helicity amplitudes due to the choice of negative helicity labels. Note that the simple factorization of helicity factor (\ref{MHV2}) is specific to MHV sector and for higher $k$-degree amplitudes it gets more complicated. Similarly, for MHV graviton amplitudes we can write
\begin{equation}
    {\cal M}_n(1^-2^-3^+\dots n^+) = \la12\ra^8\times M_n
\end{equation}
where $M_n$ has the complete permutational $S_n$ symmetry. In ${\cal N}=8$ supergravity the MHV amplitude is simply obtained by
\begin{equation}
    {\cal M}_n = \delta^{16}(Q)\times M_n
\end{equation}
where the super-momentum conserving delta function is the same (\ref{delta}), just $I=1,2,\dots 8$. In the following, we review various MHV graviton amplitudes available in the literature and put them in the context of our work.

\subsection*{BGK formula}

The closed form for the $n$-pt MHV graviton amplitude was first discovered in 1988 by Berends, Giele and Kujf \cite{Berends:1988zp} using off-shell recursion relations,
\begin{equation}
    M_n = \sum_{{\cal P}(2,\dots,n{-}2)}\frac{[12][n{-}2\,n{-}1]}{\la 1\,n{-}1\ra\la 2n\ra\la n{-}2\,n\ra}\frac{1}{\la12\ra\dots \la n1\ra}\prod_{k=3}^{n{-}3}\frac{[k|p_{k{+}1}{+}\dots {+} p_{n{-}1}|n\ra}{\la kn\ra}\label{BGK}
\end{equation}
where ${\cal P}(\dots)$ stands for the sum over permutations. The BGK formula has some notable features: (i) there are only holomorphic poles $\la ij\ra$; (ii) there is a non-trivial numerator which depends both on both holomorphic and anti-holomorphic brackets $\la ij\ra$ and $[ij]$; (iii) three legs are chosen as special and the sum is over permutations of remaining $(n{-}3)$ legs gives $(n{-}3)!$ terms. This formula was later proved to be equivalent to a more symmetric Mason-Skinner formula \cite{Mason:2008jy}
\begin{equation}
    M_n = \sum_{{\cal P}(1,\dots,n{-}3)}\frac{1}{\la n\,n{-}2\ra\la n{-}2\,n{-}1\ra\la n{-}1\,n\ra}\frac{1}{\la12\ra\dots \la n1\ra}\prod_{k=1}^{n{-}3}\frac{[k|p_{k{+}1}{+}\dots{+}p_{n{-}2}|n{-}1\ra}{\la k\,n{-}1\ra} \label{Mason}
 \end{equation}
Note that both (\ref{BGK}) and (\ref{Mason}) exhibit $S_{n{-}3}$ symmetry in labels $1,\dots,n{-}3$, which is a subset of full $S_n$ permutational symmetry of $M_n$.

\subsection*{Inverse soft factor formula}

The Parke-Taylor formula can be derived in many different ways, one particular construction uses the soft factors. The soft limit $p_n\rightarrow0$ of the color-ordered gluon amplitude ${\cal A}_n$ for positive helicity particle $n^+$ is
\begin{equation}
    {\cal A}_n \xrightarrow[]{p_n\rightarrow0} {\cal S}^{(n{-}1,1)}_n\times {\cal A}_{n{-}1}\quad \mbox{where} \quad {\cal S}^{(n{-}1,1)}_n = \frac{\la n{-1}\,1\ra}{\la n{-}1\,n\ra\la n\,1\ra}
\end{equation}
where ${\cal S}^{(n{-}1,1)}_n$ is a soft factor for the removal of the leg $n$ between legs $n{-}1$ and $1$. For MHV amplitudes this is a simple reorganization of terms in the Parke-Taylor formula,
\begin{equation}
    \frac{1}{\la 12\ra\la 23\ra\dots \la n{-}2\,n{-}1\ra\la n{-}1\,n\ra\la n\,1\ra} = \frac{\la n{-}1\,1\ra}{\la n{-}1\,n\ra\la n\,1\ra}\times \frac{1}{\la 12\ra\la 23\ra\dots \la n{-}2\,n{-}1\ra\la n{-}1\,1\ra} 
\end{equation}
In the opposite direction, we can build the $n$-pt MHV amplitude by multiplying the three-point amplitude by the soft factors ${\cal S}_n$, this is also called \emph{inverse-soft factor} construction. It is an iterative process where we add particle $n$ between $n{-}1$ and $1$ in the $(n{-}1)$-pt amplitude, starting from 3pt amplitude,
\begin{equation}
    {\cal A}_n(1^-2^-3^+\dots n^+)  =   {\cal S}^{(n{-}1,1)}_n \times {\cal S}^{(n{-}2,1)}_{n{-}1}\times \dots {\cal S}^{(3,1)}_4 \times {\cal A}_{3}(1^-2^-3^+)
\end{equation}
The same procedure can be used in gravity. The soft limit $p_n\rightarrow0$ of the positive helicity $n^+$ particle in the $n$-pt amplitude is given by
\begin{equation}
    {\cal M}_n \xrightarrow[]{p_n\rightarrow0} {\cal S}_n \times {\cal M}_{n{-}1} \quad\mbox{where}\quad {\cal S}_n = \sum_{k=1}^{n{-}1}\frac{[kn]\la ka\ra\la kb\ra}{\la an\ra\la bn\ra\la kn\ra} \label{GRsoft}
\end{equation}
where ${\cal S}_n$ is a gravity soft factor \cite{Weinberg:1965nx}, and $\lambda_a,\lambda_b$ are arbitrary reference spinors. We can use the same idea of inverse soft factors to construct the general $n$-pt MHV graviton amplitude. The practical implementation is different from the gluon case as we can ``add" particle $n$ anywhere, not necessarily between $n{-}1$ and $1$ (as there is no cyclic symmetry). It was first discovered in \cite{Bern:1998sv} using ``half-soft" factors, and later reformulated diagrammatically \cite{Nguyen:2009jk},
\begin{equation}
   M_n = \frac{1}{\la n{-}1n\ra}\sum_{\rm trees}\left(\prod_{{\rm edges}\,a,b}\frac{[ab]}{\la ab\ra}\right)\left(\prod_{{\rm vertices}\,a}(\la a n{-}1\ra\la an\ra)^{{\rm deg}\,a-2}\right)\label{MHV3}
\end{equation}
where $a$ is the degree of the vertex and we sum over all ``tree graphs" which consist of vertices of any multiplicity labeled by $1,2,\dots,n{-}2$. One of the six-point tree graphs is 
$$
\includegraphics[scale=0.92]{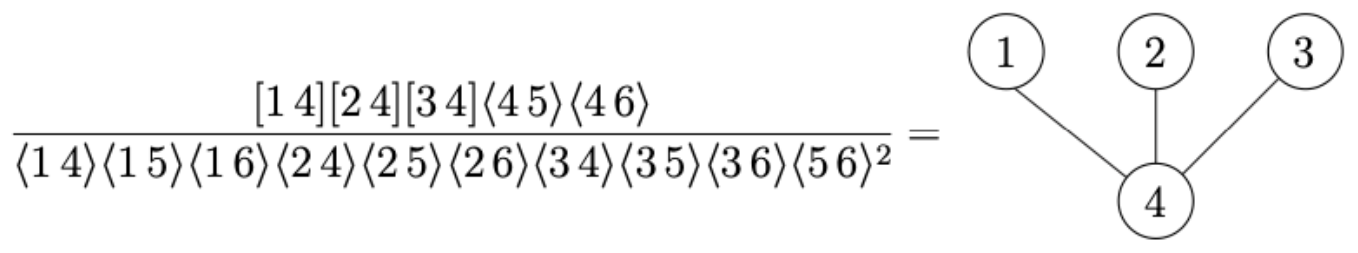}
$$
which is one of the terms in (\ref{MHV3}) for $M_6$. The representation (\ref{MHV3}) manifests the $S_{n{-}2}$ subgroup of the full permutational group $S_n$.

\subsection*{BCFW formulas}

As pointed out earlier, the graviton amplitudes can be also calculated using BCFW recursion relations. The crucial property is the improved ${\cal O}(1/z^2)$ behavior at infinity which allows to reconstruct the amplitude from the factorization channels. In the context of MHV amplitudes the BCFW calculation was first calculated in \cite{Cachazo:2005ca,Bedford:2005yy} and the general $n$-pt formula takes the form
\begin{equation}
    M_n = \sum_{ {\cal P}(3,4,\dots,n)}\frac{1}{\la12\ra^2}\frac{[1n]}{\la 1n\ra}\frac{[34]}{\la23\ra\la24\ra\la34\ra\la35\ra\la45\ra}\prod_{s=5}^{n{-}1}\frac{\la2|3{+}\dots{+}(s{-}1)|s]}{\la s\,s{+}1\ra\la 2\,s{+}1\ra} 
\end{equation}
where we sum over all permutations of labels $3,4,\dots,n$ modulo the symmetry in legs $3\leftrightarrow4$. This formula is based on the ${\cal N}=8$ recursion relations, a different formula using ${\cal N}=7$ recursion was provided in \cite{Hodges:2011wm}. 

The recursion relations have been also studied in the context of gravity on-shell diagrams \cite{Herrmann:2016qea,Heslop:2016plj,Farrow:2017eol,Armstrong:2020ljm}. In the Yang-Mills case, the on-shell diagrams directly serve as building blocks for tree-level amplitudes and BCFW recursion relations are implemented diagrammatically \cite{Arkani-Hamed:2016byb}. In gravity the due to the dimensionality of the coupling constant on-shell diagrams (calculated as cuts of loop integrands and given by the products of tree amplitudes) must be decorated by additional kinematical factors to be used for tree-level amplitudes via BCFW recursion relations \cite{Heslop:2016plj,Armstrong:2020ljm},

Inspired by both the KLT formula and the BCFW recursion relations, Elvang and Freedman found another MHV formula \cite{Elvang:2007sg}
\begin{equation}
    M_n = \sum_{{\cal P}(2,\dots,n{-}1)} \left[A_n(1,2,\dots,n)\right]^2\times s_{12} \prod_{s=2}^{n{-}3}\frac{\la s|(s{+}1)(s{+}2\dots n{-}1)|n\ra}{\la sn\ra} \label{Elvang}
\end{equation}
where each term contains the square of the Yang-Mills ordered amplitude $A_n(1,2,\dots,n)$, and the sum is over permutation of $n{-}2$ labels making $S_{n{-}2}$ manifest. This formula served as a motivation for the explicit solution of BCFW recursion relations for any N$^k$MHV amplitude which takes similar form \cite{Drummond:2009ge}.

\subsection*{Hodges formula}

In 2012, Andrew Hodges discovered a remarkable simple expression for MHV amplitudes with manifest $S_n$ permutational symmetry \cite{Hodges:2012ym},
\begin{equation}
    M_n = \frac{|\Phi_{abc}^{def}|}{(abc)(def)}\label{Hodges}
\end{equation}
where we denoted $(abc)\equiv \la ab\ra\la bc\ra\la ca\ra$, the bracket $|\dots|$ stands for determinant and $\Phi_{abc}^{def}$ is the $(n{-}3\times n{-}3)$ matrix which is obtained from the $(n\times n)$ matrix $\Phi$ with rows $a,b,c$ and columns $d,e,f$ removed. The elements of the $\Phi$ matrix are defined as
\begin{equation}
    \Phi_{i,j} = \frac{[ij]}{\la ij\ra},\qquad \Phi_{i,i} = - \sum_{j\neq i} \frac{[ij]\la ja\ra\la jb\ra}{\la ij\ra\la ia\ra\la ib\ra}\label{Hodges2}
\end{equation}
where $\lambda_a,\lambda_b$ are arbitrary reference spinors (and the result does not depend on them). Note that the diagonal element $\Phi_{i,i}$ is a gravitational soft factor. The stunning simplicity and manifest symmetry suggests it is a direct analogue of the Parke-Taylor formula. The formula has been used to find the twistor strings construction for ${\cal N}=8$ supergravity \cite{Cachazo:2012pz,Skinner:2013xp}, and can be interpreted in the context of ambitwistor strings \cite{Geyer:2014fka}. The geometric origin of (\ref{Hodges}) is still a mystery. 

\subsection*{NMHV amplitudes}

There are only few explicit formulas for NMHV gravity amplitudes in the literature, and most of them are based on the BCFW recursion relations. In the original paper on the recursion in gravity, Cachazo and Svrcek \cite{Cachazo:2005ca} calculated the 6pt NMHV amplitude ${\cal M}_6(1^-2^-3^-4^+5^+6^+)$ using the $\{3,4\}$ shift, 
\begin{align*}
    D_1 &= \frac{\la23\ra\la1|2+3|4]^7(\la1|2+3|4]\la5|3+4|2][51]+[12][45]\la51\ra s_{234})}{s_{234}[23][34]^2\la16\ra\la36\ra\la1|3+4|2]\la5|3+4|2]\la5|2+3|4]\la6|3+4|2]\la6|2+3|4]} \\
    D_2 &=\frac{\la13\ra^7\la25\ra[45]^7[16]}{s_{245}[24][25]\la16\ra\la36\ra\la1|2+5|4]\la6|2+5|4]\la3|1+6|5]\la3|1+6|2]}\\
    D_3 &=\frac{\la13\ra^8[56]^7[14](\la23\ra\la56\ra[62]\la1|3+4|5]+\la35\ra[56]\la62\ra\la1|3+4|2])}{s_{134}[25][26]\la14\ra\la34\ra\la1|3+4|2]\la1|3+4|5]\la1|3+4|6]\la3|1+4|2]\la3|1+4|5]\la3|1+4|6]}\\
    D_6 &= \frac{\la12\ra[56]\la3|1+2|4]^8}{s_{124}[21][14][24]\la35\ra\la36\ra\la56\ra\la5|1+2|4]\la6|1+2|4]\la3|5+6|1]\la3|5+6|2]}
\end{align*}
The terms $D_1$ and $D_3$ are further symmetrized in $(1\leftrightarrow2)$, and term $D_2$ is symmetrized in $(1\leftrightarrow2)$ and $(5\leftrightarrow6)$. The amplitude is then given by
\begin{equation}
    {\cal M}_6(1^-2^-3^-4^+5^+6^+) = D_1 + \overline{D_1} + D_2 + D_3 + \overline{D_3} + D_6  \label{Cach}
\end{equation}
where the $\overline{D}$ stands for the flip operation: $\la\ra\leftrightarrow[]$ and $i\leftrightarrow 7-i$. Note some interesting features: non-trivial numerators, double poles and the non-manifest symmetry in external legs due to the choose of BCFW shift. This is generally true for all BCFW expressions. While the final amplitude enjoys the improved ${\cal O}(1/z^2)$ behavior for $z\rightarrow\infty$, the individual terms scale only as ${\cal O}(1/z)$. Andrew Hodges found an alternative formula in \cite{Hodges:2011wm} using the same shift but in the context of ${\cal N}=7$ recursion. The individual pieces are:
\begin{align*}
    M_{356}&= \frac{\la12\ra^7[56]^7[12]\la56\ra}{s_{124}[35][36]\la14\ra\la24\ra\la1|2+4|3]\la2|1+4|3]\la4|1+2|5]\la4|1+2|6]}\\
    M_{315}&= \frac{\la2|4+6|5]^7\la15\ra[26]}{s_{135}[13][15][35]\la24\ra\la26\ra\la46\ra\la4|3+5|1]\la6|1+5|3]\la4|1+5|3]}\\
    M_{342}&= \frac{\la1|2+3|4]^7\la24\ra}{s_{234}[23][24][34]\la56\ra\la1|2+4|3]\la5|2+4|3]\la6|2+4|3]}\hspace{-0.1cm}\times\hspace{-0.1cm}\Bigg\{\hspace{-0.1cm}\frac{[15][36]}{\la15\ra\la6|3+4|2]} \hspace{-0.1cm}-\hspace{-0.1cm} \frac{[16][35]}{\la16\ra\la5|3+4|2]}\hspace{-0.1cm}\Bigg\}
\end{align*}
We symmetrize $M_{342}$ in $(1\leftrightarrow2)$, and $M_{315}$ in both $(1\leftrightarrow2)$ and $(5\leftrightarrow6)$. The expression for the six-point NMHV amplitude is then
\begin{equation}
     {\cal M}_6(1^-2^-3^-4^+5^+6^+) = M_{356} + M_{315} + M_{342} + \overline{M_{342}} \label{Hod}
\end{equation}
where the $\overline{M_{342}}$ again denotes the flip operation. Note there are no double poles in this formula and it  manifests ${\cal O}(1/z^2)$ behavior for certain shifts (more on this later). The same expression was also obtained using decorated gravity on-shell diagrams \cite{Armstrong:2020ljm} and momentum twistor variables (for a particular ordering) were used to expose spurious poles cancelations.

The ${\cal N}=8$ BCFW recursion relations have been also used to write the general solution for $n$-pt NMHV super-amplitudes using the $\{1,n\}$ shift. The result is given by a very compact formula \cite{Drummond:2009ge}, 
\begin{equation}
{\cal M}_n = \sum_{{\cal P}(2,\dots,n{-}1)} [A_n^{\rm MHV}(1,2,\dots,n)]^2 \sum_{i=2}^{n{-}3}\sum_{j=i{+}2}^{n{-}1}R^2_{n;ij}G_{n;ij}^{\rm NMHV} \label{Drum}
\end{equation}
where $A^{\rm MHV}_n$ is a the gluon Parke-Taylor factor, $R_{n;ij}$ is a $R$-invariant we review in the next section and $G^{\rm NMHV}_{n;ij}$ are certain kinematical factors with both numerator and denominator terms. This formula looks reminiscent of the KLT relations and contains squares of Yang-Mills building blocks. In the same paper \cite{Drummond:2009ge} authors provide all-$k$ generalization, the particular formulas are more complicated but follow the general pattern of building the graviton amplitudes from Yang-Mills building blocks multiplied by additional kinematical terms. As with all BCFW expressions also in (\ref{Cach}),(\ref{Hod}),(\ref{Drum}) the choice of $\{a,b\}$ shift makes the permutational symmetries of the amplitude in external labels non-manifest.

NMHV gravity amplitudes were also studied in \cite{Bianchi:2008pu} using the MHV vertex expansion \cite{Cachazo:2004kj} and 3-line shifts. The conclusion is that that for $n>11$ there are non-zero contributions at infinity, and we can not use this method.

\subsection*{BCFW terms, large-$z$ and path forward}

We have reviewed some remarkable progress in the calculation of tree-level graviton amplitudes. The most mysterious expression is the Hodges formula for MHV amplitudes as the closest analogue of Parke-Taylor formula and it is suggestive of deep structures yet to be discovered. If one attempts to directly attack the problem of positive geometry for graviton amplitudes, this formula is likely the best starting point. 

As mentioned earlier, the graviton amplitudes scale as ${\cal O}(1/z^2)$ for $z\rightarrow\infty$ under BCFW shifts. This means that the behavior is better than needed for the reconstruction using Cauchy formula leading to bonus relations \cite{ArkaniHamed:2008gz,Spradlin:2008bu,He:2010ab} between BCFW terms. While this improved behavior is very significant and might be smoking gun for new structures, it also means that the BCFW recursion relations are more flexible. In particular, we can deform the Cauchy formula,
\begin{equation}
    \oint dz\,\frac{A(z)(1+\alpha z)}{z} = 0
\end{equation}
and still reconstruct correctly the amplitude for any value of $\alpha$. This is in contrast with the gluon case where the ${\cal O}(1/z)$ behavior is exactly enough for the BCFW to work and the expansion is much more rigid. 

This has practical implications in the scaling of individual terms under the BCFW shift. For both Yang-Mills and gravity amplitudes there is a set of ``good shifts" $\{a,b\}$ for which amplitudes vanish at $z\rightarrow\infty$. If we line up helicities of external gluons or gravitons with two shifted legs in (\ref{shift}) in the following way,
\begin{equation}
\{a,b\} = \{+,+\},\{-,-\},\{-,+\}
\end{equation}
the amplitude scales as ${\cal O}(1/z)$ for Yang-Mills (for adjacent shifts) and ${\cal O}(1/z^2)$ for gravity. For the ``bad" shift $\{a,b\}=\{+,-\}$ the Yang-Mills amplitudes scale as ${\cal O}(z^3)$ and gravity amplitudes as ${\cal O}(z^6)$ and the shift can not be used for the reconstruction. Let us now take a BCFW formula for helicity amplitudes ${\cal A}_n$ and ${\cal M}_n$ and apply on individual terms a ``good" BCFW shift. In Yang-Mills the individual BCFW terms manifest the ${\cal O}(1/z)$ behavior of the amplitude for any good shift. This is closely related to the manifest dual conformal symmetry. In gravity the individual terms scale only as ${\cal O}(1/z)$ while the amplitude scales as ${\cal O}(1/z^2)$, and therefore they fail to make manifest this improved behavior. If we use ${\cal N}=7$ recursion \cite{Hodges:2011wm} the individual terms scale as ${\cal O}(1/z^2)$ for a subset of shifts but not for others. This suggests that ${\cal N}=7$ recursion is certainly an important part of the ultimate story, but some pieces of the puzzle are missing.

In this paper, we follow another path: we conjecture (and verify up to 10pt) completely new expressions for NMHV graviton amplitudes. Our expansion is analogous to the $R$-invariant representation of NMHV gluon amplitudes \cite{Drummond:2008vq} which does originate from BCFW and can be interpreted geometrically as the triangulation of the NMHV tree-level Amplituhedron. We leave the attempts to use our formula to discover putative Gravituhedron geometry for future work.

\section{Deconstructing gluon amplitudes}

We consider a color-ordered $n$-pt tree-level NMHV amplitude $A_{n,1}$ in ${\cal N}=4$ SYM theory with canonical ordering of external particles $(1,2,\dots,n)$. The dual conformal symmetry dictates that $A_n$ can be written as a product of two terms \cite{Drummond:2008vq},
\begin{equation}
    A_{n,1} = A_{n,0}\times R_{n,1}
\end{equation}
where $A_{n,0}$ is the MHV super-amplitude (\ref{MHVsuper}) and $R_{n,1}$ is invariant under both superconformal and dual superconformal symmetries, hence it is Yangian invariant \cite{Drummond:2009fd}. We can express $R_{n,1}$ as the linear combination of $R$-invariants \cite{Drummond:2008vq},
\begin{equation}
  R_{n,1} = \sum_{i<j} R[\ast,i{-}1,i,j{-}1,j]\label{YM1}
\end{equation}
where $\ast$ is an arbitrary index. The $R$-invariant $R[a,b,c,d,e]$ can be conveniently written using momentum twistors \cite{Hodges:2009hk},
\begin{equation}
R[a,b,c,d,e] = \frac{\delta^4(\eta_a \la bcde\ra + \eta_b \la cdea\ra + \eta_c \la deab\ra + \eta_d \la eabc\ra + \eta_e \la abcd\ra)}{\la abcd\ra\la bcde\ra\la cdea\ra\la deab\ra\la eabc\ra} \label{Rin}
\end{equation}
where $\eta_k$ are momentum twistor Grassmann variables. If we replace $\eta_a\rightarrow dZ_a$ then (\ref{Rin}) is a differential form \cite{Arkani-Hamed:2017vfh} (in momentum twistor space) with logarithmic singularities on the boundaries of the Amplituhedron \cite{Arkani-Hamed:2013jha}. For NMHV case the geometry reduces to a cyclic polytope in $\mathbb{P}^4$. Each $R$-invariant is a logarithmic differential form of a simplex in $\mathbb{P}^4$ where five vertices are given by five labels in $R[a,b,c,d,e]$,
$$
\includegraphics[scale=0.5]{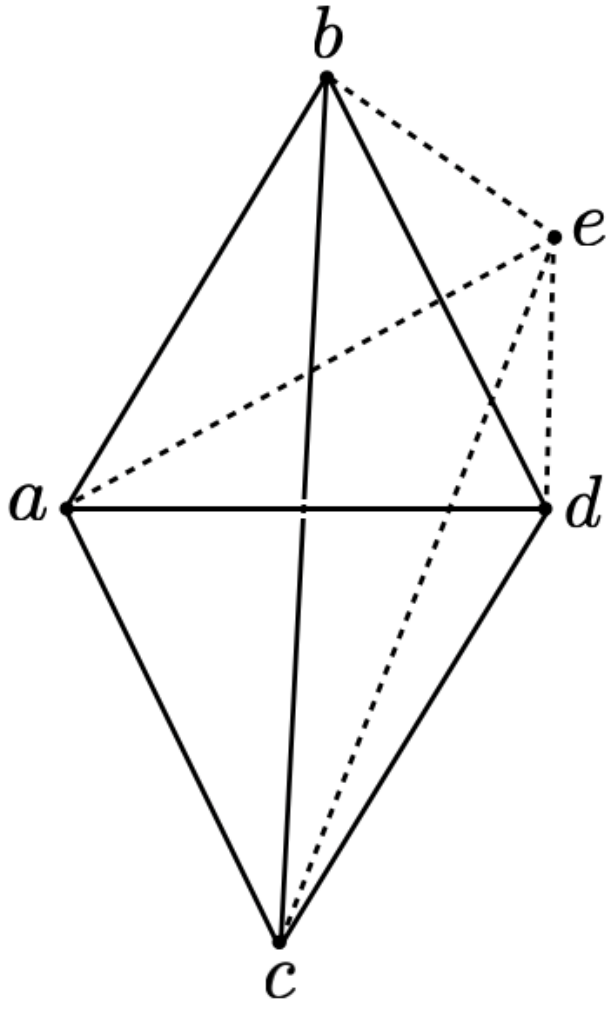}
$$
Formula (\ref{YM1}) represents a triangulation where $\ast$ is a point inside the Amplituhedron, and we use it to triangulate the whole space in terms of simplices. Each simplex is given by the point $\ast$ and four other points $i{-}1,i,j{-}1,j$ which lie on the same three-dimensional face in $\mathbb{P}^4$. 

The super-BCFW recursion relations reproduce (\ref{YM1}) where $\ast$ is one of the external labels, the exact choice depends on a particular BCFW shift \cite{ArkaniHamed:2009vw,Mason:2009qx}. At five-point we always get a single term $A_{5,1}=R[1,2,3,4,5]$, at six-point there are two different forms,
\begin{align} 
R_{6,1}&= R[1,2,3,4,5] + R[3,4,5,6,1] + R[5,6,1,2,3]\nonumber\\ 
&= R[2,3,4,5,6] + R[4,5,6,1,2] + R[6,1,2,3,4] \label{NMHV6}
\end{align}
At higher points we have $n$ different representations for $\ast=1,2,\dots,n$. Note that only if $\ast$ is one of the external labels the $R$-invariants are actually Yangian-invariant, and BCFW recursion relations is a generator of such manifest Yangian-invariant expansions. Each $R$-invariant can be associated with the cell in the positive Grassmanian $G_+(1,n)$ (or $G_+(3,n)$ in momentum space) \cite{ArkaniHamed:2009dn,ArkaniHamed:2009vw,Mason:2009qx,Arkani-Hamed:2016byb} and the Yangian invariance is an inherent property of the Grassmannian description \cite{Drummond:2010qh,Drummond:2010uq,Arkani-Hamed:2016byb}. The equality between two different representations of $A_{6,1}$ in (\ref{NMHV6}) is a consequence of the global residue theorem (GRT) on the Grassmannian integral \cite{ArkaniHamed:2009dn}.

\subsection{Reduced $R$-invariants}

Now, we make a particular choice $\ast=2$ in the $R$-invariant expansion for $R_{n,1}$,
\begin{equation}
R_{n,1} = \sum_{i=4}^{n-1}\sum_{j=i{+}2}^{n{+}1} R[2,i{-}1,i,j{-}1,j]\label{Rinv2}
\end{equation}
where we identified $n{+}1\equiv 1$. Note that while general $R$-invariant has five labels, the terms which appear in (\ref{Rinv2}) are characterized only by three labels $2,i,j$. We use the notation of \cite{Drummond:2008vq} and denote: $R_{2;ij}\equiv R[2,i{-}1,i,j{-}1,j]$. The explicit formula in the momentum space for $R_{2;ij}$ is
\begin{equation}
R_{2;ij} = \frac{\la i{-}1\,i\ra\la j{-}1\,j\ra\cdot \delta^{(4)}(\Xi_{2;ij})}{x_{ij}^2\la 2|x_{2i}x_{ij}|j\ra\la 2|x_{2i}x_{ij}|j{-}1\ra\la 2|x_{2j}x_{ji}|i\ra\la 2|x_{2j}x_{ji}|i{-}1\ra} \label{R1}
\end{equation}
where $x_{ij} = p_i + p_{i{+}1}+\dots+p_{j{-}1}$. The argument of the delta function is 
\begin{equation}
\Xi_{2;ij} = \sum_{k=j}^{n{+}1} \la 2|x_{2i}x_{ij}|k\ra\eta_k + \sum_{k=i}^{n{+}1}\la 2|x_{2j}x_{ji}|k\ra\eta_k \label{delta2}
\end{equation}
This is just a momentum space version of the general momentum twistor expression (\ref{Rin}). This notation is very compact but it heavily relies on the cyclic symmetry. We write explicitly all labels in the long brackets $\la a|xx|b\ra$. For example,
\begin{equation}
\la 2|x_{2i}x_{ij}|j\ra = \la 2|(34\dots i{-}1)(i\dots j{-}1)|j\ra \equiv \sum_{p=3}^{i{-}1}\sum_{q=i}^{j{-}1} \la2p\ra[pq]\la qj\ra \label{long2}
\end{equation}
We also use similar extended notation for shorter brackets such as
\begin{equation}
\la 2|x_{2i}|i] = \la 2|34\dots i{-}1|i]  \equiv \sum_{p=3}^{i{-}1}\la 2p\ra[pi] \label{long1}
\end{equation}
and explicitly list all labels. In the gravity case this is necessary as the lack of cyclic symmetry no longer allows to use $x_{ij}$ variables. We ware interested in the formula for $A_{n,1}$ rather than $R_{n,1}$, hence we define decorated $\widetilde{R}$-invariants which are multiplied by the Parke-Taylor factor and the super-momentum delta functions,
\begin{equation}
\widetilde{R}_{2;ij} \equiv \frac{\delta^{(8)}(Q)}{\la12\ra\la23\ra\dots \la n1\ra} R_{2;ij}\label{R2}
\end{equation}
and the amplitude $A_{n,1}$ is then given by
\begin{equation}
A_{n,1} = \sum_{i=4}^{n-1}\sum_{j=i{+}2}^{n{+}1} \widetilde{R}_{2;ij}\label{Rtil}
\end{equation}
For the six-point amplitude this reduces to the sum over three terms,
\begin{equation}
A_{6,1} = \widetilde{R}_{2,46} + \widetilde{R}_{2,41} + \widetilde{R}_{2,51}  
\end{equation}
which correspond to the second line of (\ref{NMHV6}). The explicit formula for $\widetilde{R}_{2;46}$ is
\begin{align}
\widetilde{R}_{2;46} &= \frac{\delta^{(8)}(Q)}{\la12\ra\la23\ra\la34\ra\la45\ra\la56\ra\la16\ra}\cdot \frac{\la34\ra\la56\ra\,\delta^{(4)}(\Xi_{2;46})}{s_{45}\la 2|(3)(45)|6\ra\la 2|(3)(4)|5\ra\la2|(16)(5)|4\ra\la2|(16)(45)|3\ra}\nonumber\\
&=\frac{\delta^{(2\times4)}(Q)\cdot\delta^{(4)}(\Xi_{2;46})}{\la23\ra^4\la45\ra^4\cdot s_{612}[34][45]\la12\ra\la16\ra\la6|45|3]\la2|16|5]}
\end{align}
where the longer brackets simplified and are written using simpler kinematical objects, e.g. $\la 2|(3)(4)|5\ra = \la 23\ra[34]\la45\ra$, $\la2|(16)(45)|3\ra = \la23\ra s_{612}$, etc. Using the momentum and super-momentum conservation we can write the argument of the delta function as
\begin{align}
\Xi_{2;46} &=-\la23\ra\la45\ra\cdot \Big\{[45]\eta_3 + [53]\eta_4 + [34]\eta_5 \Big\}
\end{align}
Plugging back we see that the multiple poles in the denominator cancels and we can write the result as 
\begin{equation}
\widetilde{R}_{2;46}  = \frac{\delta^{(8)}(Q)\cdot\delta^{(4)}([45]\eta_1 + [51]\eta_4 + [14]\eta_5)}{s_{612}[34][45]\la12\ra\la16\ra\la6|45|3]\la2|16|5]}
\end{equation}
The other two $\widetilde{R}$-invariants are related by simple cyclic shifts $k\rightarrow k{+}2$, $k\rightarrow k{+}4$,
\begin{align}
\widetilde{R}_{2;41} &= \frac{\delta^{(8)}(Q)\cdot\delta^{(4)}([23]\eta_1 + [31]\eta_2+[12]\eta_3)}{s_{456}[12][23]\la45\ra\la56\ra\la6|45|3]\la4|56|1]}\\
\widetilde{R}_{2;51} &= \frac{\delta^{(8)}(Q)\cdot\delta^{(4)}([56]\eta_1+[61]\eta_5+[15]\eta_6)}{s_{561}[56][61]\la23\ra\la34\ra\la2|34|5]\la4|56|1]}
\end{align}
There are many relations between kinematical invariants, e.g. $s_{612}=s_{345}$, and we can also use the supermomentum conservation to rewrite the arguments of delta functions in various equivalent ways.

\subsection{Split-helicity amplitudes}

We now focus on a split helicity gluon amplitude,
\begin{equation}
    {\cal A}_n \equiv {\cal A}_n(1^-2^-3^-4^+5^+\dots n^+) \label{split}
\end{equation}
Note that various helicity configurations are different, and split helicity is a particularly symmetric choice and leads to simpler expressions. Note that for graviton amplitudes all helicity configurations are equivalent up to relabeling, and there is no such simplification for any helicity choice. For split helicity gluon amplitudes the Amplituhedron geometry reduces to the cyclic polytopes in $\mathbb{P}^3$. At six-point, it is given by a simple polyhedron. The two representations for $R_{n,1}$ in terms of $R$-invariants correspond to two different triangulations. The expansion (\ref{Rinv2}) is one of them, and divides the polyhedron into three tetrahedra, each of them for one $R$-invariant,

\vspace{-0.5cm}
$$
\includegraphics[scale=.45]{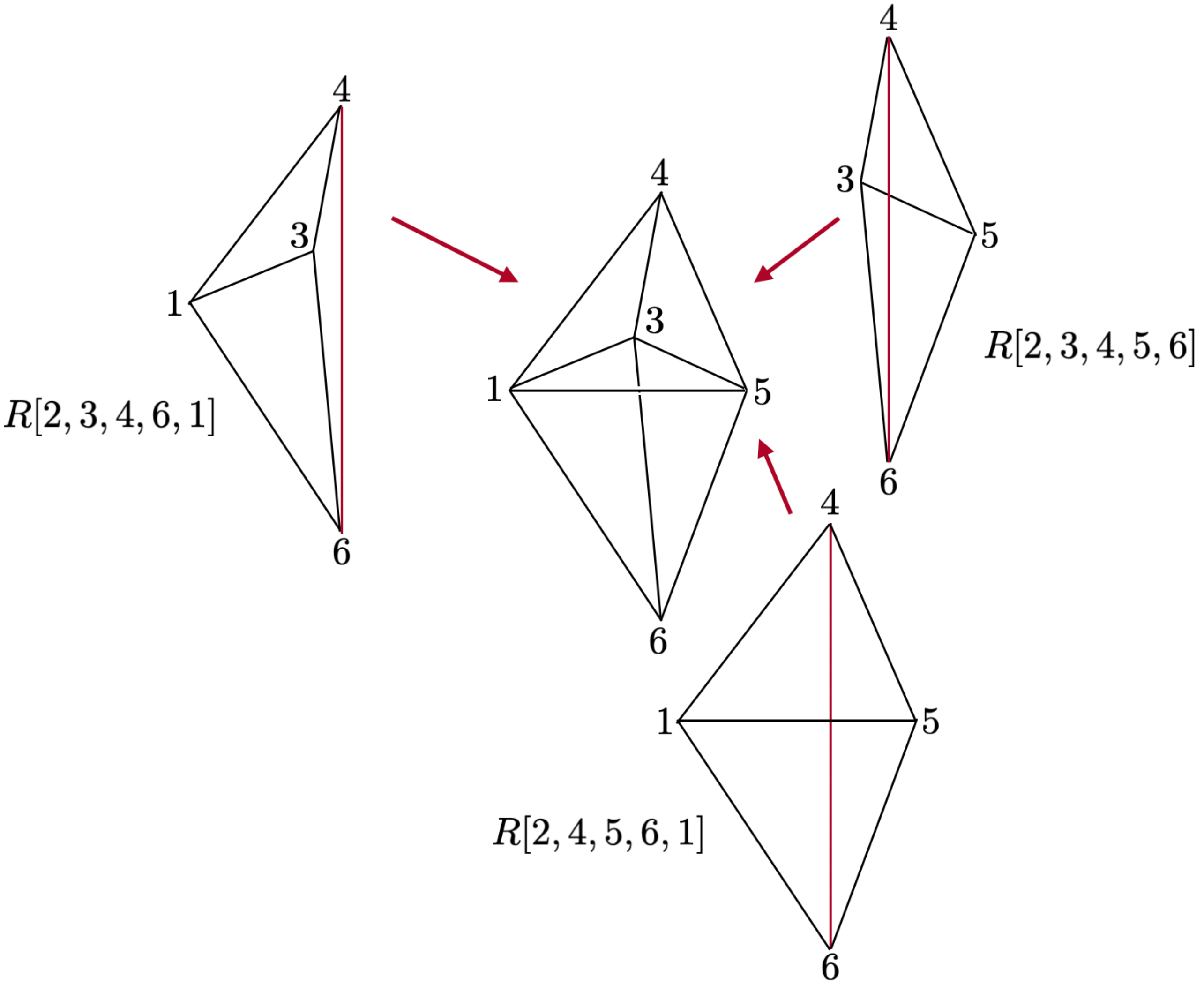}
$$
The label $2$ appears in all $R$-invariants here, but it is absent in the geometry. The $\mathbb{P}^3$ polyhedron is a projection of the $\mathbb{P}^4$ cyclic polytope through point $2$. This is $m=3$, $k=1$ tree-level Amplituhedron (also a cyclic polytope) with a particular choice of origin as cyclic polytopes in odd dimensions are in fact not cyclic. This is reflected in the split helicity amplitude (\ref{split}) where the cyclic symmetry (from the super-amplitude $A_n$) is broken by the choice of labels of negative helicity gluons.

To obtain ${\cal A}_n$ from (\ref{Rtil}) we have to integrate over Grassmann variables $\eta_1$, $\eta_2$, $\eta_3$ for negative helicity gluons. We then define the bosonic ${\cal R}$-invariants from the supersymmetric reduced $R$-invariants,
\begin{equation}
    {\cal R}_{2;ij} = \int d^4\eta_1\, d^4\eta_2\,d^4\eta_3\,\,\widetilde{R}_{2;ij} \label{bosR}
\end{equation}
and the NMHV split helicity amplitude (\ref{split}) can be written as
\begin{equation}
    {\cal A}_n = \sum_{i=4}^{n{-}1}\sum_{j=i{+}2}^{n{+}1} {\cal R}_{2;ij} \label{split2}
\end{equation}
The explicit form of ${\cal R}_{2;ij}$ can be obtained by plugging (\ref{R1}) into (\ref{R2}) and (\ref{bosR}),
\begin{align}
    {\cal R}_{2;ij} &= \frac{\la i{-}1\,i\ra\la j{-}1\,j\ra\cdot{\cal H}^4}{
    \begin{tabular}{c}
     $\la 12\ra\dots \la n1\ra  s_{i\dots j{-}1} \la 2|(34..i{-}1)(i..j{-}1)|j\ra\la 2|(34..i{-}1)(i..j{-}2)|j{-}1\ra$\\
         $\la 2|(1n..j)(j{-}1..i{+}1)|i\ra \la 2|(1n..j)(j{-}1..i)|i{-}1\ra$
    \end{tabular}}\label{Rred}
\end{align}
where ${\cal H}$ is the helicity factor from the Grassmann integration (\ref{bosR}). For the split helicity configuration,
\begin{equation}
{\cal H}^4 = \int d^4\eta_1\,d^4\eta_2\,d^4\eta_3\,\,\delta^{(8)}(Q)\,\delta^4(\Xi_{2;ij})
\end{equation}
Performing the integral explicitly using (\ref{delta}) and (\ref{delta2}) we obtain
\begin{equation}
    {\cal H} = (\la 2|x_{2i}x_{ij}|1\ra + \la 2|x_{2j}x_{ji}|1\ra)\la23\ra = (\la 2|(x_{2i}-x_{2j})x_{ij}|1\ra\la23\ra = \la12\ra\la23\ra s_{i\dots j{-}1} \label{helicity}
\end{equation}
where we used $x_{ij}^2 = s_{i\dots j{-}1}$. If we wanted to calculate the amplitude for other helicity configuration, all terms in (\ref{Rred}) would stay the same, only the helicity factor changes. If the cyclic symmetry is present, the range of labels $(i\dots j{-}1)$ can be specified by $i,j$. Note that the helicity factor ${\cal H}$ for the split helicity amplitude is a product of terms which also appear in the denominator of ${\cal R}_{2;ij}$, therefore one power of ${\cal H}$ always cancels. Similarly, the numerator terms $\la i{-}1\,i\ra\la j{-}1\,j\ra$ always cancel against the terms in the Parke-Taylor factor.
Let us provide some explicit formulas for (\ref{Rred}) which contribute to (\ref{split2}) at low points. At 5pt there is only one contributing term,
\begin{equation}
 {\cal R}_{2;41} = \frac{[45]^3}{[12][23][34][51]}
\end{equation}
where we used kinematical identities to simplify the general expressions for poles, $\la 2|(3)(45)|1\ra = - \la 2|(3)(123)|1\ra = -\la 23\ra[32]\la21\ra$, etc. Note that this is just the parity conjugate of the 5pt Parke-Taylor factor. At 6pt we have three contributions,
\begin{align}
    {\cal R}_{2;46} &=\frac{\la12\ra^3[45]^3}{s_{345}[34]\la61\ra \la6|12|3]\la2|34|5]},\qquad  {\cal R}_{2;41}=\frac{s_{456}^3}{[12][23]\la45\ra\la56\ra\la6|12|3]\la4|56|1]},\nonumber \\   &\hspace{2cm}   
    {\cal R}_{2;51} =
    \frac{\la23\ra^3[56]^3}{s_{234}[16]\la34\ra\la2|16|5]\la4|56|1]}. \label{R6pt}
\end{align}
Note that the terms ${\cal R}_{2;46}$ and ${\cal R}_{2;51}$ are related by relabeling $1\leftrightarrow3$, $4\leftrightarrow 6$, which corresponds to a particular permutation: $(1,2,3,4,5,6)\rightarrow (6,5,4,3,2,1)$. We can write the expression for ${\cal A}_6$ as
\begin{equation}
{\cal A}_6 = \sum_{\cal P}  {\cal R}_{2;51}+ {\cal R}_{2;41}\label{A6YM}
\end{equation} 
where the sum over ${\cal P}$ stands for two permutations: $(1,2,3,4,5,6)$ and $(3,2,1,6,5,4)$. At 7pt we get a sum of six terms,
\begin{align}
    &\hspace{-0.7cm} {\cal R}_{2;46} =\frac{\la12\ra^3[45]^3}{s_{345}[34]\la67\ra\la71\ra \la7|12|3]\la2|34|5]},\,\,
    {\cal R}_{2;47} =\frac{\la12\ra^3s_{456}^3}{s_{127}\la45\ra\la56\ra\la67\ra\la7|12|3]\la6|45|3]\la2|(17)(56)|4\ra}\nonumber\\
    &\hspace{-0.7cm} {\cal R}_{2;41} = \frac{s_{4567}^3}{[12][23]\la45\ra\la56\ra\la67\ra\la 7|12|3]\la4|23|1]},\,\, {\cal R}_{2;51} = \frac{\la23\ra^3s_{567}^3}{s_{234}\la34\ra\la56\ra\la67\ra\la 2|(34)(56)|7\ra\la4|23|1]\la5|67|1]},\nonumber\\
    &\hspace{-0.7cm} {\cal R}_{2;57} = \frac{\la12\ra^3\la23\ra^3[56]^3}{\la34\ra\la71\ra\la2|17|6]\la2|34|5]\la2|(34)(56)|7\ra\la2|(17)(56)|4\ra},\,\,{\cal R}_{2;61} = \frac{\la23\ra^3[67]^3}{s_{167}[17]\la34\ra\la45\ra\la2|17|6]\la5|67|1]}.\nonumber \\
  \end{align}
and again we can recognize that the pair of ${\cal R}_{2;46}$ and ${\cal R}_{2;61}$, resp. ${\cal R}_{2;47}$ and ${\cal R}_{2;51}$ are related by the similar relabeling $1\leftrightarrow3$, $4\leftrightarrow7$, $5\leftrightarrow6$, which is a permutation: $(1,2,3,4,5,6,7)\rightarrow (3,2,1,7,6,5,4)$, or said differently $i\rightarrow 11{-}i$ with identification $7{+}k\equiv k$. The amplitude then takes the form
\begin{equation}
{\cal A}_7 = \sum_{{\cal P}} {\cal R}_{2;61} + \sum_{{\cal P}} {\cal R}_{2;51} + {\cal R}_{2;57} + {\cal R}_{2;41} \label{A7YM}
\end{equation}
The general split helicity amplitude ${\cal A}_n$ has the symmetry in external labels,
\begin{equation}
i\rightarrow n{+}4{-}i\quad\mbox{with $n{+}k\equiv k$} \label{YMsym}
\end{equation}
We can then organize the expansion for ${\cal A}_n$ as the sum over permutations over non-equivalent terms, same as we did in (\ref{A6YM}) and (\ref{A7YM}), where ${\cal P}$ is the set of two permutations,
\begin{equation}
{\cal P}:\Big\{(1,2,3,4,\dots,n),(3,2,1,n,\dots,4)\Big\} \label{YMsym2}
\end{equation}
Note that some terms are invariant under (\ref{YMsym}), and therefore appear in the result with no sum over ${\cal P}$, such as the second term in (\ref{A6YM}) and last two terms in (\ref{A7YM}). This organization of terms does not simplify the result much, but it will be very useful in gravity case where ${\cal P}$ will contain larger set of permutations.

\subsection{Factorization diagrams}

For each bosonic ${\cal R}$-invariant ${\cal R}_{2;ij}$ we define a {\bf factorization diagram} which represents the division of all $n$ labels into two sets,
$$
{\cal R}_{2;ij}:\hspace{0.2cm} \raisebox{-48pt}{\includegraphics[scale=.37]{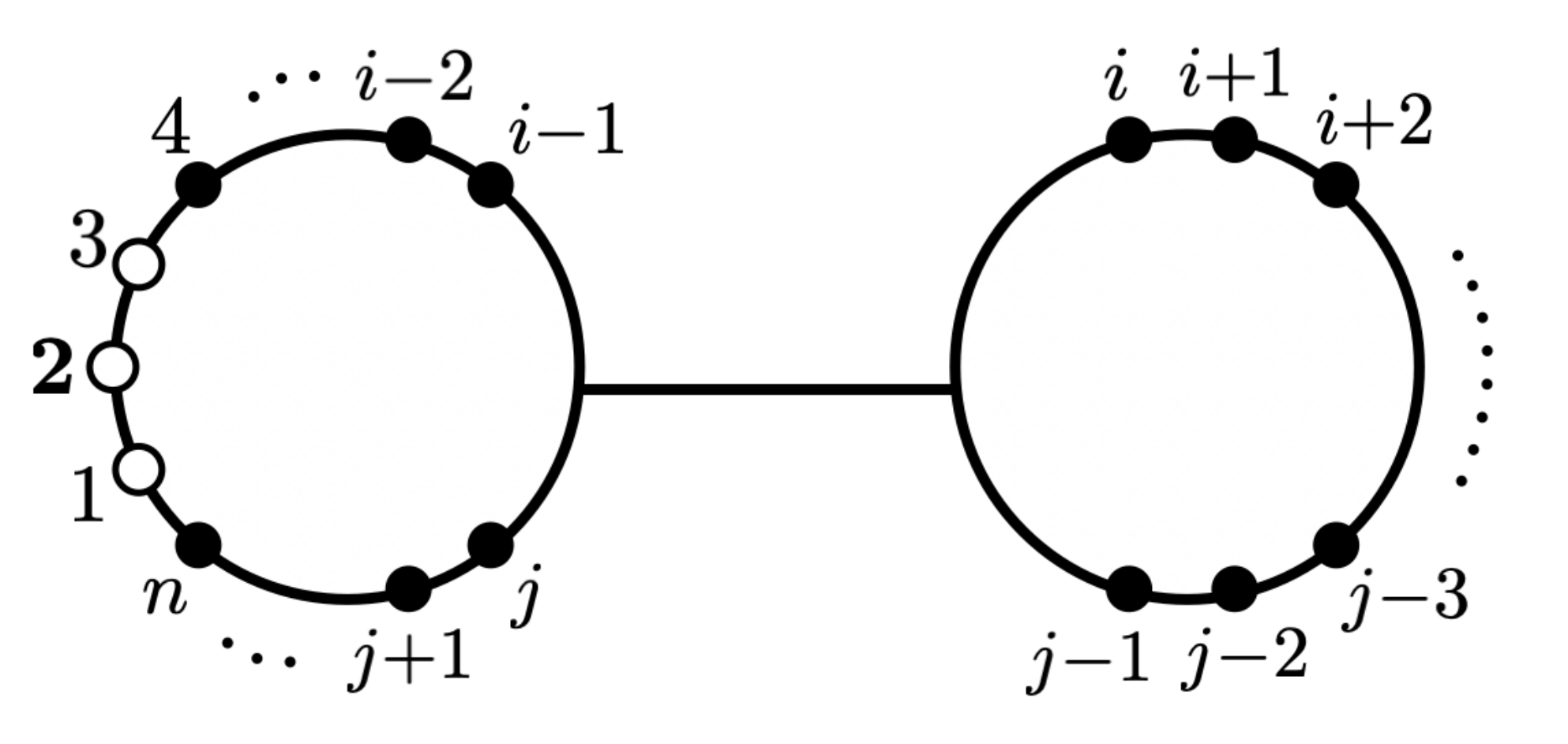}}
$$
We call the left part of the diagram a {\bf left blob}. It contains special labels $1,{\bf 2},3$ (denoted by empty circles) which correspond to negative helicity gluons, the rest of the labels refer to positive helicity gluons (full circles). We denoted the central index ${\bf 2}$ in bold which appears in the labeling of ${\cal R}_{2;ij}$. The left blob also contains the clockwise labels $4,5,\dots,i{-}1$, and anti-clockwise labels $n,n{-}1,\dots,j$. The ordering of all labels is given by the cyclic symmetry. We refer to the right part of the diagram as a {\bf right blob}, and it contains labels $i,i{+}1,\dots,j{-}1$. The formula for the $n$-point NMHV amplitude (\ref{split2}) corresponds to the set of all factorization diagrams. 

As an example, we inspect the six-point amplitude ${\cal A}_6$ given by the following set of factorization diagrams,
$$
\includegraphics[scale=.35]{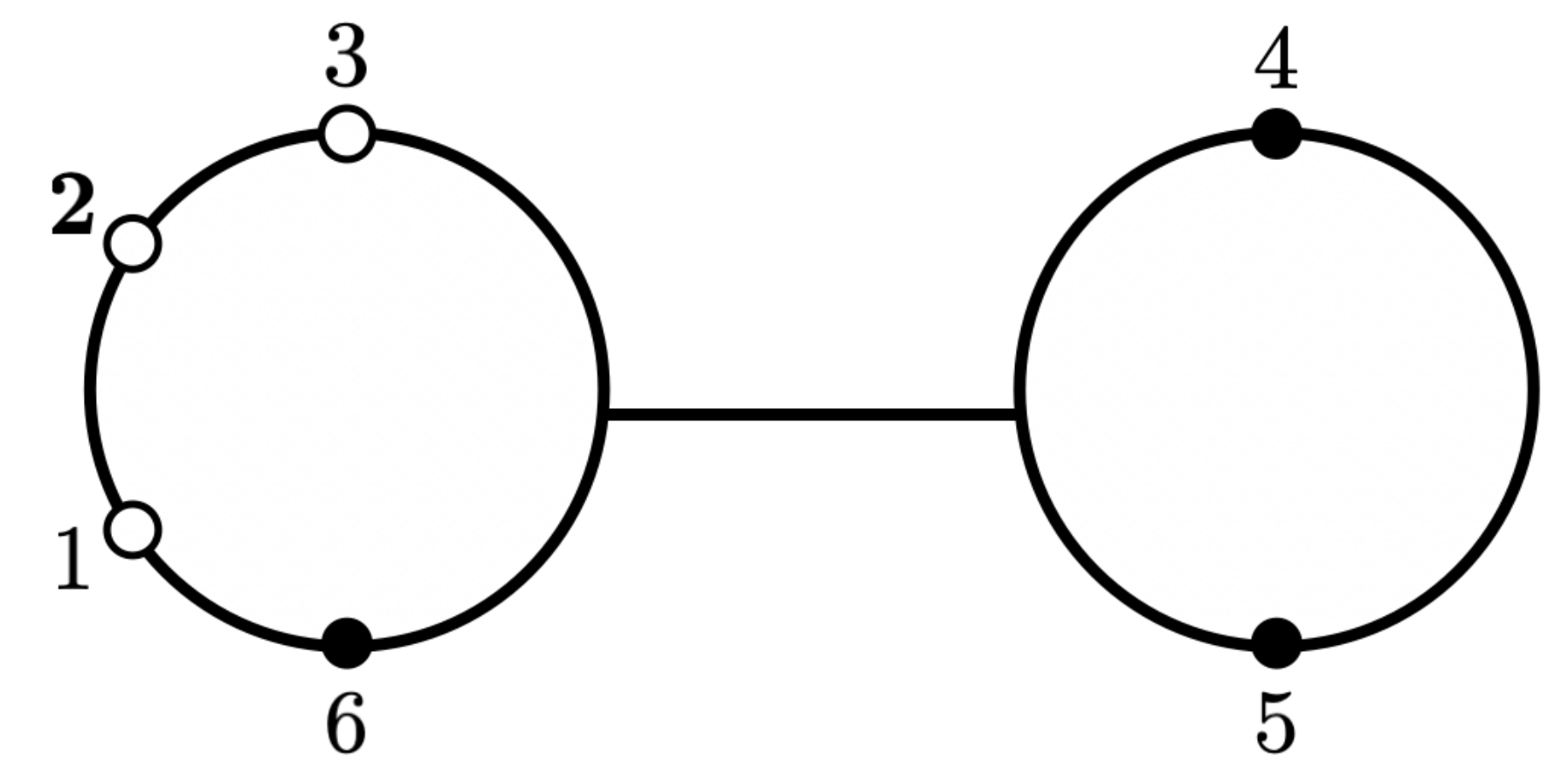}\hspace{1.5cm}\includegraphics[scale=.35]{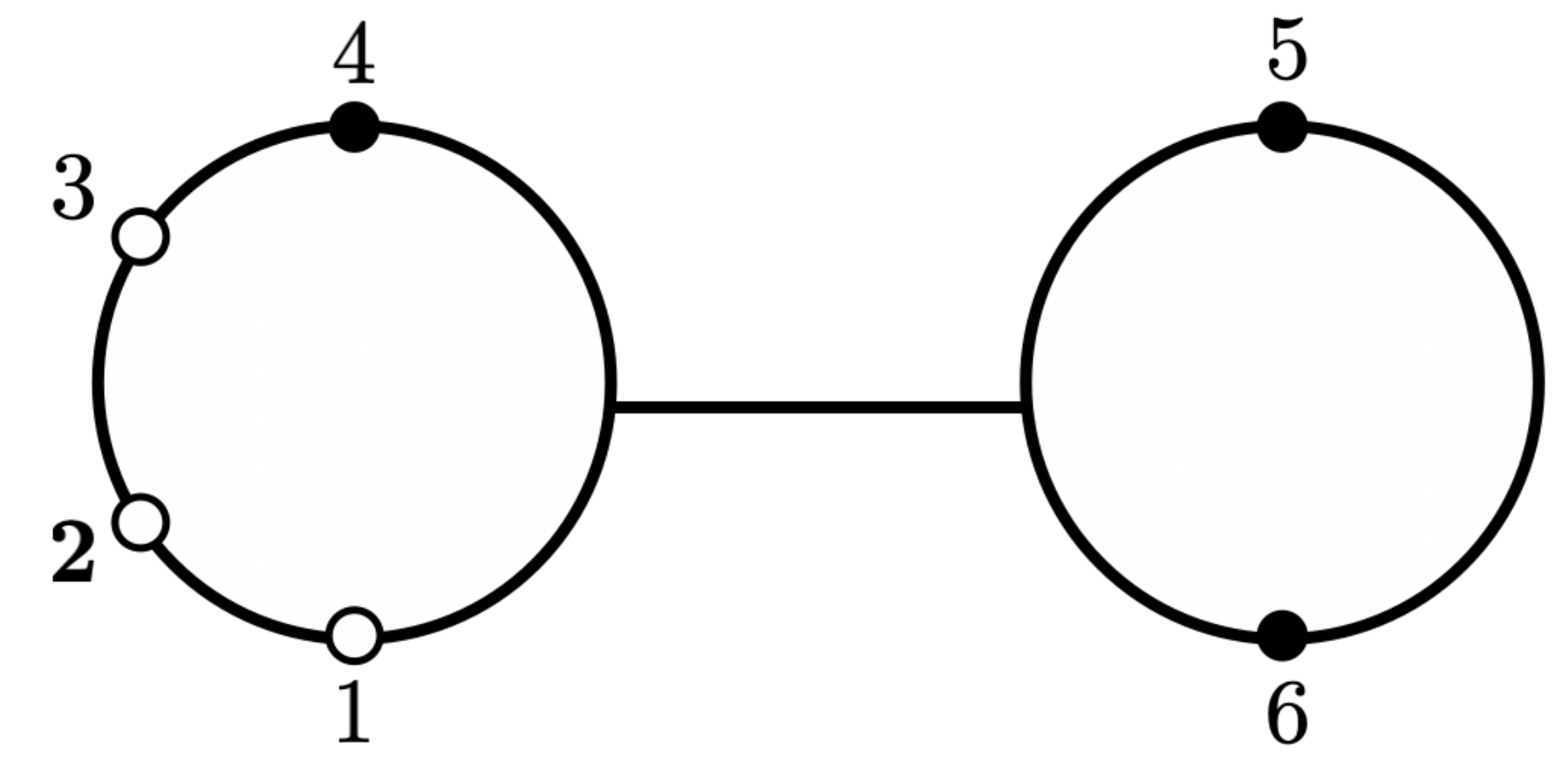}
$$
$$
\includegraphics[scale=.35]{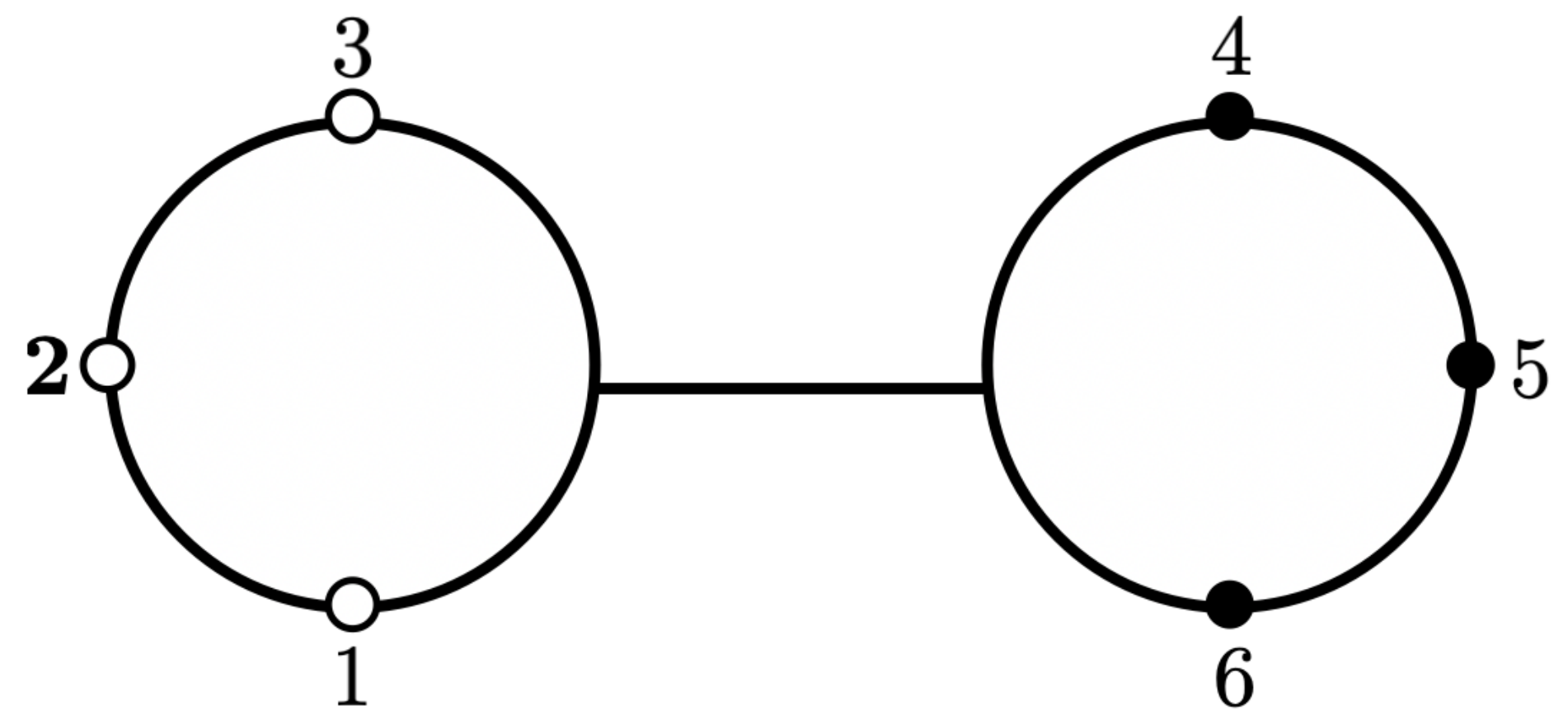}
$$
Note that this representation is not ``sum over all factorization channels". In fact, we sum over ``all channels" which do not involve labels $1,2,3$. In fact, none of them are present in the split-helicity amplitude ${\cal A}_n$ as we will discuss later. In any case, this is nothing else than a certain graphical representation of the ${\cal R}$-invariant expansion (\ref{split2}). Now we revert the story and use factorization diagrams to re-build the formula (\ref{Rred}). First, we write the expression for ${\cal R}_{2;ij}$ as
\begin{equation}
   {\cal R}_{2;ij}= \frac{{\cal H}^4}{s_{i\dots j{-}1}}\times (P_L\otimes P_R) \label{Rexp}
\end{equation}
where $P_L$ and $P_R$ are left and right blob functions and $\otimes$ refers to merging $P_L$ and $P_R$. 
\begin{align}
    P_L&={\color{blue} \frac{1}{\la j{+}1\,j{+}2\ra\dots \la12\ra\la23\ra\dots \la i{-}1\,i\ra}}\times {\color{OliveGreen} \frac{1}{\la 2|(34..i)(P)|j{+}1\ra\la 2|(1n..j{+}1)(P)|i\ra}} \nonumber\\
    &\hspace{2cm}\times {\color{red}\frac{1}{[\la 2|(34..i)\ast_1][\la 2|(1n..j{+}1)\ast_2]}}\label{PL1a}\\
    P_R&= {\color{blue} \frac{1}{\la i{+}1\,i{+}2\ra\dots \la j{-}1\,j\ra}}\times {\color{red}\frac{1}{[\ast_1(P)|j\ra][\ast_2(P)|i{+}1\ra]}}\label{PR1}
\end{align}
where $P=p_{i}+\dots+p_{j{-}1}$ is the sum of momenta in the right blob. The merging procedure takes a {\bf half-pole} $\la a|(P_1) \ast$ and combines it with the other half pole $\ast(P_2)|b\ra$ by creating a longer bracket,
\begin{equation}
    [\la a|(P_1)\ast]\otimes[\ast(P_2)|b\ra] \rightarrow \la a|(P_1)(P_2)|b\ra \label{merge}
\end{equation}
There are three types of poles in ${\cal R}_{2;ij}$ apart from the factorization pole $s_{i\dots j{-}1}$, we will use colors to distinguish them. This will be especially useful later when we will use the same framework in the gravity case, 
\begin{itemize}
    \item {\color{Blue} \bf Blue} : $\la k\,k{+}1\ra$ with $k\in (j,j{+}1,..,1,2,3,..,i{-}2)$ for the left blob\\ and $k\in(i,..,j{-}2)$ for the right blob.
    \item {\color{Red} \bf Red}: ``clockwise" $\la 2|(34..i{-}1)(P)|j{-}1\ra$ and ``anti-clockwise" $\la 2|(1n..j)(P)|i\ra$.
    \item {\color{Green} \bf Green}: ``clockwise" $\la 2|(34..i{-}1)(P)|j\ra$ and ``anti-clockwise" $\la 2|(1n..j)(P)|i{-}1\ra$.
\end{itemize}
The blue poles are simple holomorphic poles, while the red and green poles are (in general) spurious and are represented by the long brackets. Note that all poles start at $\la 2|$, and then pick the clockwise factor $(34..i{-}1)$ or the anti-clockwise factor $(1n..j)$, and the factor $(P)$. This is ``sandwiched" from the other side by left blob boundary labels $i{-}1,j$ in green poles and right blob boundary labels $i,j{-}1$ in red poles. Note that the long brackets are completely cyclic. Here we use some trivial relations such as $(P)|j{-}1\ra \equiv (i..j{-}1)|j{-}1\ra =(i..j{-}2)|j{-}1\ra$. While the blue and green poles are associated with left or right blobs, the red poles are products of the merging procedure (\ref{merge}). The $\ast_1$ and $\ast_2$ symbols in (\ref{PL1a}) and (\ref{PR1}) denote appropriate pairing. We get
\begin{align}
    {\cal R}_{2;ij} & = \frac{{\cal H}^4}{s_{i\dots j{-}1}}\times {\color{blue} \frac{1}{\la12\ra\dots \la i{-}1\,i\ra\la i{+}1\,i{+}2\ra\dots \la j{-}1\,j\ra\la j{+}1\,j{+}2\ra\dots \la n1\ra}}\nonumber\\
  &\hspace{-1.5cm}  \times {\color{red} \frac{1}{\la 2|(34..i)(P)|j\ra\,\la2|(1n..j{+}1)(P)|i{+}1\ra}}    {\color{OliveGreen} \times \frac{1}{\la 2|(34..i)(P)|j{+}1\ra\,\la2|(1n..j{+}1)(P)|i\ra}} \label{RYMform}
\end{align}
in agreement with (\ref{Rred}). Let us demonstrate it on a six-point example. For ${\cal R}_{2;41}$ we get a factorization diagram,
$$
\includegraphics[scale=.37]{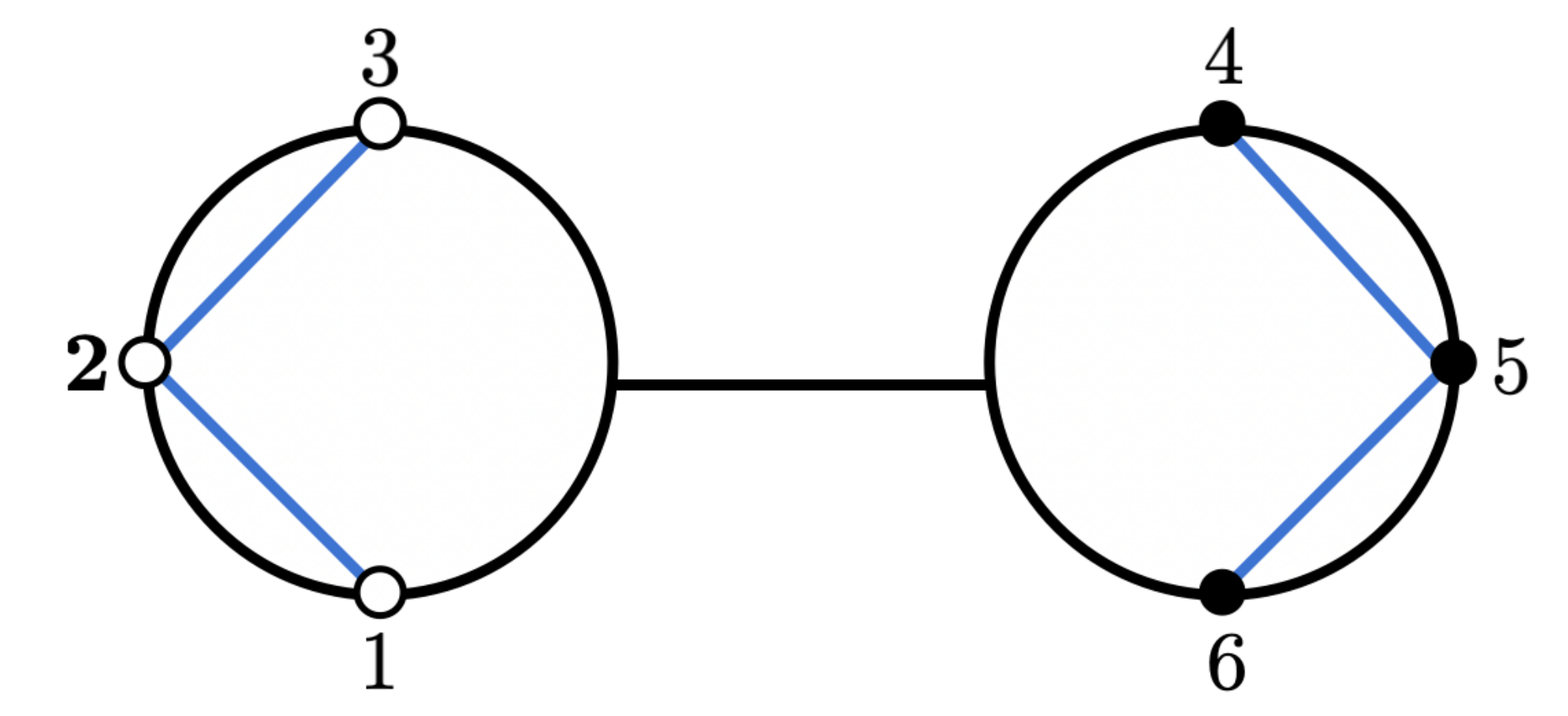}
$$
where we already denoted the blue poles for each blob. The $P_L$ and $P_R$ functions are
\begin{align}
    P_L&={\color{blue} \frac{1}{\la12\ra\la23\ra}}\times {\color{red}\frac{1}{[\la2|(3)\ast_1][\la 2|(1)\ast_2]}}
    \times {\color{OliveGreen} \frac{1}{\la 2|(3)(456)|1\ra\la 2|(1)(456)|3\ra}}\\
    P_R&={\color{blue}\frac{1}{\la45\ra\la56\ra}}\times {\color{red}\frac{1}{[\ast_1(45)|6\ra][\color{Red} \ast_2(65)|4\ra]}}
\end{align}
After merging the red poles we get an expression for ${\cal R}_{2;41}$,
\begin{align}
    {\cal R}_{2;41} &=  \frac{{\cal H}^4}{s_{456}}\times{\color{blue} \frac{1}{\la12\ra\la23\ra\la45\ra\la56\ra}} {\color{red}\frac{1}{\la 2|(3)(45)|6\ra\la2|(1)(65)|4\ra}} {\color{OliveGreen}\frac{1}{\la2|(3)(456)|1\ra\la 2|(1)(654)|3\ra}}\nonumber\\ 
    &= \frac{{\cal H}^4}{s_{456}\la12\ra^4\la23\ra^4\la45\ra\la56\ra[12][23]\la4|56|1]\la6|12|3]}
\end{align}
where we simplified the expression on the first line using various kinematical relations. The helicity factor for the split helicity configuration (\ref{helicity}) is ${\cal H}=\la12\ra\la23\ra s_{456}$ and we reproduce the second formula in (\ref{R6pt}). Note that the helicity factor canceled the multiple poles $\la12\ra^4\la23\ra^4$ in the denominator (which originated from red and green poles). This is a general feature: whenever the multiple poles are generated in the denominator, the numerator always cancels them.
We prefers to think about the red poles as a product of merger, but together with green poles they can be interpreted graphically as certain clockwise and anti-clockwise procedure starting with the central label ${\bf 2}$. In particular the pair of red poles are associated with following pictures,
$$
\includegraphics[scale=.36]{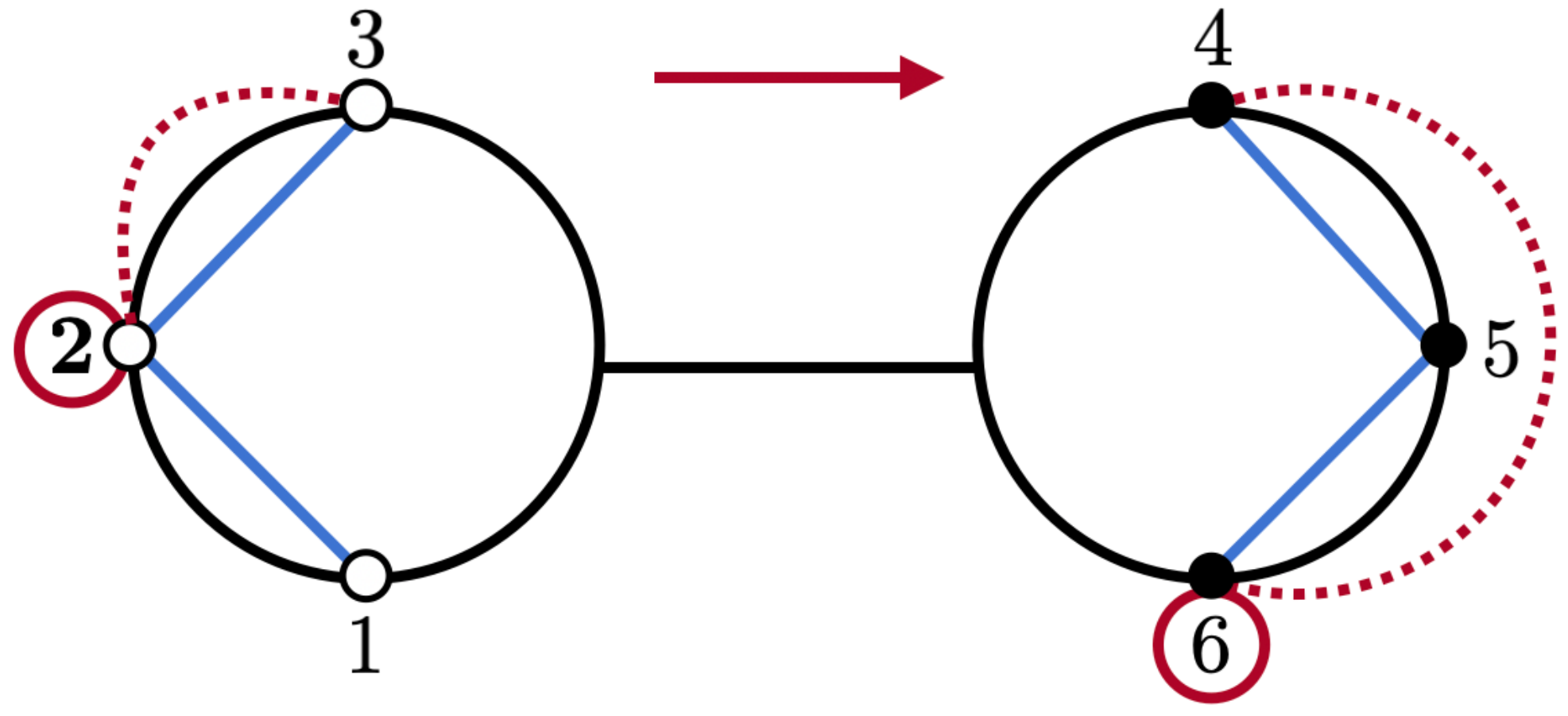}\hspace{1.2cm}
\raisebox{0pt}{\includegraphics[scale=.36]{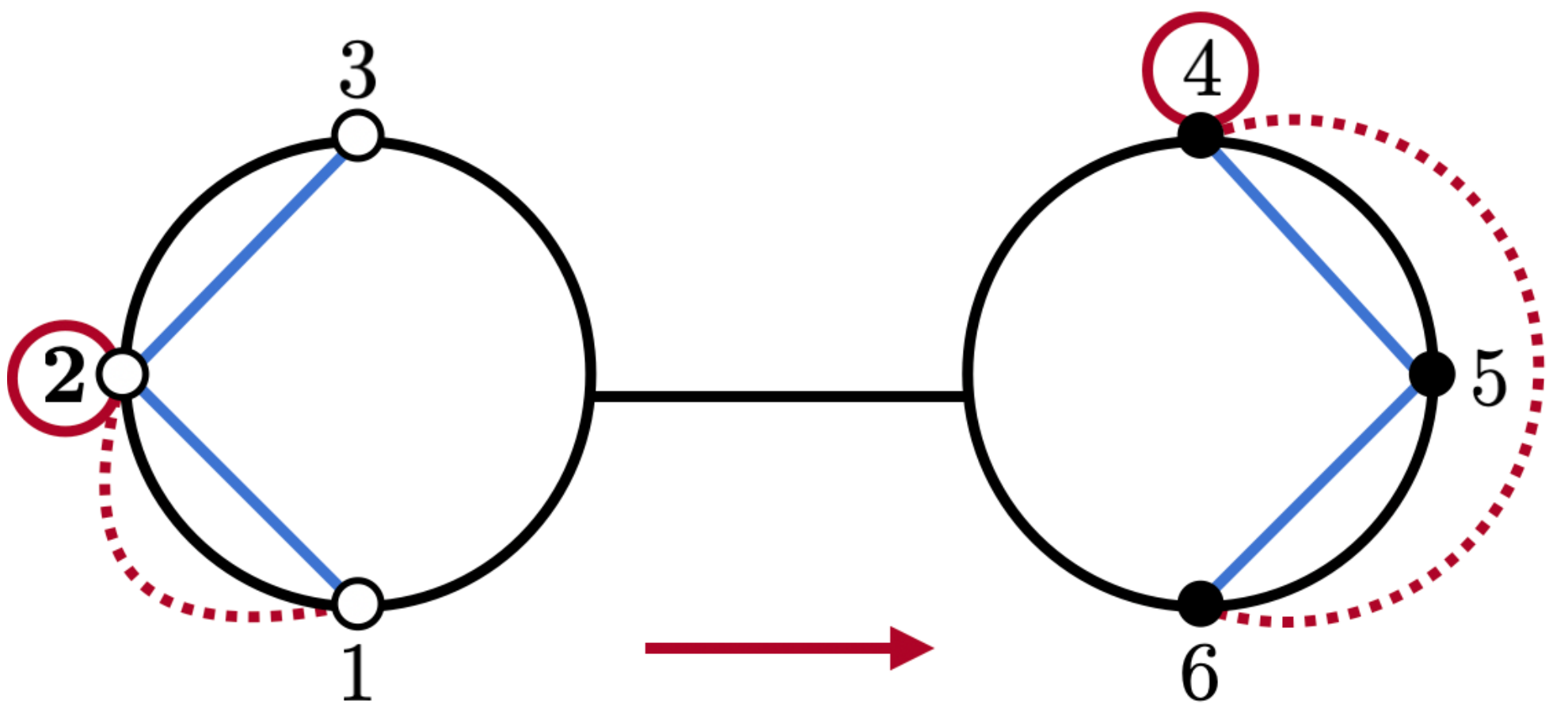}}
$$
The picture provides a multi-step process to generate the pole, the arrow denoted in which direction we go. For the first term
\vspace{-0.1cm}
\begin{enumerate}
\item Start with the central index {\bf 2}, write $\la 2|$.
\vspace{-0.2cm}
\item Then keep all indices until the next stop, which is the end of the left blob, here 3, and write $\la 2|(3)$.
\vspace{-0.2cm}
\item Keep again all indices until the next stop, which is the next-to-last index in the right blob, here 5, and write $\la 2|(3)(45)$.
\vspace{-0.2cm}
\item Finish with the index of the last index in the right blob, ${\color{red} \la 2|(3)(45)|6\ra}$.
\end{enumerate}
The other red pole is ${\color{red} \la2|(1)(65)|3\ra}$. Note that both poles can be also written with the entry $(456)$ as $\la 2|(3)(456)|6\ra$, $\la 2|(1)(456)|4\ra$, this is notation which will be later used in the graviton case. The green poles are similar just the final stop is different, and it is the first index in the left blob (in a given direction):
$$
\includegraphics[scale=.36]{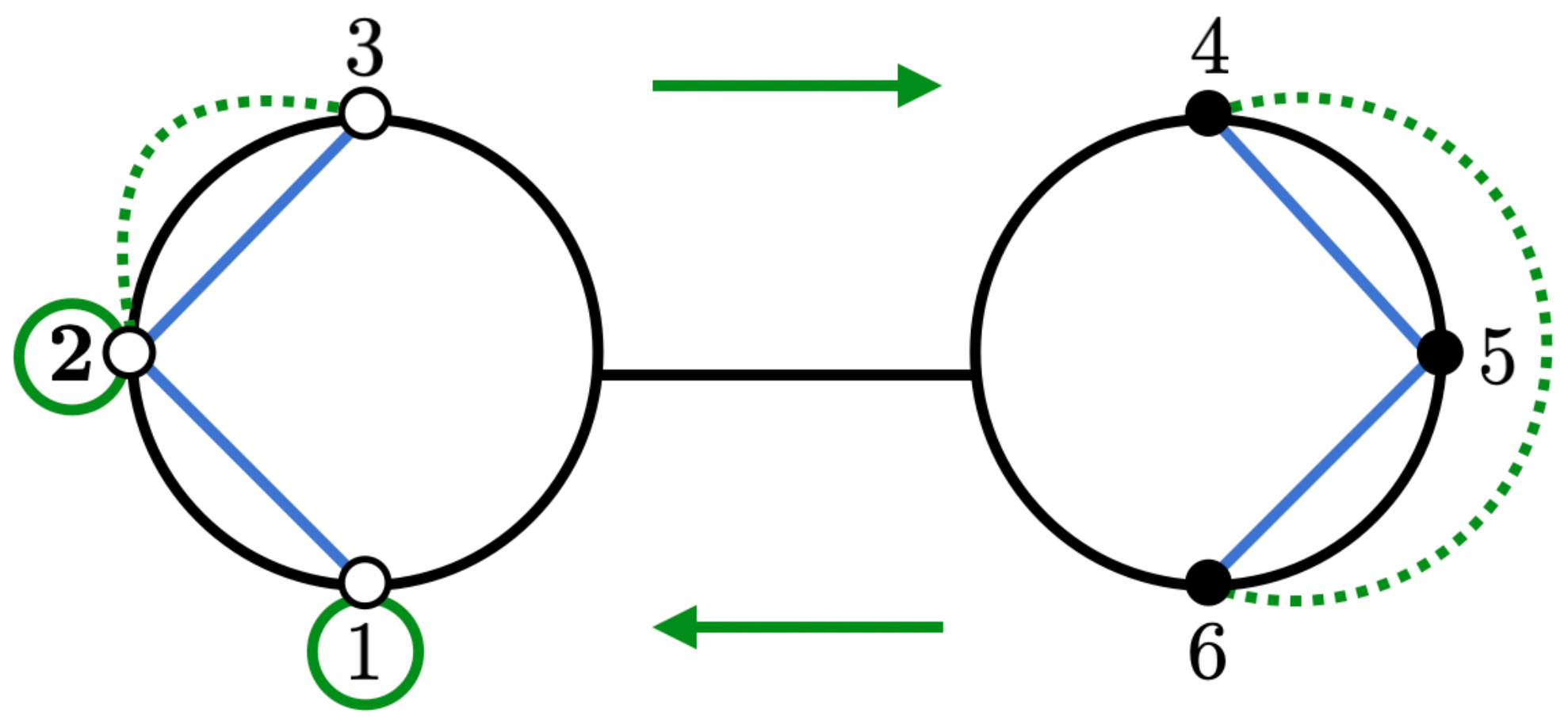}\hspace{1.2cm}
\raisebox{3pt}{\includegraphics[scale=.36]{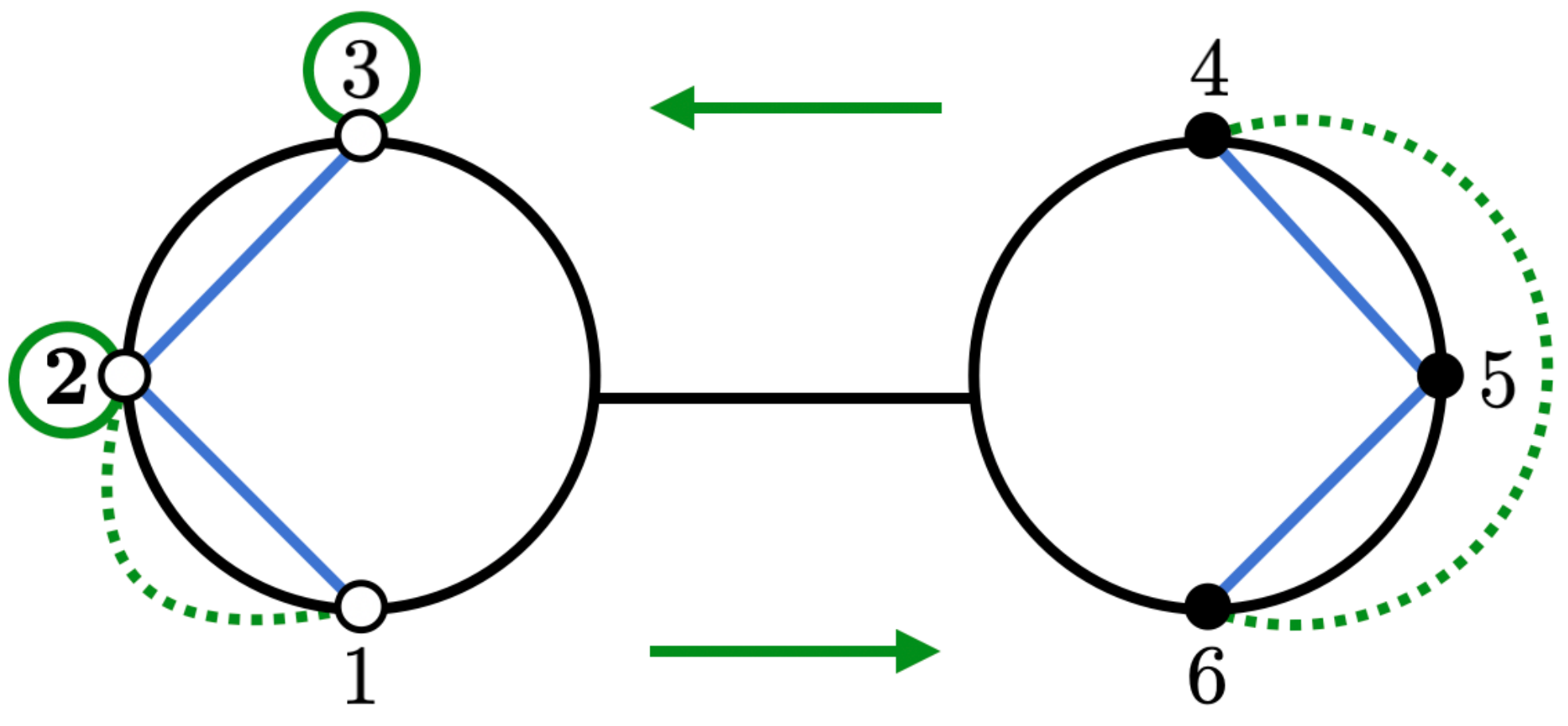}}
$$
generate the poles ${\color{OliveGreen}\la2|(3)(456)|1\ra}$, ${\color{OliveGreen} \la 2|(1)(654)|3\ra}$. We use either $(456)$ or $(654)$ to preserve clockwise or anti-clockwise cyclic ordering, but obviously they are the same.

The next term in ${\cal A}_6$ is the ${\cal R}$-invariant ${\cal R}_{2;51}$. The factorization diagram is
$$
\includegraphics[scale=.37]{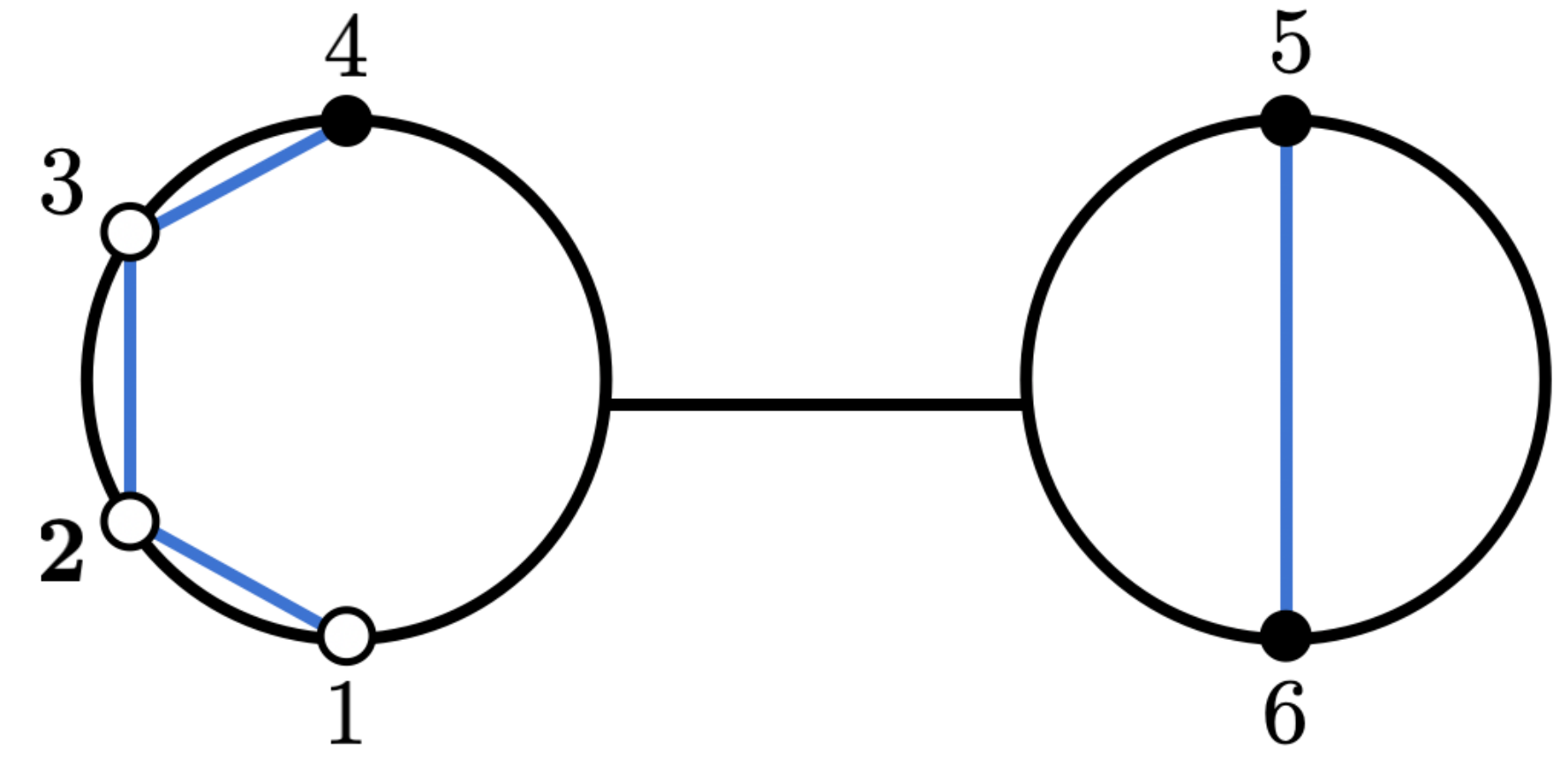}
$$
with blue poles indicated in the picture. The left and right blob functions are
\begin{align}
    P_L&={\color{blue} \frac{1}{\la12\ra\la23\ra\la34\ra}}\times {\color{red}\frac{1}{[\la2|(34)\ast_1][\la 2|(1)\ast_2]}}
    \times {\color{OliveGreen} \frac{1}{\la 2|(34)(56)|1\ra\la 2|(1)(65)|4\ra}}\\
    P_R&={\color{blue}\frac{1}{\la56\ra}}\times {\color{red}\frac{1}{[\ast_1(5)|6\ra][\color{red} \ast_2(6)|5\ra]}}
\end{align}
Plugging into the general expressions (\ref{RYMform}) and merging the red half-poles we get,
\begin{align}
{\cal R}_{2;51} &= \frac{{\cal H}^4}{s_{56}}\times {\color{blue}\frac{1}{ \la12\ra\la23\ra\la34\ra\la56\ra}}
{\color{red}\frac{1}{\la 2|(34)(5)|6\ra\la2|(1)(6)|5\ra}} {\color{OliveGreen} \frac{1}{\la2|(34)(56)|1\ra\la2|(1)(65)|4\ra}}\nonumber\\
&=\frac{{\cal H}^4}{s_{234}\la12\ra^4\la56\ra^4\la23\ra\la34\ra[56][16]\la2|34|5]\la4|56|1]}
\end{align}
For the split helicity factor ${\cal H}=\la12\ra\la23\ra s_{56}$ we reconstruct the third expression in (\ref{R6pt}). The red and green poles can be again visualized as
$$
\raisebox{0pt}{\includegraphics[scale=.35]{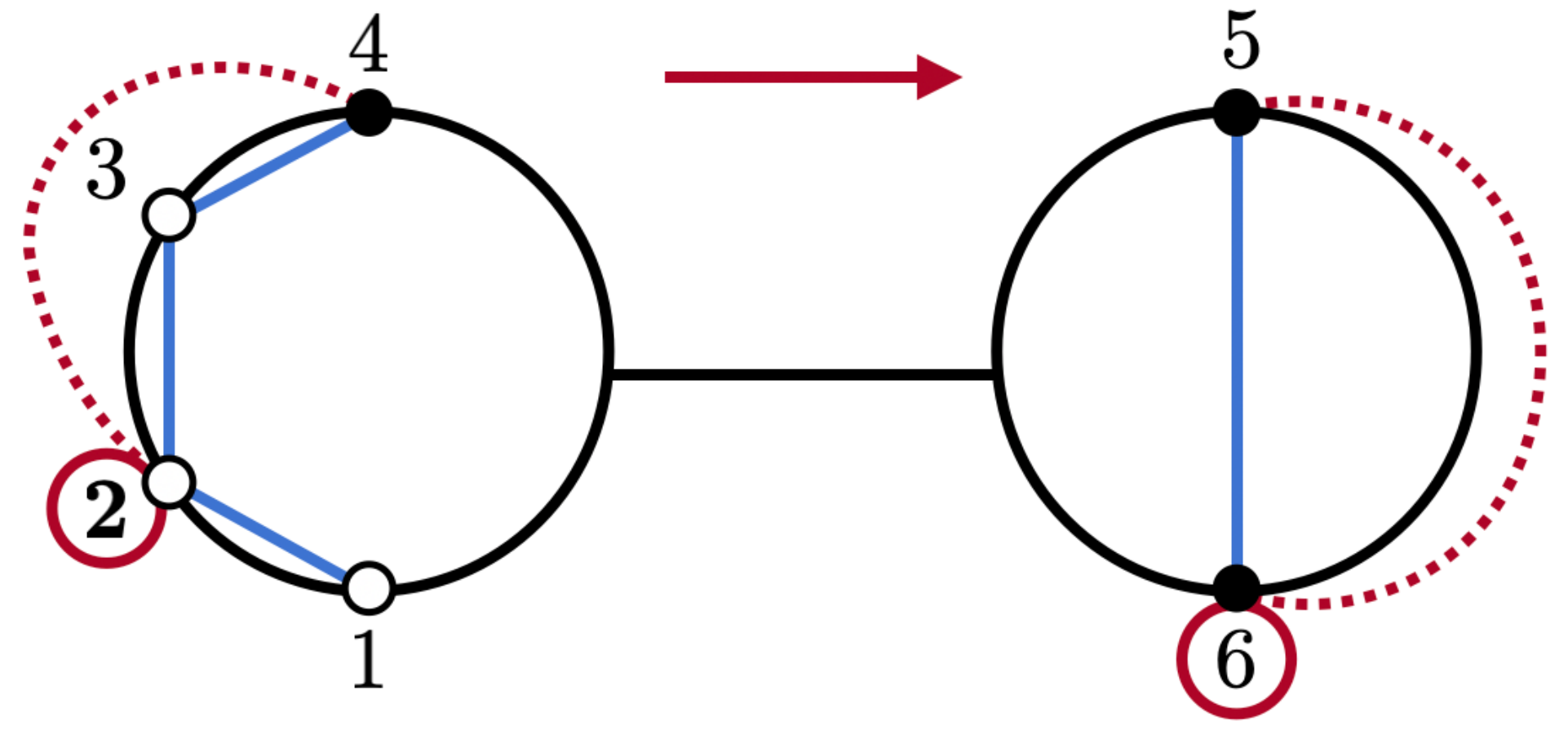}}\hspace{1.5cm}
\raisebox{0pt}{\includegraphics[scale=.35]{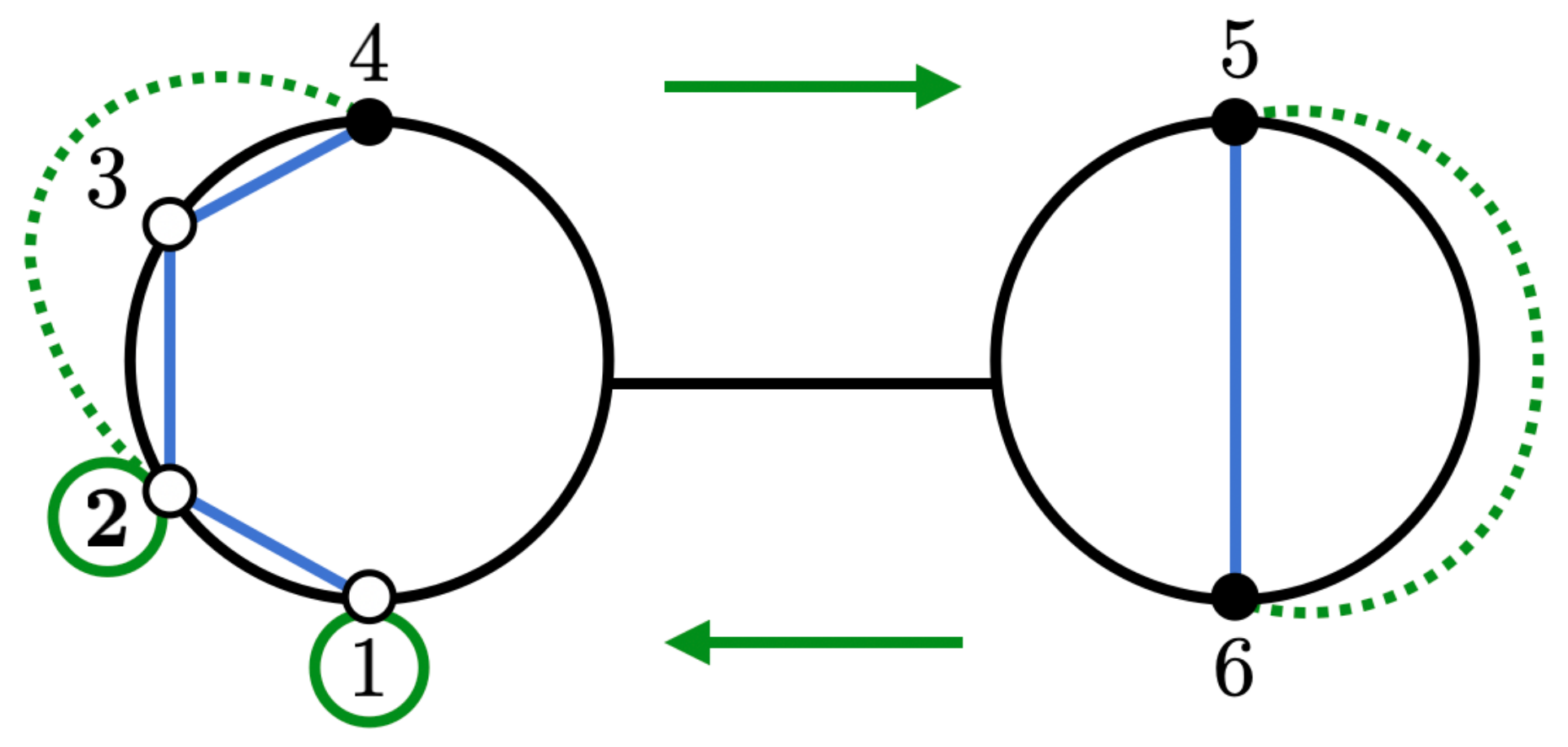}}
$$
which generates ${\color{red} \la2|(34)(5)|6\ra}$, ${\color{OliveGreen} \la2|(34)(56)|1\ra}$ and similar for others. The last factorization diagram corresponds to ${\cal R}_{2;46}$
$$
\includegraphics[scale=.37]{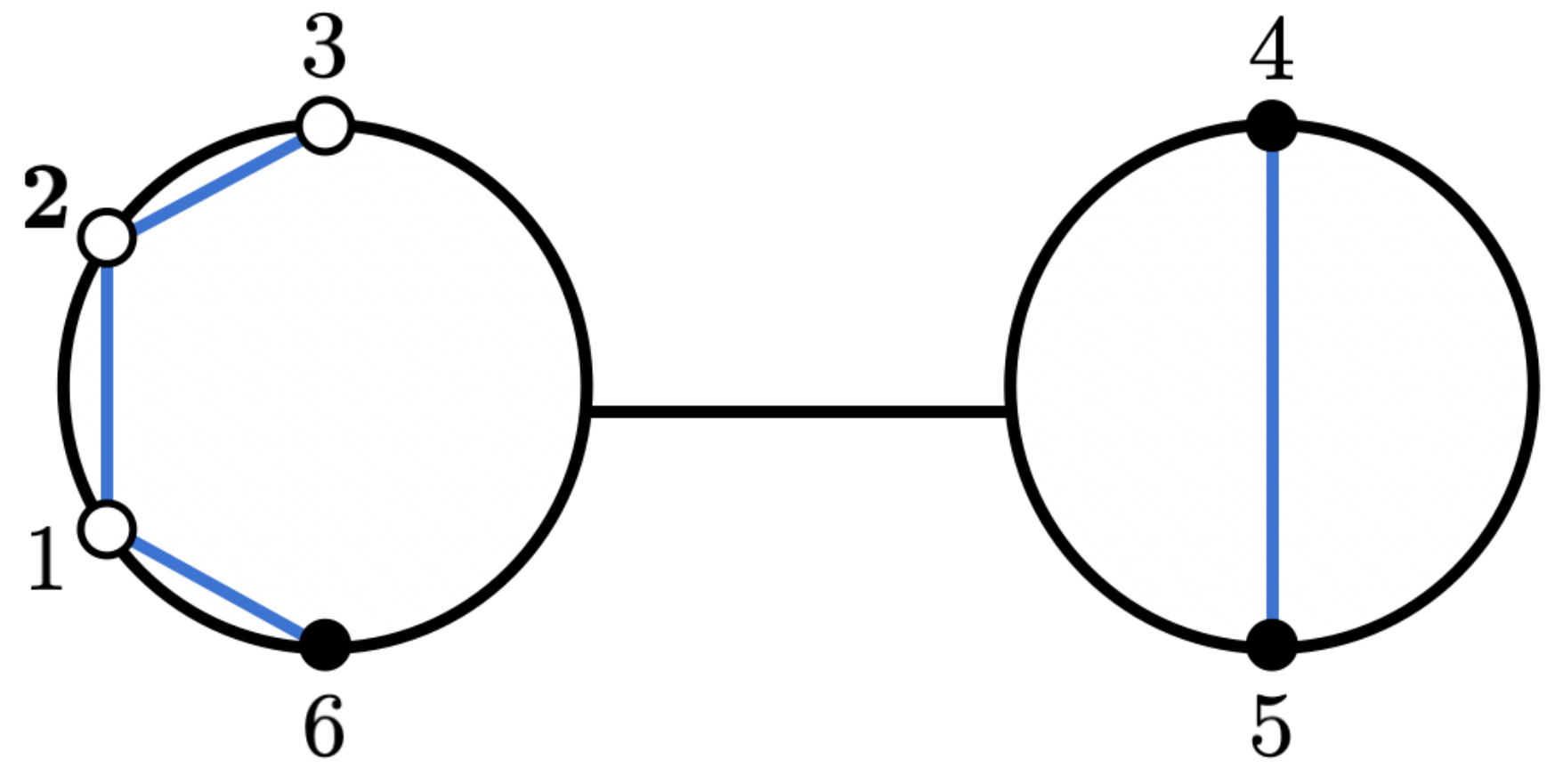}
$$
and it is related to ${\cal R}_{2;51}$ by the relabeling symmetry (\ref{YMsym}). Same rules for $P_L$, $P_R$ and merging give the first expression in (\ref{R6pt}) completing the six-point calculation. Finally, we evaluate one factorization diagram for a 9pt term ${\cal R}_{2;59}$,
$$
\includegraphics[scale=.37]{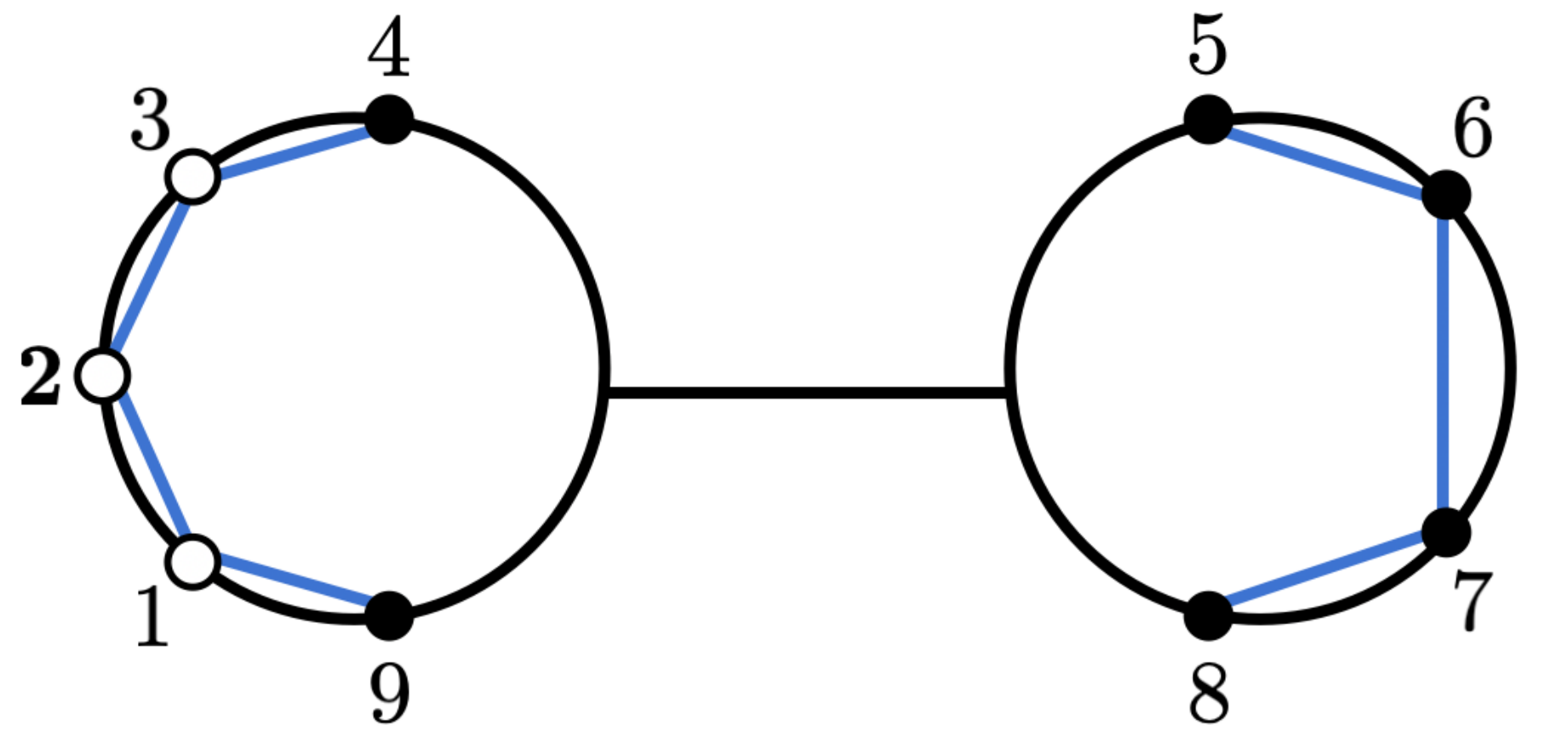}
$$
Reading off red and green poles using the same rules we get
\begin{align}
    {\cal R}_{2;59}&= \\
    & \hspace{-1cm}\frac{{\cal H}^4}{s_{5678}{\color{blue}\la91\ra\la12\ra\la23\ra\la34\ra\la56\ra\la67\ra\la78\ra}{\color{red}\la2|(34)(567)|8\ra\la2|(19)(876)|5\ra}{\color{OliveGreen}\la2|(19)(8765)|4\ra\la2|(34)(5678)|9\ra}}\nonumber
\end{align}
The helicity factor for the split helicity configuration is ${\cal H}_{5678}=\la12\ra\la23\ra s_{5678}$ and we get the contribution to the 9pt amplitude
\begin{equation}
{\cal R}_{2;59} = \frac{\la12\ra^3\la23\ra^3s_{5678}^3}{\la91\ra\la34\ra\la56\ra\la67\ra\la78\ra\la2|(34)(567)|8\ra\la2|(19)(876)|5\ra\la2|(19)(8765)|4\ra\la2|(34)(5678)|9\ra}
\end{equation}
Note that none of the long poles simplify now -- this is indeed a generic case. As a preparation for the graviton case we now use extended labeling to denote ${\cal R}_{2;ij}$
\begin{equation}
    {\cal R}_{2;ij} = {\cal R}(1,{\bf 2},3,\{Q_1\},\{P\},\{Q_2\})\label{LabExt}
\end{equation}
where we denoted $\{Q_1\}=\{4,5,\dots,i{-}1\}$, $\{Q_2\} = \{j,j{+}1,\dots,n\}$, two sets of labels in the left blob, and $\{P\}=\{i,i{+}1,\dots,j{-}1\}$ all labels in the right blob. 
$$
\includegraphics[scale=.41]{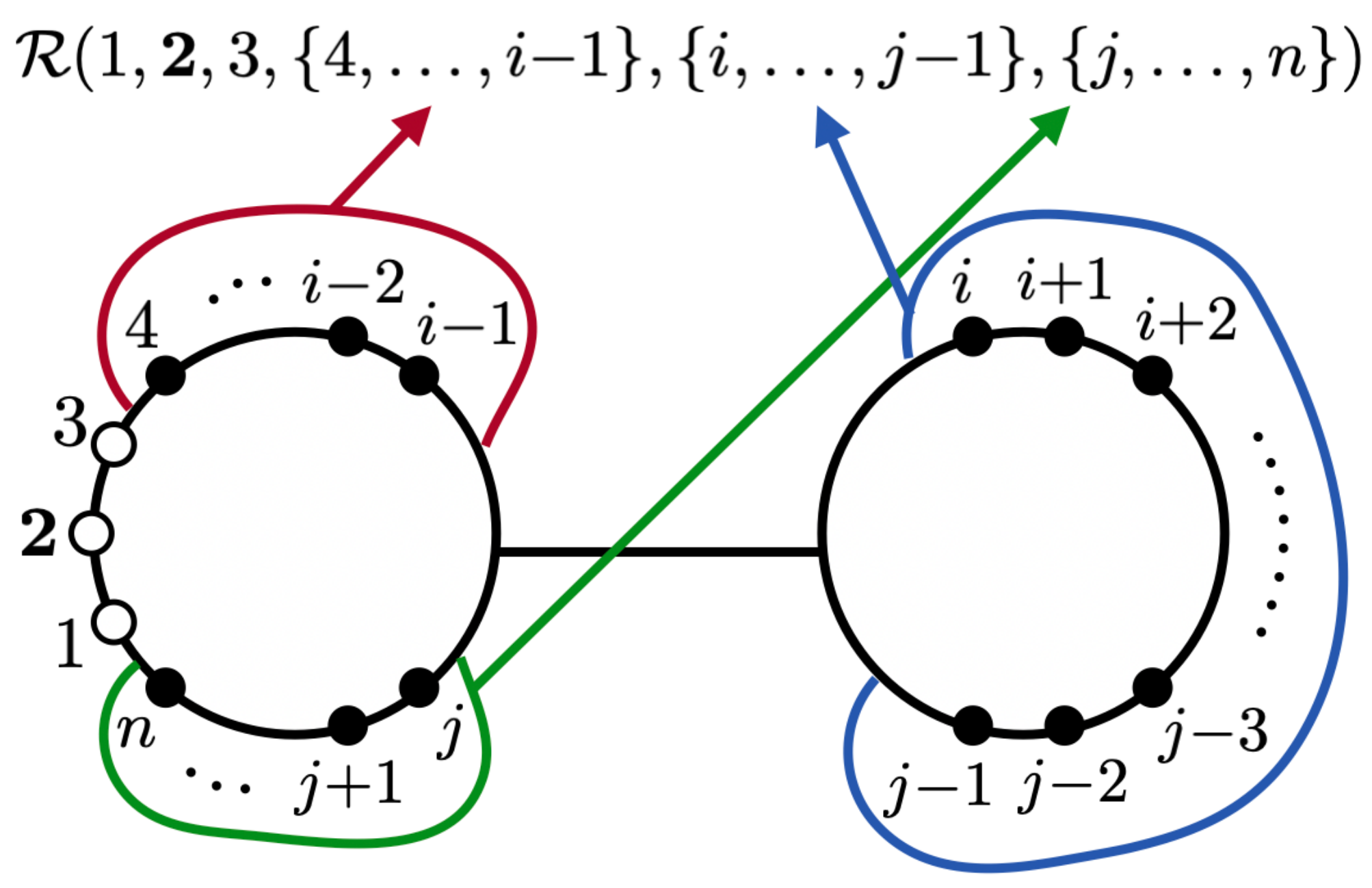}
$$
Note that this labeling is redundant for the Yang-Mills case as specifying indices $2,i,j$ already fixes all other labels due to cyclic symmetry. 

\section{New formulas for gravity amplitudes}

The main idea of this paper is to use the same factorization diagrams in the context of graviton amplitudes. We focus on the $n$-pt NMHV graviton helicity amplitude
\begin{equation}
    {\cal M}_n\equiv {\cal M}_n(1^-2^-3^-4^+5^+\dots n^+)
\end{equation}
Note that while we choose $1,2,3$ to be negative helicity gravitons, there is no notion of split helicity amplitude as gravitons are not ordered. In other words, this is a generic graviton amplitude and all other NMHV helicity amplitudes are related by a simple relabeling. The ${\cal N}=8$ superamplitude enjoys the full $S_n$ permutational symmetry in all labels. The helicity amplitude is invariant under its subgroup: $S_3\times S_{n{-}3}$, corresponding to separate permutations in three negative helicity labels $1,2,3$ and $(n{-}3)$ positive helicity labels $4,5,\dots,n$.

\subsection{${\cal G}$-invariants and main conjecture}

We define a ${\cal G}$-invariant to be the expression associated with one gravity factorization diagram. We use the same labeling as we introduced for ${\cal R}$-invariants,
\begin{equation}
    {\cal G}(1,{\bf 2},3,\{Q_1\},\{P\},\{Q_2\}) \label{Ginv}
\end{equation}
but the labels in $Q_1$, $P$ and $Q_2$ are now not ordered, and the formula (\ref{Ginv}) is permutational invariant in all labels $m\in \{Q_1\}$, $k\in \{P\}$ and $q\in \{Q_2\}$. We will often use the canonical ordering when drawing factorization diagrams but there is no cyclic symmetry present and the result must be symmetrized in these labels. As an example, we take a particular 12pt factorization diagram and the corresponding ${\cal G}$-invariant,
$$
\includegraphics[scale=.4]{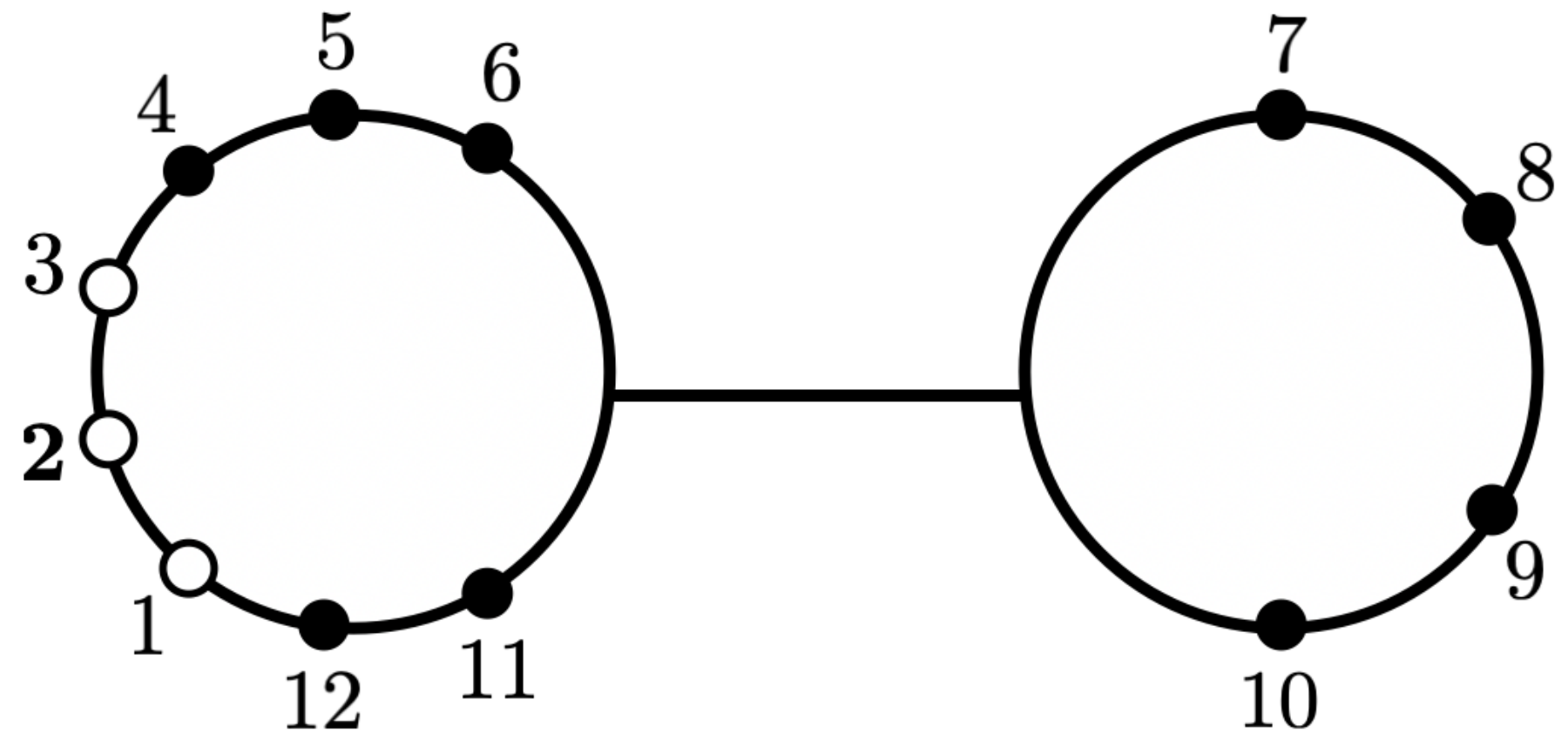}
$$
$$
{\cal G}(1,{\bf 2},3,\{4,5,6\},\{7,8,9,10\},\{11,12\})
$$
The ${\cal G}$-invariant is symmetric separately in labels $(4,5,6)$, $(7,8,9,10)$ and $(11,12)$, and therefore preserves $S_3\times S_4\times S_2$ subgroup of the $S_9$ permutation group in positive helicity labels, while the $S_3$ permutation group of negative helicity labels is broken completely. The labels $1,{\bf 2},3$ are again special and we refer to ${\bf 2}$ as the origin. In graviton case the ordering of the negative helicity labels is not fixed and we consider ${\cal G}$-invariants which differ from (\ref{Ginv}) by permuting $(1,2,3)$. 

We conjecture that the expression for the ${\cal G}$-invariant can be written in an analogous form as (\ref{Rexp}), namely
\begin{equation}
{\cal G}(1,{\bf 2},3,\{Q_1\},\{P\},\{Q_2\}) = \frac{{\cal H}^8}{(s_P)^2}\times (P_L\otimes P_R)\label{Ggen}
\end{equation}
where ${\cal H}$ is the helicity factor, $s_P=(\sum_{k\in P}p_k)^2$ is the kinematical invariant of the right blob, and $P_L$, $P_R$ are left and right blob functions. The helicity factor ${\cal H}$ is the same as in the Yang-Mills case (\ref{helicity}) if the indices are canonically ordered, ie. $Q_1=\{4,\dots,i{-}1\}$, $P=\{i,\dots,j{-}1\}$ and $Q_2=\{j,\dots,n\}$. We can always relabel and for any ordering we get 
\begin{equation}
{\cal H} = \la12\ra\la23\ra s_P \label{helicity2}
\end{equation}
The double pole $s_P$ cancels and we are left with 
\begin{equation}
{\cal G}(1,{\bf 2},3,\{Q_1\},\{P\},\{Q_2\}) =\la12\ra^8\la23\ra^8s_P^6\times (P_L\otimes P_R)\label{Ggen2}
\end{equation}
The main task is to write down the blob functions $P_L$ and $P_R$ and the merging rules and calculate arbitrary ${\cal G}$-invariant. While for gluons this was merely an interpretation of known expressions for $R$-invariants, for gravitons this is completely new. For writing down explicit formulas, we need to define more general spinor brackets, generalizations of (\ref{long2}), with arbitrary number of insertions of $(..)$ between the left and right spinors,
\begin{equation}
   \la a|(b^{(1)}_1\dots b^{(1)}_{n_1})(b^{(2)}_1\dots b^{(2)}_{n_2})\dots
    (b^{(m)}_1\dots b^{(m)}_{n_m})|c\ra \equiv \sum_{j_1,..j_m} \la a\,b^{(1)}_{j_1}\ra[b^{(1)}_{j_1} b^{(2)}_{j_2}]\dots \la b^{(m)}_{j_m} c\ra\label{longB}
\end{equation}
In (\ref{longB}) the index $m$ is even in order to start and end with holomorphic spinor $\la a|$ and $|c\ra$. Note that we can not longer used the simplified notation of (\ref{long1}), (\ref{long2}) using $x_{ij}$ as our labels are no longer ordered. We conjecture that the $n$-pt NMHV amplitude ${\cal M}_n$ is given by the sum over all $G$-invariants,

\vspace{0.05cm}

\begin{equation}
  {\cal M}_n =  \sum_{{\cal P}'(1,2,3)}\sum_{P,Q_1,Q_2}  {\cal G}(1,{\bf 2},3,\{Q_1\},\{P\},\{Q_2\}) \label{GRamp}
\end{equation}

\vspace{0.15cm}

\noindent  where the first sum is over all non-equivalent permutations of labels $1,2,3$, and the second sum is over all distributions of labels $4,5,\dots,n$ into three subsets. Note that $Q_1,Q_2$ can be empty, while $P$ needs to have at least two labels. This is a precise analogue of (\ref{split2}) using labeling of (\ref{LabExt}) in the Yang-Mills case.

\subsection{Starter: five and six-point amplitudes}

As a warm-up we will guess new expressions for few low-points NMHV amplitudes using the factorization diagram framework. The simplest NMHV amplitude is 5pt and can be obtained as the parity conjugate of the 5pt MHV expression (\ref{GR5}),
\begin{equation}
{\cal M}_5 = \frac{[45]^7\cdot(\la 12\ra[23]\la34\ra[41]-[12]\la23\ra[34]\la41\ra)}{[12][13][14][15][23][24][25][34][35]}\label{GR5s}
\end{equation}
Our general formula (\ref{GRamp}) reduces at 5pt to
\begin{equation}
    {\cal M}_5 = \sum_{{\cal P}'(1,2,3)}{\cal G}(1,{\bf 2},3,\{\},\{4,5\},\{\})\label{our5}
\end{equation}
where ${\cal P}'(1,2,3)$ denotes a sum over permutations of labels 1,2,3 modulo identical terms. Because the ${\cal G}$-invariant in (\ref{our5}) is symmetric in $1\leftrightarrow3$, the set of permutations ${\cal P}'(1,2,3)$ only contains three terms ${\cal P}'(1,2,3) = \Big\{(1,2,3),(1,3,2),(2,1,3)\Big\}$.
The ${\cal G}$-invariant in (\ref{our5}) is symmetric $4,5$, and hence no sum over positive helicity labels. This also means that there is really only one distinct factorization diagram,
$$
\includegraphics[scale=.37]{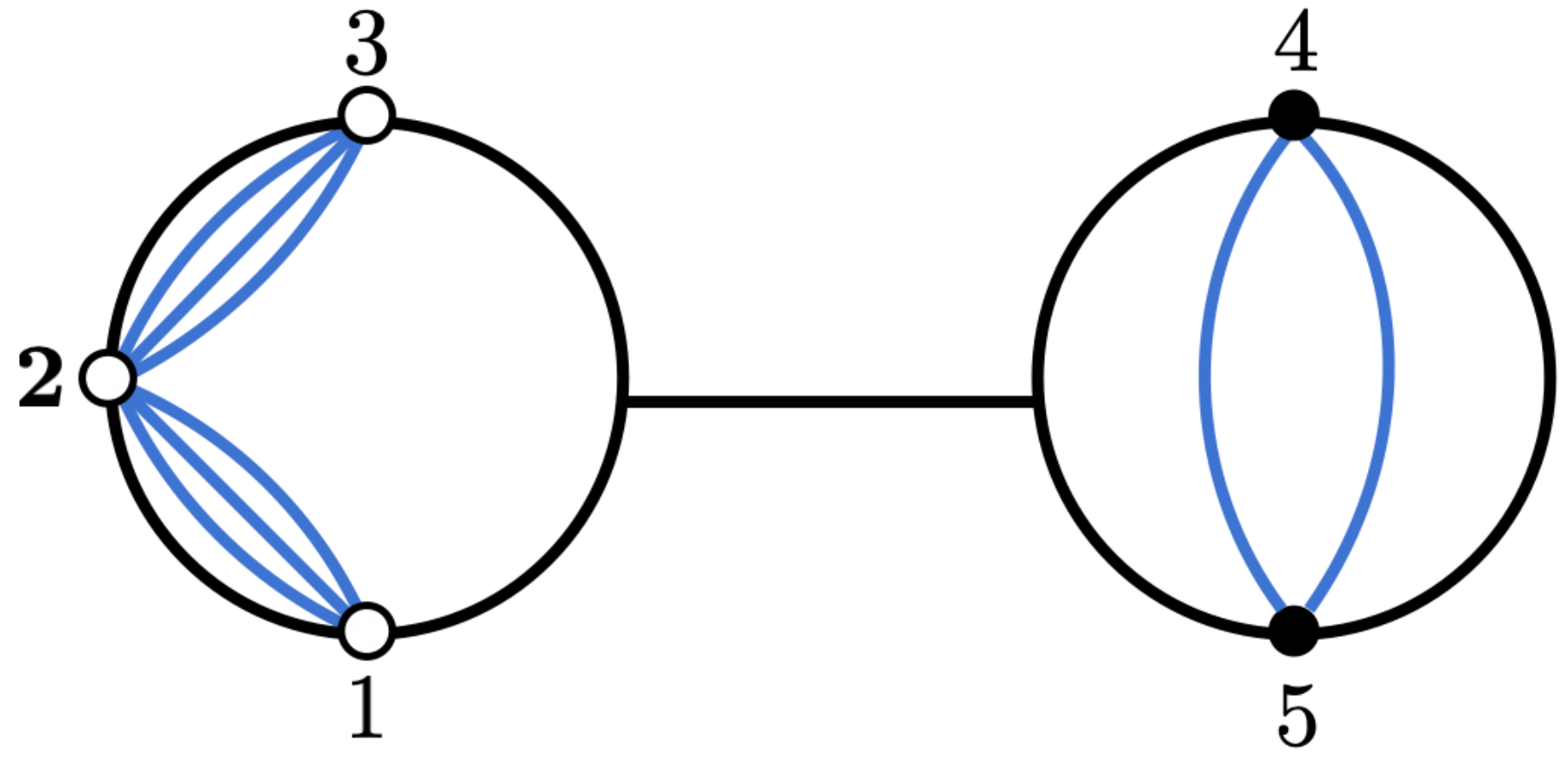}
$$
where we already added blue poles. There are third powers of poles $\la12\ra$ and $\la23\ra$ and the second power of $\la45\ra$. The left and right blob functions $P_L$ and $P_R$ are
\begin{align}
    P_L &=  {\color{blue}  \frac{1}{\la12\ra^3\la23\ra^3}} \times {\color{red} \frac{1}{[\la2|(3)\ast_1][\la2|(3)\ast_2][\la2|(1)\ast_3][\la2|(1)\ast_4]}}\times {\color{OliveGreen} \frac{1}{\la 2|(3)(45)|1\ra\,\la 2|(1)(45)|3\ra}}\nonumber\\
    P_R &= {\color{blue}  \frac{1}{\la45\ra^2}} \times {\color{red} \frac{1}{[\ast_1(4)|5\ra][\ast_2(5)|4\ra][\ast_3(4)|5\ra][\ast_4(5)|4\ra]}}
\end{align}
The green poles in $P_L$ are the same as in the Yang-Mills case, the number of red poles is doubled as they can ``end" on either index $4$ or $5$ in both clockwise and anti-clockwise directions. We have four mergers using the same rules (\ref{merge}), and we get from (\ref{Ggen2}),
\begin{align}
     {\cal G}(1,{\bf 2},3,\{\},\{4,5\},\{\}) &\\ & \hspace{-4cm}=\frac{{\cal H}^8}{s_{45}^2{\color{blue} \la12\ra^3\la23\ra^3\la45\ra^2}{\color{red} \la2|(3)(4)|5\ra\la2|(3)(5)|4\ra\la2|(1)(4)|5\ra\la2|(1)(5)|4\ra}{\color{OliveGreen}\la2|(3)(45)|1\ra\la2|(1)(45)|3\ra}}\nonumber
\end{align}
After plugging for the helicity factor ${\cal H}= \la12\ra\la23\ra s_{45}$ from (\ref{helicity2}) and simplifying the brackets due to kinematic relations we get
\begin{equation}
      {\cal G}(1,{\bf 2},3,\{\},\{4,5\},\{\}) = \frac{[45]^7\cdot \la12\ra\la23\ra}{[12][14][15][23][34][35]},
\end{equation}
and the permutational sum (\ref{our5}) gives 
\begin{align}
    {\cal M}_5 &=   {\cal G}(1,{\bf 2},3,\{\},\{4,5\},\{\}) +   {\cal G}(2,{\bf 1},3,\{\},\{4,5\},\{\}) +   {\cal G}(1,{\bf 3},2,\{\},\{4,5\},\{\})\nonumber  \\
    &= \frac{[45]^6\cdot \la12\ra\la23\ra}{[12][14][15][23][34][35]} + \frac{[45]^6\cdot \la12\ra\la13\ra}{[12][13][24][25][34][35]} + \frac{[45]^6\cdot \la13\ra\la23\ra}{[13][14][15][23][24][25]}
\end{align}
which is equal to (\ref{GR5s}) after few manipulations. This representation is different from any other available in literature, but 5pt is too special to draw any conclusions.

\medskip

The first really non-trivial case is the six-point NMHV amplitude. As reviewed earlier, this particular amplitude has been a subject of extensive studies in the past using BCFW recursion relations. There are two types of factorization diagrams
$$
\includegraphics[scale=.35]{YM56skel.pdf}\hspace{1.5cm}\includegraphics[scale=.35]{YM456skel.pdf}
$$
which is an analogue of two types of contributions to the Yang-Mills amplitude (\ref{A6YM}). The formula (\ref{GRamp}) reduces to 
\begin{equation}
    {\cal M}_6 =  \sum_{{\cal P}(1,2,3)}\sum_{{\cal P}'(4,5,6)}{\cal G}(1,{\bf 2},3,\{4\},\{5,6\},\{\}) + \sum_{{\cal P}'(1,2,3)}{\cal G}(1,{\bf 2},3,\{\},\{4,5,6\},\{\})\label{our6}
\end{equation}
The first term in (\ref{our6}) has $5\leftrightarrow6$ symmetry and
\begin{equation}
{\cal P}'(4,5,6) = \Big\{(4,5,6),(5,4,6),(6,4,5)\Big\}
\end{equation}
while the second term is symmetric in $1\leftrightarrow3$, 
\begin{equation}
{\cal P}'(1,2,3) = \Big\{(1,2,3),(2,1,3),(3,2,1)\Big\}
\end{equation}
and in total we get $3\times 6+3=21$ terms. To evaluate the first ${\cal G}$-invariant we add blue poles to the diagram,
$$
\raisebox{0pt}{\includegraphics[scale=.37]{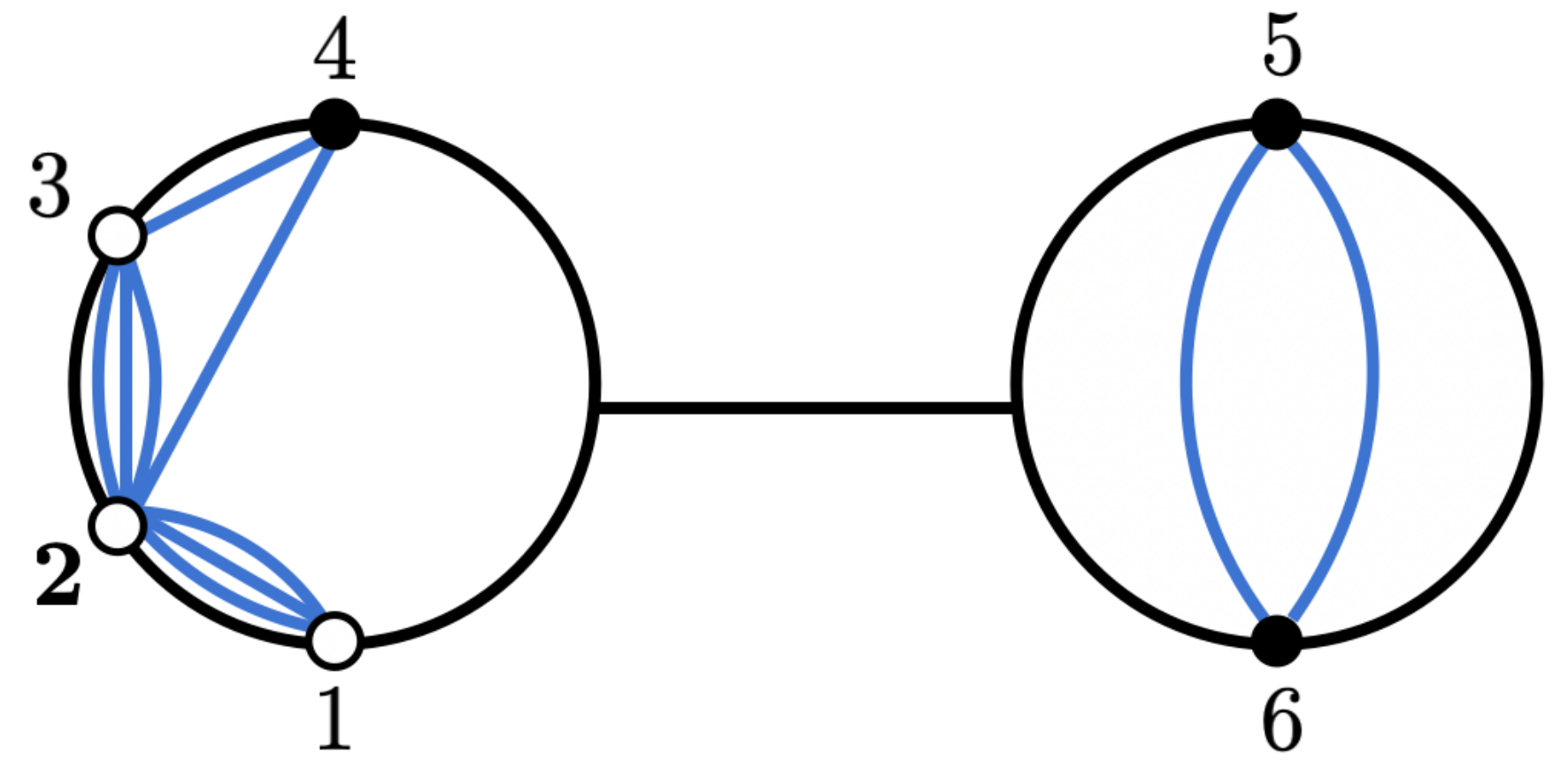}}
$$
and write the left and right blob functions,
\begin{align}
    P_L &=  {\color{blue}  \frac{\la2|3|4]}{\la12\ra^3\la23\ra^3\la34\ra\la24\ra}} \times {\color{red} \frac{1}{[\la2|(34)\ast_1][\la2|(34)\ast_2][\la2|(1)\ast_3][\la2|(1)\ast_4]}}\times {\color{OliveGreen} \frac{1}{\la 2|(34)(56)|1\ra\la 2|(1)(56)|4\ra}}\nonumber\\
    P_R &= {\color{blue}  \frac{1}{\la56\ra^2}} \times {\color{red} \frac{1}{[\ast_1(5)|6\ra][\ast_2(6)|5\ra][\ast_3(5)|6\ra][\ast_4(6)|5\ra]}} \label{PL1234}
\end{align}
The differences from the 5pt case are: (i) addition of few new blue poles, (ii) shift in the arguments of brackets $(3)\rightarrow (34)$, and (iii) new numerator associated with the left blob, not present in the Yang-Mills construction. Merging $P_L$ and $P_R$ in (\ref{Ggen}) gives
\begin{align}
     {\cal G}(1,{\bf 2},3,\{4\},\{5,6\},\{\}) &=\\
     &\hspace{-4.5cm}\frac{{\cal H}^8\times {\color{blue} \la2|3|4]}}{s_{56}^2{\color{blue}\la12\ra^3\la23\ra^3\la34\ra\la24\ra\la56\ra^2}{\color{red} \la 2|(34)(5)|6\ra\la 2|(34)(6)|5\ra\la 2|(1)(5)|6\ra\la 2|(1)(6)|5\ra}{\color{OliveGreen} \la2|(34)(56)|1\ra\la2|(1)(56)|4\ra}}\nonumber
\end{align}
The helicity factor (\ref{helicity2}) is ${\cal H} = \la 12\ra\la23\ra s_{56}$ and after some simplifications we get
\begin{equation}
    {\cal G}(1,{\bf 2},3,\{4\},\{5,6\},\{\}) = \frac{\la23\ra^6[56]^6\cdot \la12\ra[34]}{s_{234}[15][16]\la24\ra\la34\ra\la4|56|1]\la2|34|5]\la2|34|6]}\label{G56b}
\end{equation}
The second factorization diagram with blue poles is
$$
\includegraphics[scale=.37]{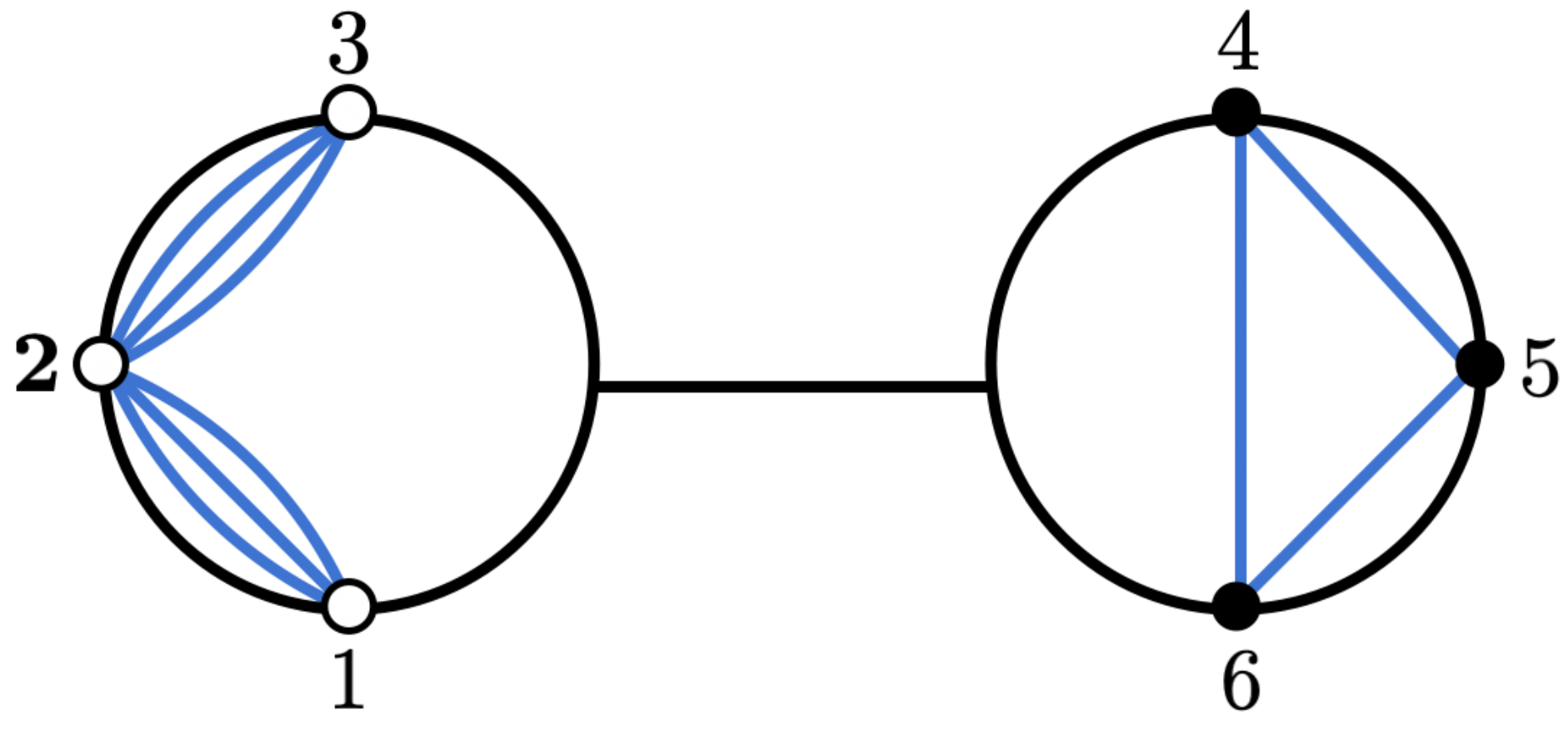}
$$
The left and right functions are now a bit more complicated,
\begin{align}
    P_L &=  {\color{blue}  \frac{1}{\la12\ra^3\la23\ra^3}} \times {\color{red} \frac{[\la\la2|(3)\diamond(1)|2\ra\ra]}{[\la2|(3)\ast_1][\la2|(3)\ast_2][\la2|(3)\ast_3][\la2|(1)\ast_4][\la2|(1)\ast_5][\la2|(1)\ast_6]}}\nonumber\\
    &\hspace{2cm}\times {\color{OliveGreen} \frac{1}{\la 2|(3)(456)|1\ra\la 2|(1)(456)|3\ra}}\\
    P_R &= {\color{blue}  \frac{1}{\la45\ra\la46\ra\la56\ra}} \times {\color{red} \frac{[\diamond(456)(6)(5)(456)\diamond]}{[\ast_1(45)|6\ra][\ast_2(46)|5\ra][\ast_3(56)|4\ra][\ast_4(45)|6\ra][\ast_5(46)|5\ra][\ast_6(56)|4\ra]}} \label{PL123}
\end{align}
The green poles are unchanged, as well as the blue poles in the left blob. However, there are now six red poles instead of four, three clockwise ${\color{red} \la2|(3)(456)|a\ra}$ and three anti-clockwise ${\color{red} \la 2|(1)(456)|a\ra}$ for $a=4,5,6$. There is also a non-trivial red numerator which needs to me merged between both blobs,
\begin{equation}
{\color{red} [\la\la 2|(3)\diamond(1)|2\ra\ra]}\otimes{\color{red} [\diamond(456)(6)(5)(456)\diamond]} = {\color{red} \la\la2|(3)(456)(6)(5)(456)(1)|2\ra\ra}\label{num456}
\end{equation}
The symbol $\diamond$ inserts the numerator expression from $P_R$ into the expression in $P_L$. Here we defined the new kinematical object, the {\bf double-angle bracket} $\la\la\dots\ra\ra$. It can be expanded in terms of standard kinematical invariants in multiple ways
\begin{align}
    \la\la 2|(3)(456)(6)(5)(456)(1)|2\ra\ra&=\la 2|(3)(456)(6)(5)(456)(1)|2\ra + s_{456}\la2|(3)(5)(6)(1)|2\ra\nonumber\\
    &\hspace{-3cm}= \la 2|(3)(5)(456)(6)(456)(1)|2\ra - \la 2|(3)(6)(456)(5)(456)(1)|2\ra\nonumber\\
    &\hspace{-3cm}= \la 2|(3)(456)(5)(456)(6)(1)|2\ra - \la 2|(1)(456)(6)(456)(5)(1)|2\ra \label{longB2}
\end{align}
using the long brackets we defined earlier (\ref{longB}). Importantly, (\ref{num456}) is antisymmetric in labels $4,5,6$ and symmetric $1,3$. This is not manifest in any particular expansion (\ref{longB2}) which suggests we should consider the new bracket $\la\la 2|(1)(456)(5)(6)(456)(3)|2\ra\ra$ as one object and do not really expand it. As a result we obtain,
\begin{align}
 {\cal G}(1,{\bf 2},3,\{\},\{4,5,6\},\{\}) &= \label{G456a}\\
 &\hspace{-2cm}\frac{{\cal H}^8\times {\color{red}\la\la2|(3)(456)(6)(5)(456)(1)|2\ra\ra}}{\begin{tabular}{c}
$s_{456}^2 {\color{blue} \la12\ra^3\la23\ra^3\la45\ra\la46\ra\la56\ra}{\color{OliveGreen} \la2|(3)(456)|1\ra\la 2|(1)(456)|3\ra}{\color{red} \la 2|(3)(45)|6\ra}$\\
${\color{red}\la2|(3)(46)|5\ra\la2|(3)(56)|4\ra\la2|(1)(45)|6\ra\la2|(1)(46)|5\ra\la2|(1)(56)|4\ra}$\end{tabular}}\nonumber
\end{align}
Simplifying (\ref{G456a}) and plugging for the helicity factor ${\cal H}_{456} = \la12\ra\la23\ra s_{456}$ gives
\begin{align}
 {\cal G}(1,{\bf 2},3,\{\},\{4,5,6\},\{\}) &= \label{G456b}\\
 &\hspace{-2cm}\frac{s_{456}^6\cdot \la\la 2|(3)(456)(6)(5)(456)(1)|2\ra\ra}{[12][23]\la45\ra\la46\ra\la56\ra\la6|45|3]\la5|46|3]\la4|56|3]\la6|45|1]\la5|46|1]\la4|56|1]}\nonumber
\end{align}
The full six-point NMHV amplitude is given by the sum (\ref{our6}) over permutations of (\ref{G56b}) and (\ref{G456b}). We checked the formula to be numerically equal to other expressions in the literature, though it looks quite different and does not seem to be related to any BCFW or KLT representations. It makes manifest the complete $S_3\times S_3$ symmetry of this helicity amplitude. There are some new features which distinguishes the structure of ${\cal G}$-invariants from ${\cal R}$-invariants:
\begin{itemize}
\item {\bf Prefactor}: Contains the helicity factor and factorization pole in (\ref{Ggen}, and it is a square of the Yang-Mills prefactor (\ref{Rexp}). 
\item {\bf Structure of poles}:  Green poles are the same as in Yang-Mills, bu there are more blue poles (some of them quadratic or cubic) and more red poles, though the merging procedure is the same. 
\item {\bf Non-trivial numerators}: Beyond the simplest case there are non-trivial numerators associated with both left and right blobs. 
\end{itemize}
The six-point amplitude is still too special and the general rules become more apparent when we calculate the seven-point amplitude.

\subsection{Seven-point amplitude}

The seven-point amplitude ${\cal M}_7(1^-2^-3^-4^+5^+6^+7^+)$ is given by the sum over four types of factorization diagrams as reflected in the formula (\ref{GRamp}) in terms of ${\cal G}$-invariants,
\begin{align}
{\cal M}_7 &= \sum_{{\cal P}(1,2,3)}\sum_{{\cal P}'(4,5,6,7)}  {\cal G}(1,{\bf 2},3,\{4\},\{5,6\},\{7\}) + \sum_{{\cal P}(1,2,3)}\sum_{{\cal P}'(4,5,6,7)} {\cal G}(1,{\bf 2},3,\{4,5\},\{6,7\},\{\})   \nonumber\\
&\hspace{0.1cm}+ \sum_{{\cal P}(1,2,3)}\sum_{{\cal P}'(4,5,6,7)}  {\cal G}(1,{\bf 2},3,\{4\},\{5,6,7\},\{\}) + \sum_{{\cal P}(1,2,3)}{\cal G}(1,{\bf 2},3,\{\},\{4,5,6,7\},\{\}) \label{GR7}
\end{align}
The number of terms in the sums depends on the symmetries of the ${\cal G}$-invariants. The first term is symmetric in $5,6$ and exchange $3,4 \leftrightarrow 1,7$; the second term is symmetric separately in $4,5$ and $6,7$; the third term is symmetric in $5,6,7$; the last term is symmetric in $1,3$ and $4,5,6,7$. These symmetries are obvious from the labelings in (\ref{GR7}). In total we get $36 + 36 + 24 + 3 = 99$ terms in ${\cal M}_7$. 

\medskip

\noindent $\bullet$ The first term in (\ref{GR7}) corresponds to the factorization diagram,
$$
\includegraphics[scale=.37]{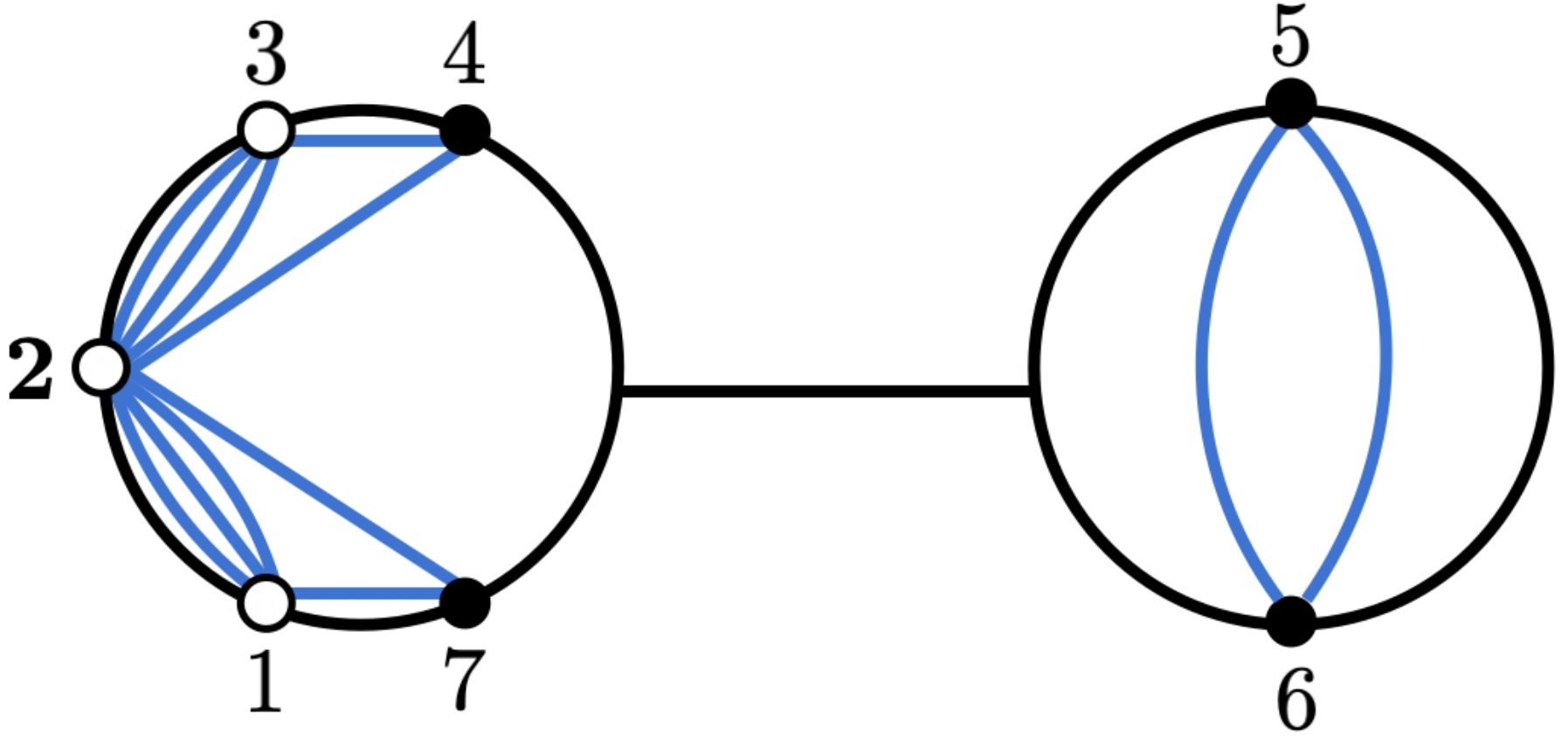}
$$
The left and right blob functions in (\ref{Ggen}) are
\begin{align}
    P_L &=  {\color{blue}  \frac{\la2|3|4]\times \la2|1|7]}{\la12\ra^3\la23\ra^3\la34\ra\la24\ra\la17\ra\la27\ra}} \times {\color{red} \frac{1}{[\la2|(34)\ast_1][\la2|(34)\ast_2][\la2|(17)\ast_3][\la2|(17)\ast_4]}}\nonumber\\
    &\hspace{0.5cm}\times {\color{OliveGreen} \frac{1}{\la 2|(34)(56)|7\ra\la 2|(17)(56)|4\ra}}\\
    P_R &= {\color{blue}  \frac{1}{\la56\ra^2}} \times {\color{red} \frac{1}{[\ast_1(5)|6\ra][\ast_2(6)|5\ra][\ast_3(5)|6\ra][\ast_4(6)|5\ra]}}
\end{align}
where the blue part of the left blob is basically two copies of (\ref{PL1234}). The red and green poles follow are unchanged. Merging the $P_L$ and $P_R$ terms we get
\begin{align}
(P_L\otimes P_R) &=\frac{{\color{blue} \la2|3|4]\la2|1|7]}}{
\begin{tabular}{c}
${\color{blue} \la12\ra^3\la23\ra^3\la34\ra\la24\ra\la17\ra\la27\ra\la56\ra^2}{\color{red} \la2|(34)(5)|6\ra\la2|(34)(6)|5\ra}$\\${\color{red} \la2|(17)(5)|6\ra\la2|(17)(6)|5\ra} {\color{OliveGreen} \la 2|(34)(56)|7\ra\la 2|(17)(56)|4\ra}$\end{tabular}}\label{P71234}
\end{align}
Plugging for the helicity factor ${\cal H}=\la12\ra\la23\ra s_{56}$ into (\ref{Ggen}) and simplifying some of the long brackets we get
\begin{align}
{\cal G}(1,{\bf 2},3,\{4\},\{5,6\},\{7\})  &=\\
&\hspace{-3cm}-\frac{\la12\ra^6\la23\ra^6[56]^6\cdot [34][17]}{\la24\ra\la34\ra\la17\ra\la27\ra\la2|34|5]\la2|34|6]\la2|17|5]\la2|17|6]\la2|(34)(56)|7\ra\la2|(17)(56)|4\ra}\nonumber
\end{align}

\medskip

\noindent $\bullet$ The second factorization diagram is
$$
\includegraphics[scale=.37]{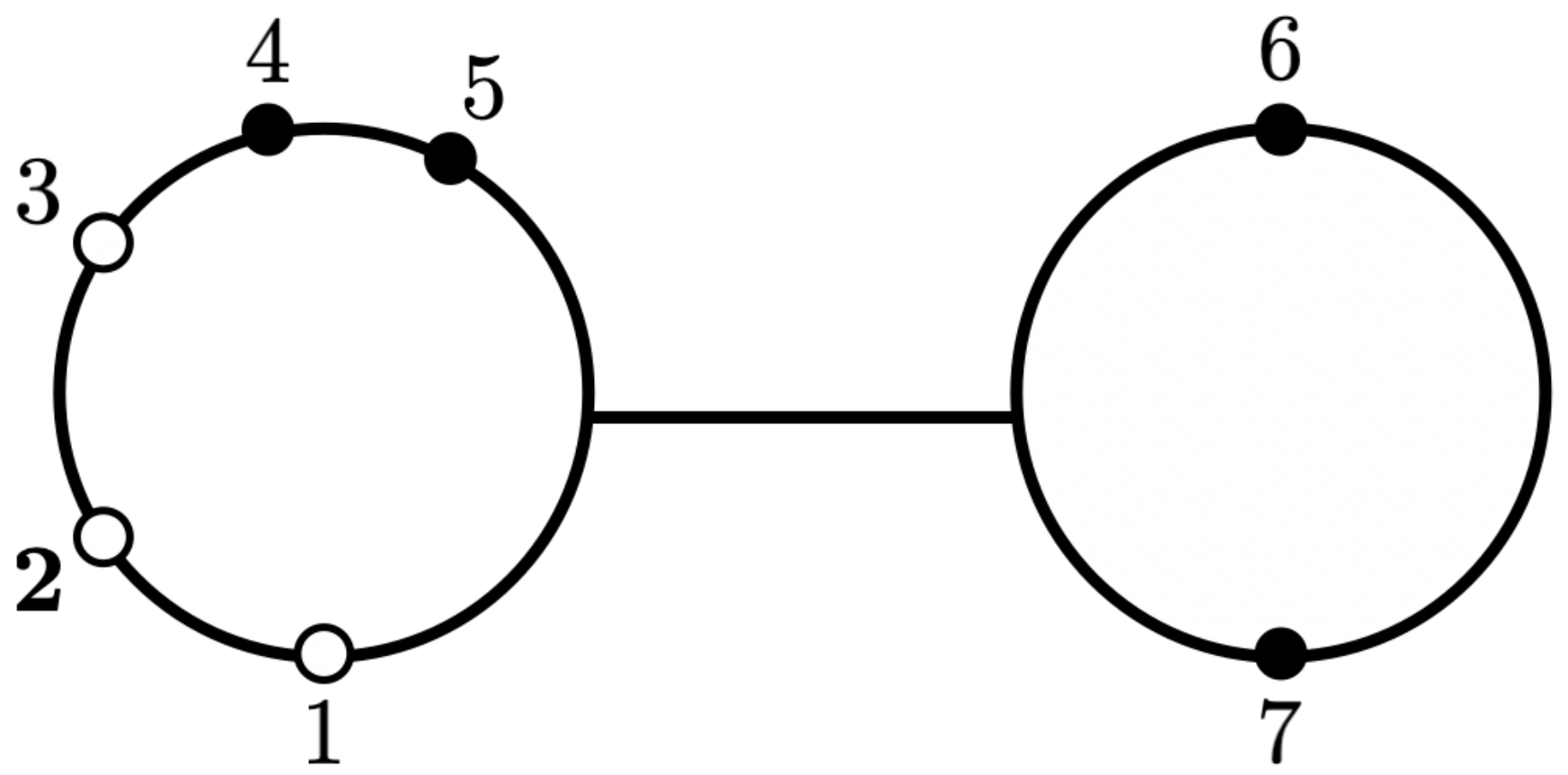}
$$
We did not include the blue poles in the diagram above because that requires fixing the ordering of labels $4,5$ in the left blob. There are two options, and we have to symmetrize over both of them. For a fixed ordering of $4,5$ we get a {\it partial factorization diagram},
$$
\includegraphics[scale=.37]{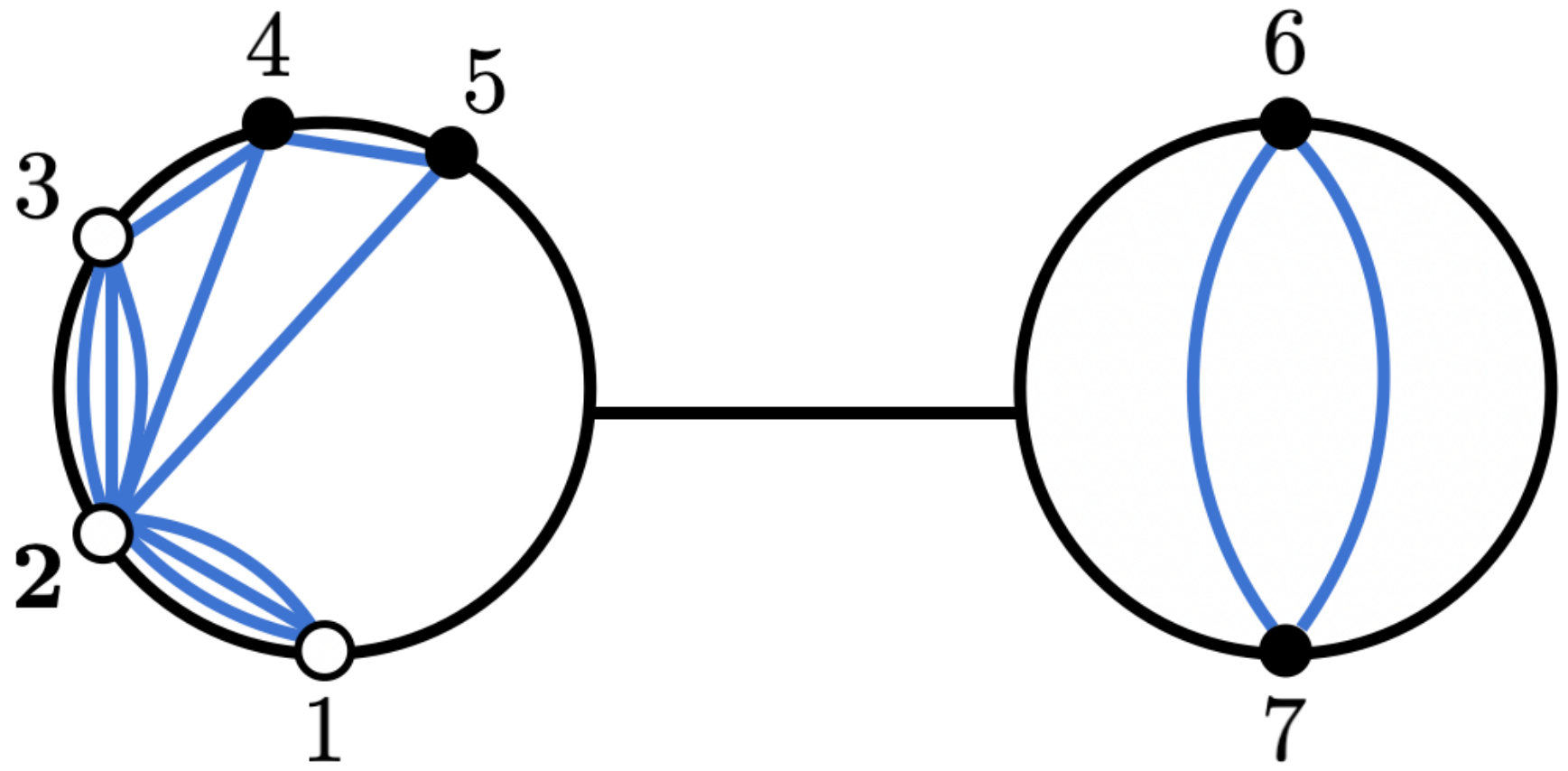}
$$
The left blob function for this diagram with fixed ordering of $4,5$ is 
\begin{align}
    P_L^{(45)} &=  {\color{blue}  \frac{\la2|3|4]\times \la2|34|5]}{\la12\ra^3\la23\ra^3\la34\ra\la45\ra\la24\ra\la25\ra}} \times {\color{red} \frac{1}{[\la2|(345)\ast_1][\la2|(345)\ast_2][\la2|(1)\ast_3][\la2|(1)\ast_4]}}\nonumber\\
    &\hspace{0.5cm}\times {\color{OliveGreen} \frac{1}{\la 2|(345)(67)|1\ra\la 2|(1)(67)|5\ra}}
\end{align}
We can clearly see some pattern for $P_L$. The blue poles are just given by cyclic holomorphic poles ${\color{blue} \la i\,i{+}1\ra}$, with third power of ${\color{blue} \la12\ra}$, ${\color{blue} \la23\ra}$, in addition to ${\color{blue} \la 2k\ra}$ poles. The numerator of the left blob is is given by the product of terms ${\color{blue}\la 2|P_1|k]}$ where $P_1$ is the sum of momenta between $2$ and $k$ in clockwise or anti-clockwise direction. The green and red poles are given by simple shifts in previous cases. The full left blob function is then correspond to the symmetrization over $4,5$.
\begin{equation}
P_L = \sum_{{\cal P}(4,5)} P_L^{(45)} \label{PL12345}
\end{equation}
The right blob function $P_R$ is identical to (\ref{P71234}) up to a simple relabeling
\begin{equation}
P_R = {\color{blue}  \frac{1}{\la67\ra^2}} \times {\color{red} \frac{1}{[\ast_1(6)|7\ra][\ast_2(7)|6\ra][\ast_3(6)|7\ra][\ast_4(7)|6\ra]}} \label{PR12345}
\end{equation}
Plugging (\ref{PL12345}) and (\ref{PR12345}) into (\ref{Ggen}) we get
\begin{equation}
{\cal G}(1,{\bf 2},3,\{4,5\},\{6,7\},\{\})  = \frac{{\cal H}^4}{s_{67}^2}\times (P_L\otimes P_R)
\end{equation}
where the helicity factor is ${\cal H}=\la12\ra\la23\ra s_{67}$ and 
\begin{align}
(P_L\otimes P_R) &=\frac{{\color{blue} \la2|3|4]\la2|34|5]}}{
\begin{tabular}{c}
${\color{blue} \la12\ra^3\la23\ra^3\la34\ra\la45\ra\la24\ra\la25\ra\la67\ra^2}{\color{red} \la2|(345)(6)|7\ra\la2|(345)(7)|6\ra}$\\${\color{red} \la2|(1)(6)|7\ra\la2|(1)(7)|6\ra} {\color{OliveGreen} \la 2|(345)(67)|1\ra\la 2|(1)(67)|5\ra}$\end{tabular}}
\end{align}
After simplifications we get
\begin{equation}
{\cal G}(1,{\bf 2},3,\{4,5\},\{6,7\},\{\})  = \sum_{{\cal P}(4,5)}\frac{\la23\ra^6[67]^6\cdot \la12\ra[34]\la2|34|5]}{s_{167}[16][17]\la24\ra\la25\ra\la34\ra\la45\ra\la2|17|6]\la2|16|7] \la5|67|1]}
\end{equation}

\medskip

\noindent $\bullet$ The third term in (\ref{GR7}) is represented by the factorization diagram
$$
\includegraphics[scale=.37]{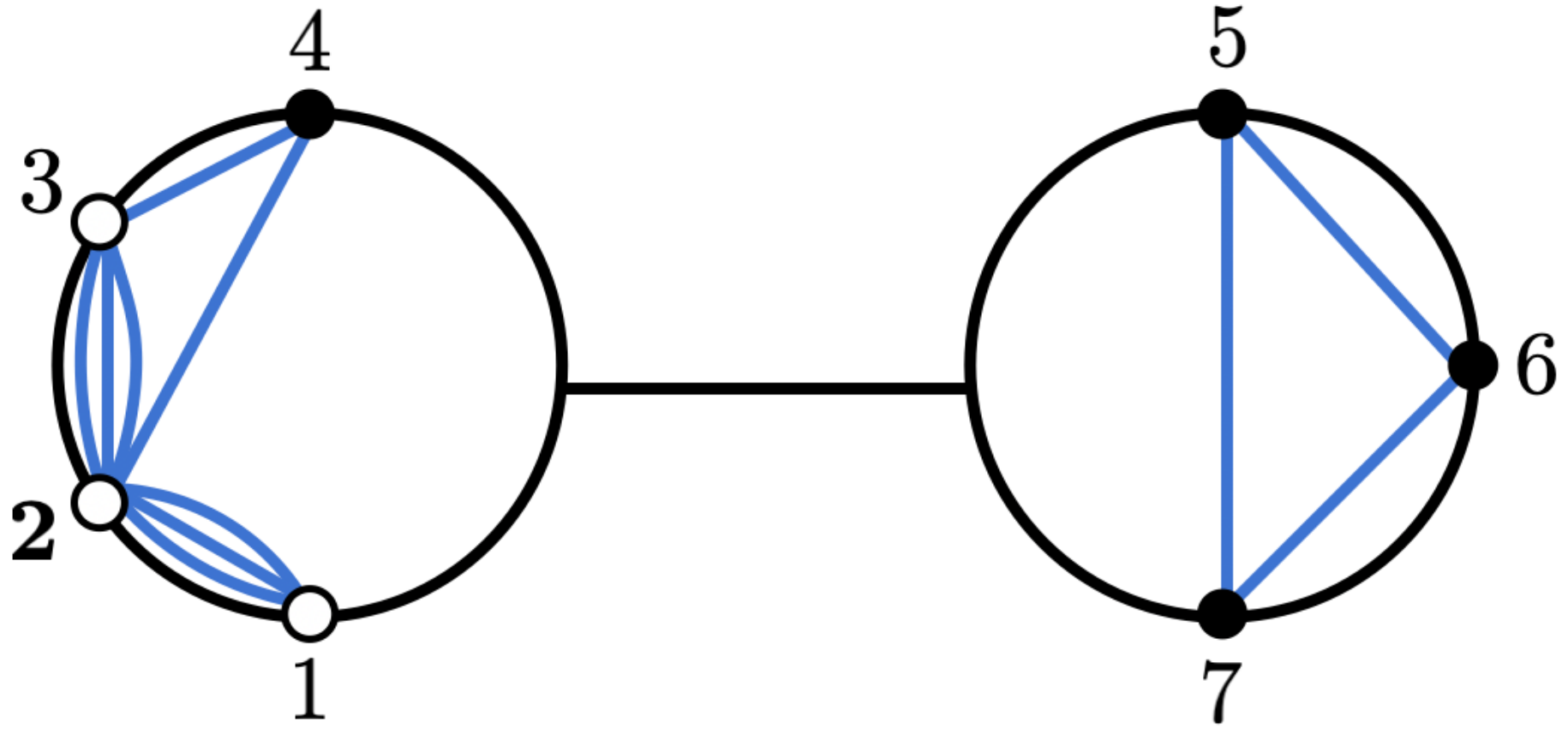}
$$
The blob functions are generalizations of (\ref{PL1234}) and (\ref{PL123}),
\begin{align}
    P_L &=  {\color{blue}  \frac{ \la2|3|4]}{\la12\ra^3\la23\ra^3\la34\ra\la24\ra}} \times {\color{red} \frac{[\la\la2|(34)\diamond(1)|2\ra\ra]}{[\la2|(34)\ast_1][\la2|(34)\ast_2][\la2|(34)\ast_3][\la2|(1)\ast_4][\la2|(1)\ast_5][\la2|(1)\ast_6]}}\nonumber\\
    &\hspace{2cm}\times {\color{OliveGreen} \frac{1}{\la 2|(34)(567)|1\ra\la 2|(1)(567)|4\ra}}\\
    P_R &= {\color{blue}  \frac{1}{\la56\ra\la57\ra\la67\ra}} \times {\color{red} \frac{[\diamond(567)(6)(5)(567)\diamond]}{[\ast_1(56)|7\ra][\ast_2(57)|6\ra][\ast_3(67)|5\ra][\ast_4(56)|7\ra][\ast_5(57)|6\ra][\ast_6(67)|5\ra]}}
\end{align}
which differ from (\ref{PL123}) by simple relabeling and adding blue poles and numerator in $P_L$. Merging $P_L$ and $P_R$ produces
\begin{align}
(P_L\otimes P_R) &=\frac{{\color{blue} \la2|3|4]}\cdot{\color{red}\la\la2|(34)(567)(6)(5)(567)(1)|2\ra\ra}}{
\begin{tabular}{c}
${\color{blue} \la12\ra^3\la23\ra^3\la34\ra\la24\ra\la56\ra\la57\ra\la67\ra}{\color{red} \la2|(34)(56)|7\ra\la2|(34)(57)|6\ra\la2|(34)(67)|5\ra}$\\${\color{red} \la2|(1)(56)|7\ra\la2|(1)(57)|6\ra\la2|(1)(67)|5\ra} {\color{OliveGreen} \la 2|(34)(567)|1\ra\la 2|(1)(567)|4\ra}$\end{tabular}} \label{PLR1}
\end{align}
Plugging into (\ref{Ggen}) for (\ref{PLR1}) and ${\cal H} = \la12\ra\la23\ra s_{567}$ gives after few manipulations
\begin{align}
{\cal G}(1,{\bf 2},3,\{4\},\{5,6,7\},\{\})  &=\\
&\hspace{-5.5cm} -\frac{\la23\ra^6s_{567}^6\cdot[34]\la\la2|(34)(567)(6)(5)(567)(1)|2\ra\ra}{
s_{234}\la24\ra\la34\ra\la56\ra\la57\ra\la67\ra\la4|567|1]\la5|67|1]\la6|57|1]\la7|56|1]\la2|(34)(56)|7\ra\la2|(34)(57)|6\ra\la2|(34)(67)|5\ra}\nonumber
\end{align}
Note that this formula is symmetric in $5,6,7$ labels. The double-angle bracket in the numerator is anti-symmetric in $5,6,7$ and it can be expanded in various ways in terms of more standard kinematical objects as in (\ref{longB2}), e.g.
\begin{equation*}
\la\la2|(34)(567)(6)(5)(567)(1)|2\ra\ra = \la2|(34)(567)(6)(5)(1)|2\ra + s_{567}\la2|(34)(5)(6)(1)|2\ra
\end{equation*}

\medskip

\noindent $\bullet$ The last term in (\ref{GR7}) is new and we have to guess additional rules for the right blob. We start with the general formula,
\begin{equation}
{\cal G}(1,{\bf 2},3,\{\},\{4,5,6,7\},\{\})  = \frac{{\cal H}^8}{s_{4567}^2}\times (P_L\otimes P_R) \label{G4567}
\end{equation}
where the left blob function $P_L$ is given by
\begin{align}
    P_L &=  {\color{blue} \frac{1}{\la12\ra^3\la23\ra^3}} \times {\color{red} \frac{[\la\la2|(3)\diamond(1)|2\ra\ra]}{[\la2|(3)\ast_1][\la2|(3)\ast_2][\la2|(3)\ast_3][\la2|(1)\ast_4][\la2|(1)\ast_5][\la2|(1)\ast_6]}}\nonumber\\
    &\hspace{0.5cm}\times {\color{OliveGreen} \frac{1}{\la 2|(3)(4567)|1\ra\la 2|(1)(4567)|3\ra}}\label{PL7}
\end{align}
This is basically the same expression as we associated with the second 6pt factorization diagram. We already saw a significant difference between the case with two and three points in the right blob. With two points we had only two red half-poles in clockwise and two in anti-clockwise direction, while for three points we had 3+3 half-poles. Now, we have four points in the right blob and naively we would expect 4+4 half-poles, and more with increasing number of points. However, our $P_L$ expression (\ref{PL7}) has only 3+3 half-poles again. The claim is that this is a generic situation, and in fact the two-point right blob was a boundary case.

On the other hand, the right blob contains four labels $4,5,6,7$ and $P_R$ can have 4+4 red half-poles of the form ${\color{red} [\ast (4567)|x\ra}$ for $x\in\{4,5,6,7\}$ to be merged with their counterparts in $P_L$, so how to choose them? The resolution is that $P_R$ must be expanded as a sum of terms. We have to choose the origin of the right blob, for example label $7$, but any other point would work as well and the final result does not depend on this choice. Then we express the right blob as a sum of three terms 
$$
\raisebox{-39pt}{\includegraphics[scale=.37]{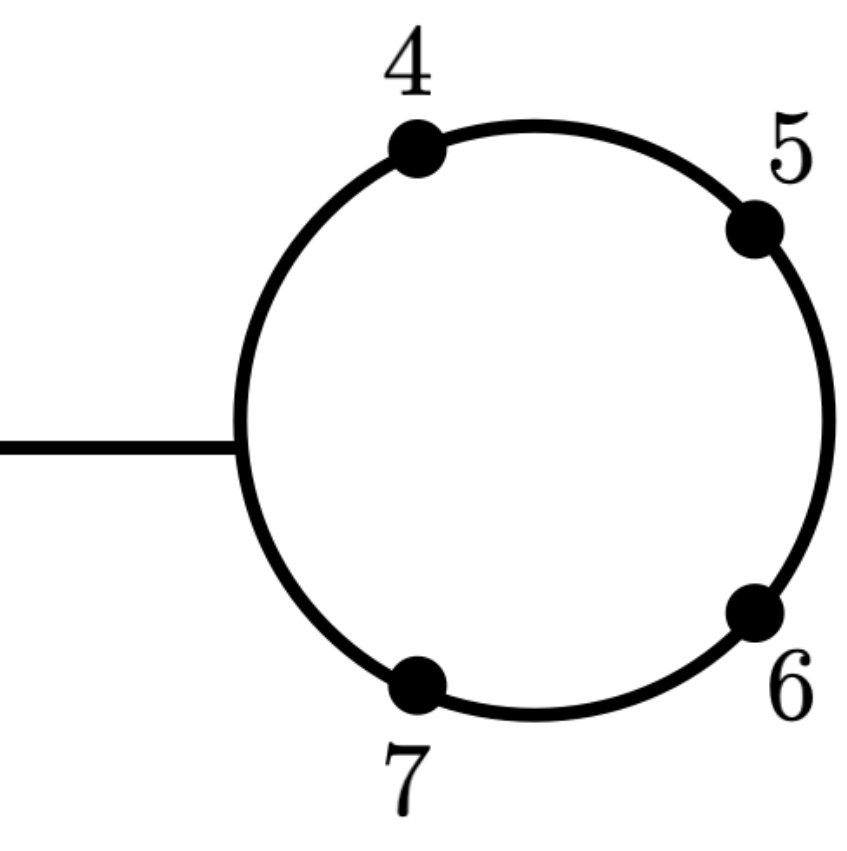}}\hspace{0.3cm} = \hspace{0.3cm} \raisebox{-40pt}{\includegraphics[scale=.37]{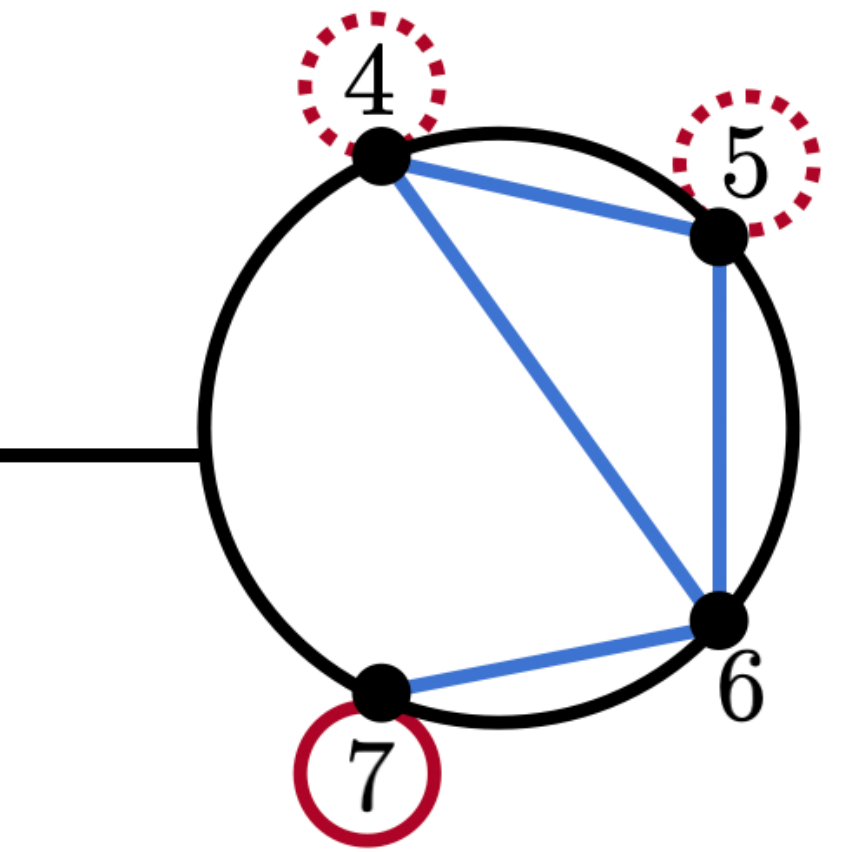}}\hspace{0.3cm} + \hspace{0.3cm} \raisebox{-40pt}{\includegraphics[scale=.37]{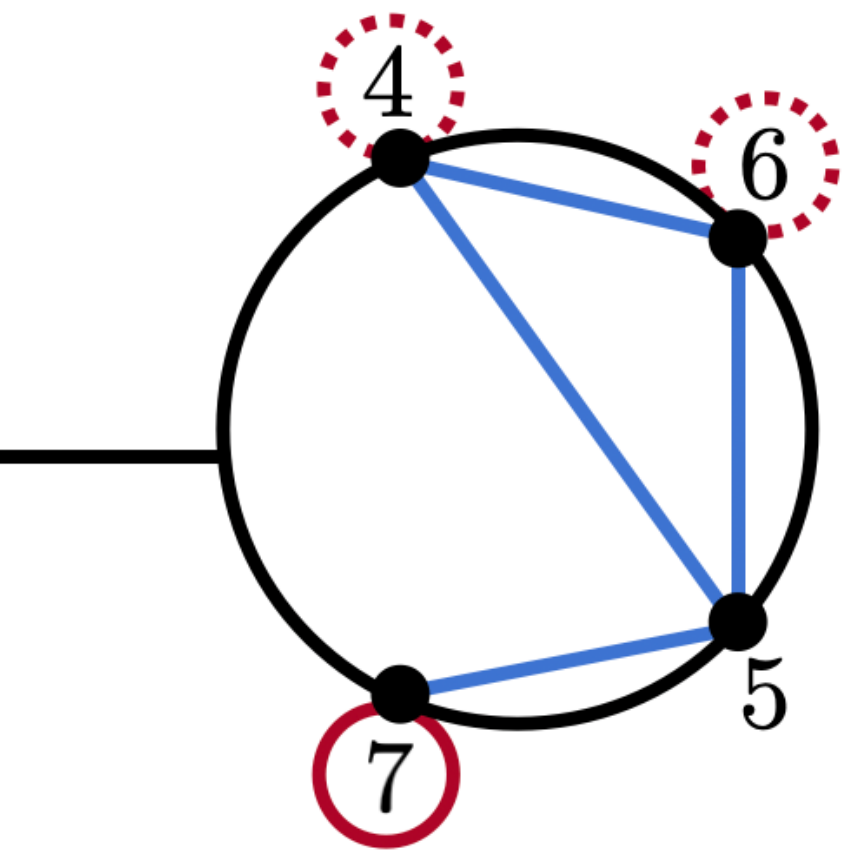}}\hspace{0.3cm} + \hspace{0.3cm} \raisebox{-41pt}{\includegraphics[scale=.37]{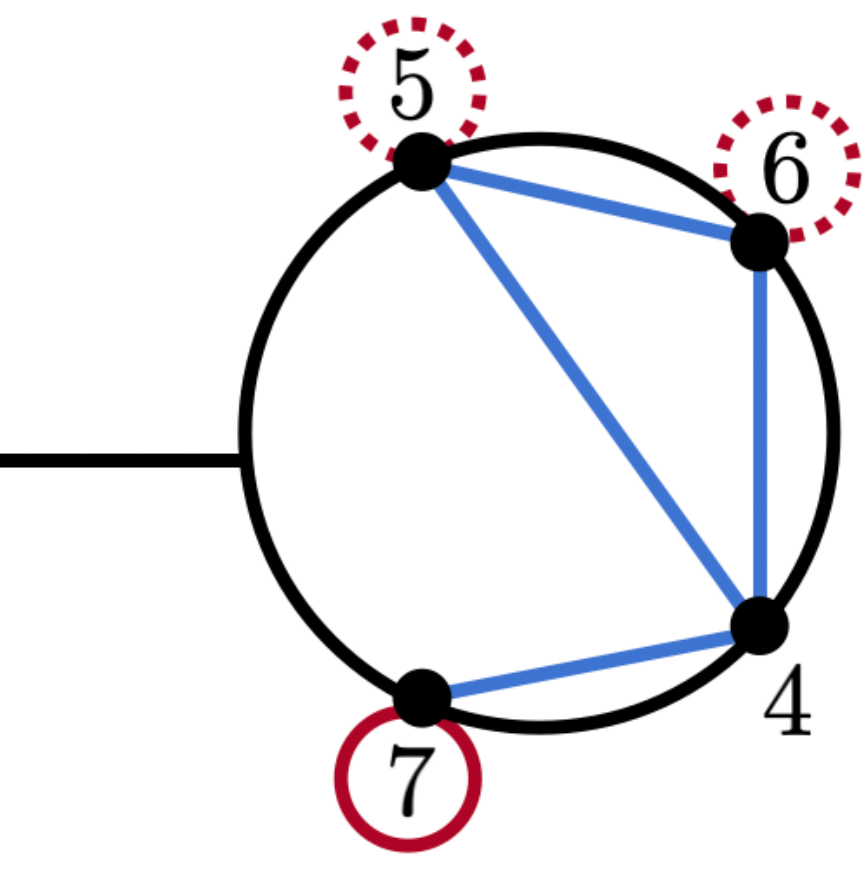}}
$$
Each term has three labels encircled: $7$ (full circle as origin) and two other labels (dotted circles). These circles denote labels which will appear in the red half-poles. We also added blue poles which are different for each term in the expansion. Note that each diagram is symmetric in two labels with dotted circles, so we have only three terms in the expansion rather than six. All three terms are related by relabeling and we can write the full right blob as sum over partial right blobs,
$$
\raisebox{-39pt}{\includegraphics[scale=.37]{GR7pt4aa.pdf}}\hspace{0.3cm} = \sum_{{\cal P}'(4,5,6)}\raisebox{-40pt}{\includegraphics[scale=.37]{GR7pt4a.pdf}}
$$
where ${\cal P}'(4,5,6)=\{(4,5,6),(4,6,5),(5,6,4)\}$. We decompose $P_R$ as
\begin{equation}
P_R = \sum_{{\cal P}'(4,5,6)} P_R^{(456)} \label{PR4567}
\end{equation}
where we denoted $P_R^{(456)}$ the right blob function for the first term with fixed ordering and encircled labels $4,5$ in addition to $7$,
\begin{equation}
P_R^{(456)} = {\color{blue} \frac{[67]}{\la45\ra\la46\ra\la56\ra\la67\ra}}\times {\color{red} \frac{[\diamond(4567)(5)(4)(4567)\diamond]}{[\ast_1(456)|7\ra][\ast_2(467)|5\ra][\ast_3(567)|4\ra][\ast_5(456)|7\ra][\ast_6(467)|5\ra][\ast_7(567)|4\ra]}}
\end{equation}
There is an extra numerator ${\color{blue} [67]}$ which was not present in the four-label case. We labeled this numerator with blue color because it only depends on the right blob labels, while the red poles and red numerator need to be merged with the left blob. Merging $P_L$ and $P_R$ leads to
\begin{equation}
(P_L\otimes P_R) =\hspace{-0.3cm} \sum_{{\cal P}'(4,5,6)}\hspace{-0.3cm} \frac{{\color{blue}[67]}\cdot{\color{red}\la\la2|(3)(4567)(5)(4)(4567)(1)|2\ra\ra}}{
\begin{tabular}{c}
${\color{blue} \la12\ra^3\la23\ra^3\la45\ra\la46\ra\la56\ra\la67\ra}{\color{red} \la2|(3)(456)|7\ra\la2|(3)(567)|4\ra\la2|(3)(467)|5\ra}$\\${\color{red} \la2|(1)(456)|7\ra\la2|(1)(567)|4\ra\la2|(1)(467)|5\ra} {\color{OliveGreen} \la 2|(3)(4567)|1\ra\la 2|(1)(4567)|3\ra}$\end{tabular}}
\end{equation}
Collecting all terms in (\ref{G4567}) we get
\begin{align}
{\cal G}(1,{\bf 2},3,\{\},\{4,5,6,7\},\{\})  &=\\
&\hspace{-4.6cm} 
\sum_{{\cal P}'(4,5,6)}\frac{s_{4567}^6\cdot [67]\cdot \la\la2|(3)(4567)(5)(4)(4567)(1)|2\ra\ra}{[12][23]\la45\ra\la46\ra\la56\ra\la67\ra\la7|456|3]\la4|567|3]\la5|467|3]\la7|456|1]\la4|567|1]\la5|467|1]}\nonumber
\end{align}
This expression is symmetric in labels $4,5,6,7$ but this is not completely manifest due to the choice of $7$ as the origin in the right blob. This concludes the 7pt calculation, and indeed the formula (\ref{GR7}) agrees numerically with BCFW expression. Note that the expansion (\ref{GR7}) is structurally identical to the Yang-Mills formula (\ref{A7YM}), as we sum over the same factorization diagrams, but ${\cal R}$- and ${\cal G}$-invariants differ by different permutational sums and the explicit expressions for $P_L$ and $P_R$.

\subsection{Blob functions}

We are now ready to write the general formulas for the left and right blob functions and complete the determination of the general ${\cal G}$-invariant, and hence the $n$-pt NMHV amplitude. We start with the left blob function as the rules for $P_L$ are simpler. Before writing a general formula, let us do one more example,
$$
\raisebox{-37pt}{\includegraphics[scale=.37]{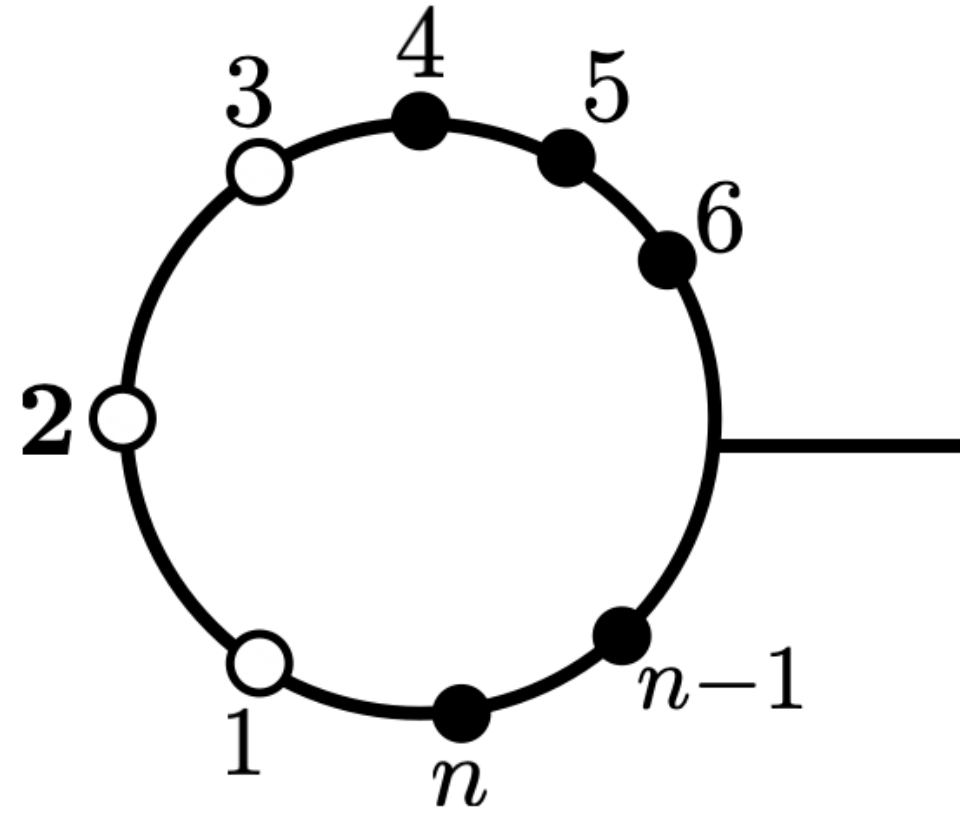}} \hspace{0.3cm} = \sum_{{\cal P}(4,5,6)}\sum_{{\cal P}(n{-}1,n)}\hspace{0.1cm}\raisebox{-39pt}{\includegraphics[scale=.37]{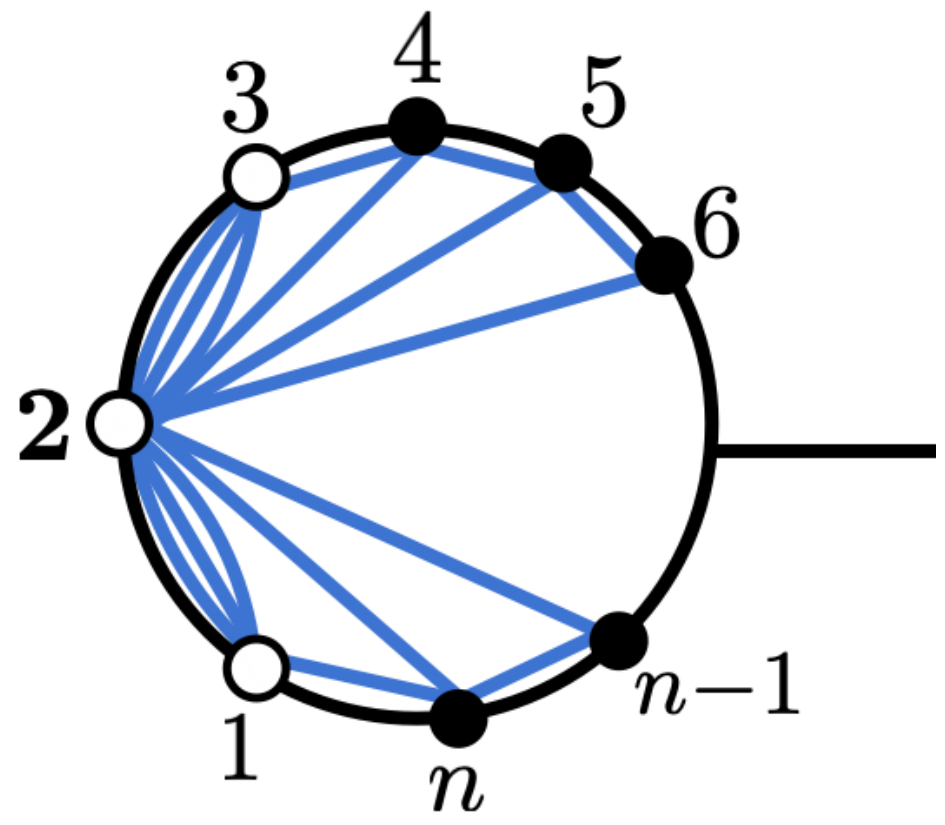}}
$$
For the {\it partial left blob} with fixed ordering, we see the cyclic structure of blue poles, and also point 2 is connected to all other points and the lines 12 and 23 are tripled. The associated numerator is a product of mixed brackets. The blue part of the partial left blob with canonical ordering is
\begin{equation}
{\color{blue}\frac{\la 2|3|4]\la 2|34|5]\la 2|345|6]\times \la2|1|n]\la 2|1n|n{-}1]}{\la23\ra^3\la 34\ra\la45\ra\la56\ra\la24\ra\la25\ra\la26\ra \times \la12\ra^3\la 1n\ra\la n{-}1n\ra\la2n\ra\la2n{-}1\ra}}
\end{equation}
where we separated terms coming from the clockwise and anti-clockwise contributions. It is easy to see the general pattern here. For the completely generic left blob with $Q_1=\{4,\dots,i{-}1\}$ and $Q_2=\{j,\dots,n\}$ we get
$$
\raisebox{-37pt}{\includegraphics[scale=.37]{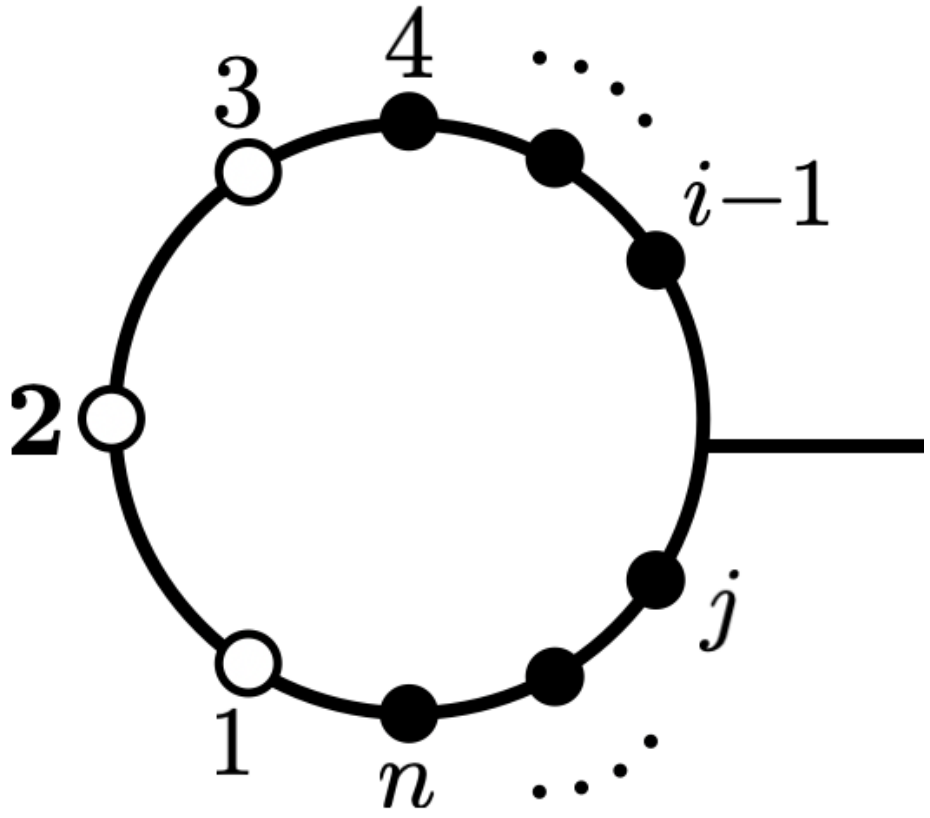}} \hspace{0.3cm} = \sum_{{\cal P}(4,\dots,i{-}1)}\sum_{{\cal P}(j,\dots,n)}\hspace{0.1cm}\raisebox{-41pt}{\includegraphics[scale=.37]{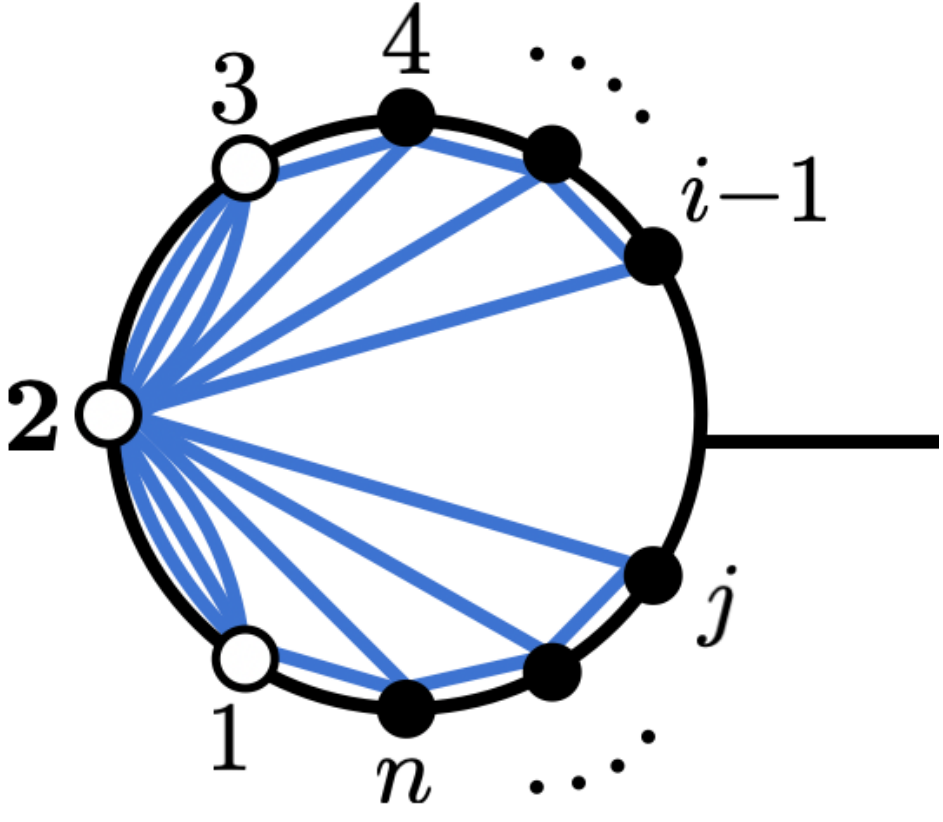}}
$$
Obviously, any other choice of labels for $Q_1$ and $Q_2$ is related by trivial relabeling. The {\color{blue} blue part} of the partial left blob with canonical ordering is
\begin{equation}
 P_L^{(1)} =   {\color{blue} \frac{\prod_{m} \la 2|\widetilde{Q}_{1m}|m] \times\prod_{q} \la 2|\widetilde{Q}_{2q}|q]}{ \la12\ra^3\la23\ra^3\times (\prod_{m} \la 2m\ra \la m{-}1\,m\ra)\times(\prod_{q}\la 2q\ra\la q\,q{+}1\ra)}}\label{PL1}
\end{equation}
where the indices $m=4,\dots,i{-}1$ and $q=j,\dots,n$. Note that for $q=n$ the denominator term gives $\la 2n\ra\la 1n\ra$ rather than $\la 2n\ra \la n1\ra$, so all brackets are ordered. We also defined
\begin{equation}
\widetilde{Q}_{1m} = p_3+\dots+p_{m{-}1},\qquad \widetilde{Q}_{2q} = p_{q{+}1}+\dots+p_n+p_1
\end{equation}
We also get {\color{red} red half-poles} in the left blob, there are three clockwise and three anti-clockwise directions, and one clockwise and one anti-clockwise {\color{OliveGreen} green poles}. There is also a non-trivial numerator factor which is merged via $\diamond$ with $P_R$. We saw all these components already in the 6pt and 7pt examples, so there is nothing new here. For the partial left blob with canonical ordering we get 
\begin{equation}
 P_L^{(2)} = \frac{{\color{red} \la\la 2|(\widetilde{Q}_1)\diamond(\widetilde{Q}_2)|2\ra\ra}}{\begin{tabular}{c}${\color{red} [\la 2|(\widetilde{Q}_1)\ast_1][\la 2|(\widetilde{Q}_1)\ast_2][ \la 2|(\widetilde{Q}_1)\ast_3][\la 2|(\widetilde{Q}_2)\ast_4][\la 2|(\widetilde{Q}_2)\ast_5][\la 2|(\widetilde{Q}_2)\ast_6]}$\\
 ${\color{OliveGreen} [\la 2|(\widetilde{Q}_1)(P)|j\ra][\la 2|(\widetilde{Q}_2)(P)|i{-}1\ra}$\end{tabular}}\label{PL2}
\end{equation}
where we denoted 
\begin{equation}
\widetilde{Q}_1=p_3+ Q_1 = p_3+\dots p_{i{-}1}, \qquad \widetilde{Q}_2 = p_1+Q_2  = p_j+\dots p_n+p_1
\end{equation}
and $P$ is the sum of momenta in the right blob, here $P=p_{i}+\dots+p_{j{-}1}$. Note that the red half-poles do not depend on the ordering of labels in the left blob while the green poles do depend on the endpoints. The complete left blob function is then given by the product of (\ref{PL1}) and (\ref{PL2}), and the sum over permutations of labels in $Q_1$ and $Q_2$,
\begin{equation}
P_L = \sum_{{\cal P}(4,\dots,i{-}1)}\sum_{{\cal P}(j,\dots,n)} (P_L^{(1)} \times P_L^{(2)}) \label{PLgen}
\end{equation}
The red part of $P_L^{(2)}$ simplifies if the the right blob has only two labels,
\begin{equation}
 P_L^{(2)} = \frac{1}{{\color{red} [\la 2|(\widetilde{Q}_1)\ast_1][\la 2|(\widetilde{Q}_1)\ast_2][ \la 2|(\widetilde{Q}_2)\ast_3][\la 2|(\widetilde{Q}_3)\ast_4]}{\color{OliveGreen} [\la 2|(\widetilde{Q}_1)(P)|j\ra][\la 2|(\widetilde{Q}_2)(P)|i{-}1\ra}}\label{PL2b}
\end{equation}
which leads to 2+2 rather than 3+3 red poles. This can be thought as the numerator factor in (\ref{PL2}) canceling two denominator factors. 

\medskip

To determine the right blob function $P_R$ is more involved and it includes multiple steps. Three simplest examples we already discussed before are
$$
\raisebox{-45pt}{\includegraphics[scale=.27]{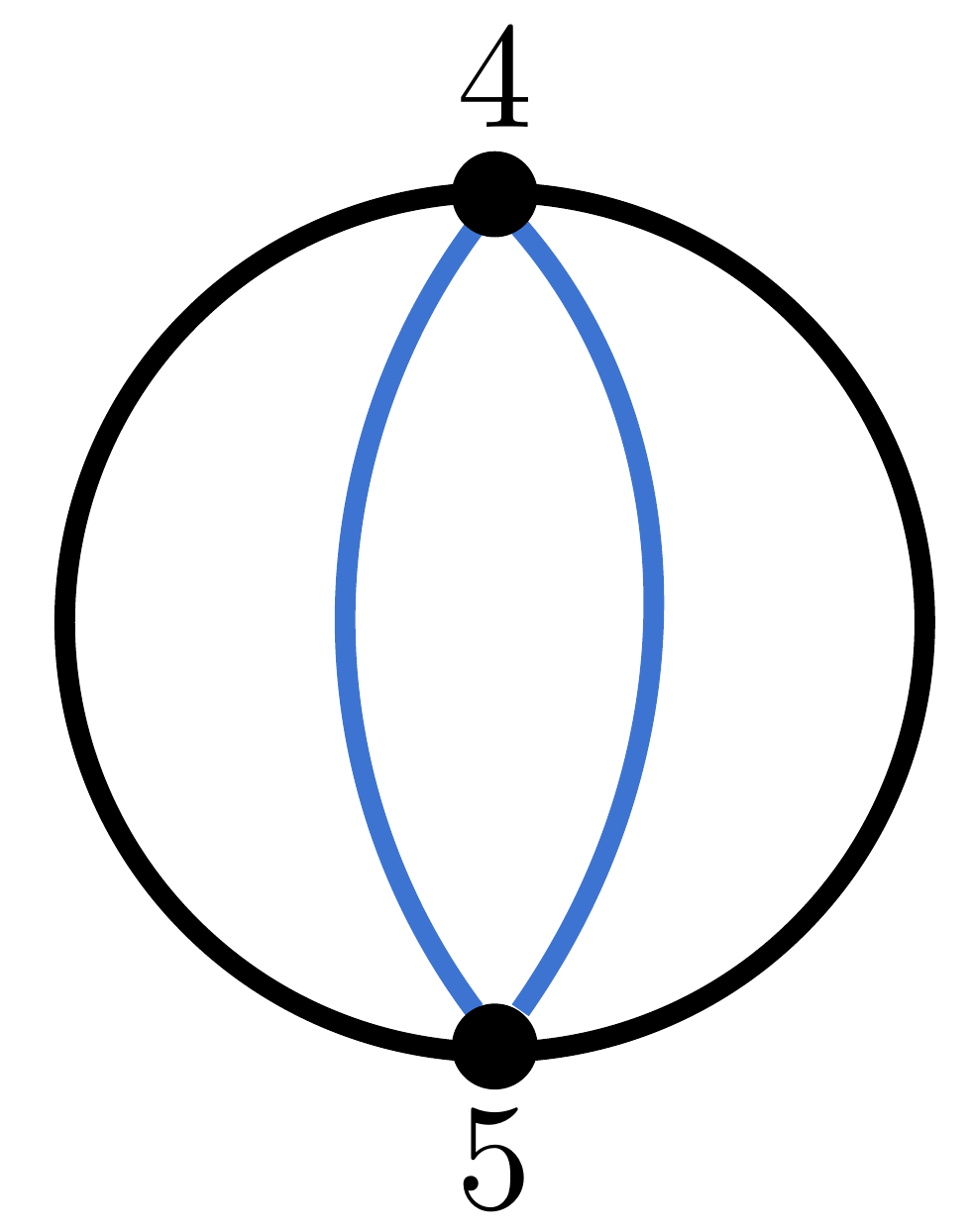}}\hspace{1.5cm}
\raisebox{-40pt}{\includegraphics[scale=.27]{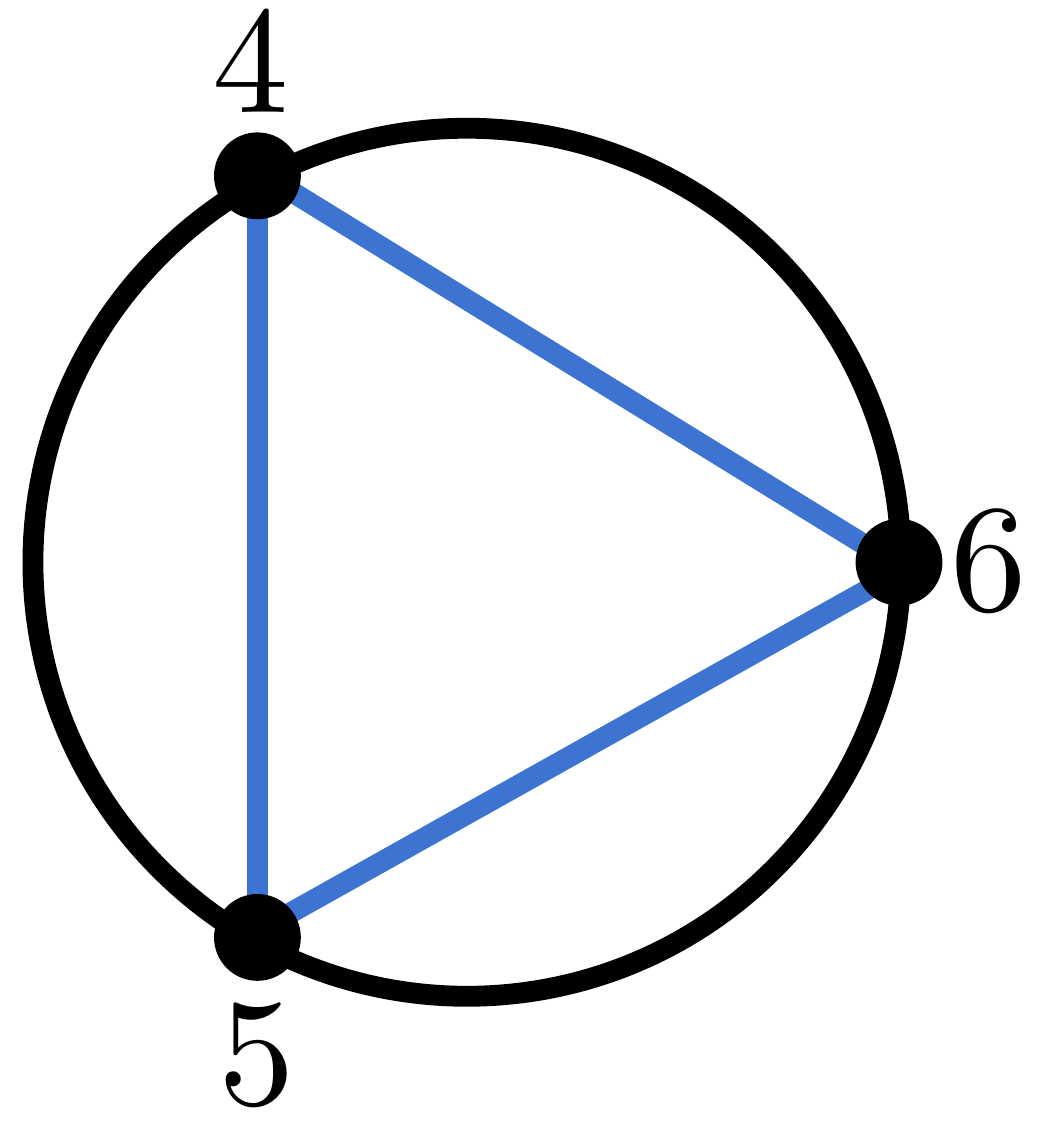}}\hspace{1.5cm}
\raisebox{-40pt}{\includegraphics[scale=.4]{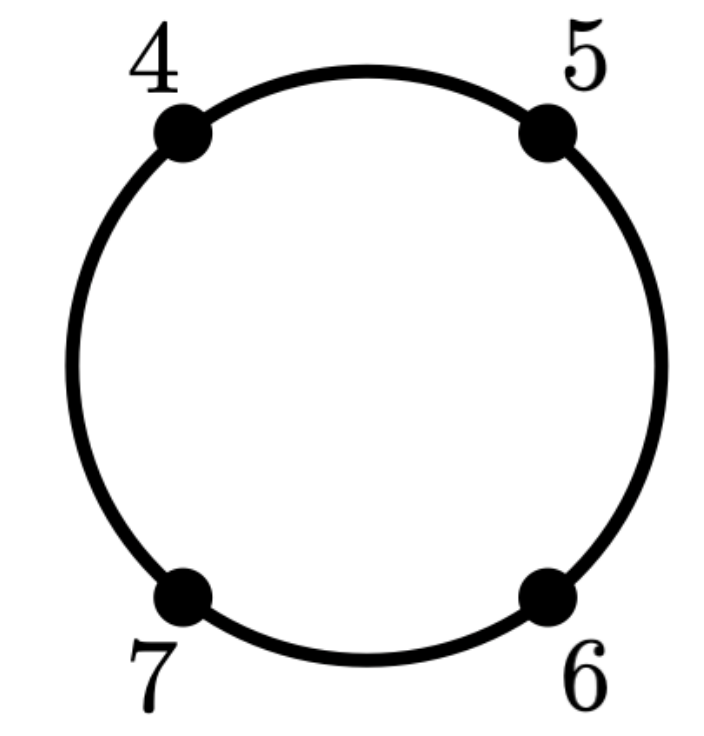}}
$$
The first diagram with two labels is considered a boundary term. The blue pole is doubled and there are only 2+2 half red-poles with no numerator
\begin{equation}
P_R = \frac{1}{{\color{blue}\la45\ra^2}\times{\color{red}[\ast_{1}(4)|5\ra][\ast_{2}(5)|4\ra][\ast_{3}(4)|5\ra][\ast_{4}(5)|4\ra]}}
\end{equation}
For the second diagram we got an expression with 3+3 red half-poles and non-trivial numerator factor,
\begin{equation}
P_R = \frac{{\color{red} [\diamond (456)(5)(4)(456)\diamond]}}{{\color{blue}\la45\ra\la46\ra\la56\ra}\times{\color{red}[\ast_{1}(45)|6\ra][\ast_{2}(46)|5\ra][\ast_{3}(56)|4\ra][\ast_{4}(45)|6\ra][\ast_{5}(46)|5\ra][\ast_{6}(56)|4\ra]}}\label{PR3}
\end{equation}
which is merged with the general expression for the left blob (\ref{PL2}). The last diagram has to be expanded in terms of three partial right blobs. We chose label $7$ as the origin in (\ref{PR4567}) and got the expression 
\begin{equation}
\hspace{-0.5cm} P_R = \sum_{{\cal P}'(4,5,6)}\frac{{\color{blue} [67]}\times{\color{red}[\diamond(4567)(5)(4)(4567)\diamond]}}{\begin{tabular}{c} ${\color{blue} \la45\ra\la46\ra\la56\ra\la67\ra}\times{\color{red} [\ast_1(456)|7\ra][\ast_2(467)|5\ra][\ast_3(567)|4\ra]}$\\ ${\color{red} [\ast_5(456)|7\ra][\ast_6(467)|5\ra][\ast_7(567)|4\ra]}$\end{tabular}}\label{PR4}
\end{equation}
Next, we increase the number of points in the right blob to five, and will immediately start to see a general pattern. We consider points $P=\{4,5,6,7,8\}$ but obviously other cases are trivially related by relabeling. We use the last point 8 as he origin and expand
\begin{equation}
\raisebox{-40pt}{\includegraphics[scale=.37]{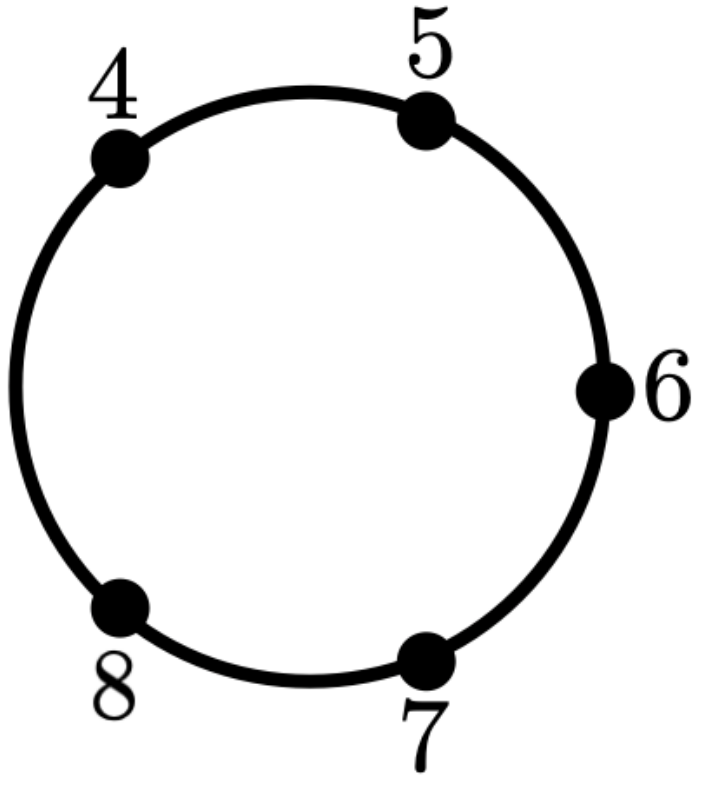}}\hspace{0.3cm} = \hspace{0.1cm} \sum_{{\cal P}'(4,5,6,7)}\hspace{0.2cm}\raisebox{-40pt}{\includegraphics[scale=.37]{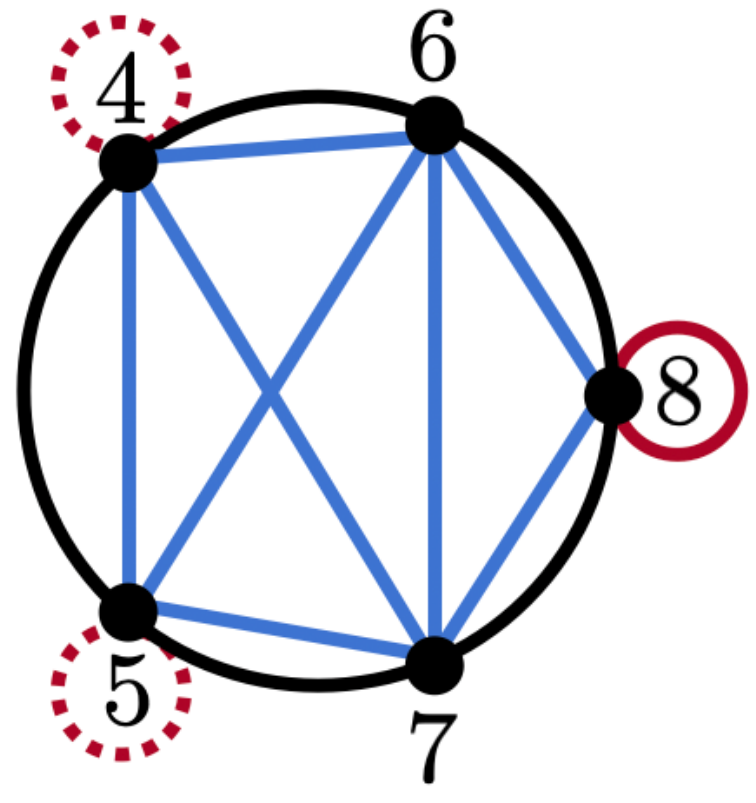}} \label{right5}
\end{equation}
where the sum is over all permutation of 4,5,6,7 modulo symmetries $(4\leftrightarrow5)$ and $(6\leftrightarrow7)$, so the sum in fact contains only six terms: 
$$
{\cal P}'(4,5,6,7) = \{(4,5,6,7),(4,6,5,7),(4,7,5,6),(5,6,4,7),(5,7,4,6),(6,7,4,5)\}
$$
We ordered points along the circle in a way that would make it easier to see the general pattern, with the origin on the right and adjacent points $k,k{+}1$ along vertical line. The function for the partial right blob is 
\begin{equation}
\raisebox{-42pt}{\includegraphics[scale=.37]{GRright3a.pdf}}\hspace{0.1cm} = \hspace{0.05cm} \frac{{\color{blue}  \la\la 4|(678)(6)(7)(678)|5\ra\ra}\times{\color{red}[\diamond (P)(5)(4)(P)\diamond]}}
{\begin{tabular}{c}
${\color{blue}\la 45\ra\la46\ra\la47\ra\la56\ra\la57\ra\la67\ra\la68\ra\la78\ra}\times
{\color{red} [\ast_1(P)|4\ra] [\ast_2(P)|5\ra]}$\\${\color{red} [\ast_3(P)|8\ra][\ast_4(P)|4\ra] [\ast_5(P)|5\ra] [\ast_6(P)|8\ra]}$\end{tabular}}\label{right1}
\end{equation}
where we denoted $P=p_4+p_5+\dots+p_8$, the sum of momenta in the blob. Note the red part stayed exactly the same, but the blue part has more poles and the numerator has the double-angle bracket ${\color{blue} \la\la 4|(678)(6)(7)(678)|5\ra\ra}$. This bracket is symmetric in $4,5$ and antisymmetric in $6,7$ while the red double angle bracket after merging is antisymmetric in $4,5$ (and trivially symmetric in $6,7$). Together with denominator this makes the whole expression symmetric in $4,5$ and symmetric in $6,7$ as indicated earlier. We calculate two more functions for the partial right blobs with six and seven labels,
\begin{equation}
\raisebox{-42pt}{\includegraphics[scale=.37]{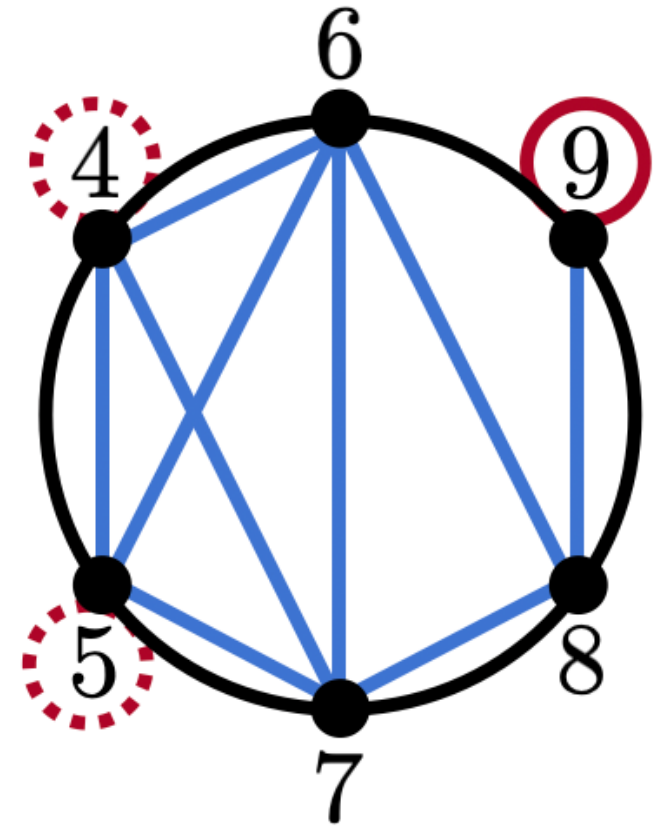}}\hspace{0.1cm} = \hspace{0.05cm} \frac{{\color{blue} \la\la 4|(6789)(6)(7)(6789)|5\ra\ra\cdot[89]}\times{\color{red} [\diamond(P)(5)(4)(P)\diamond]}}{
\begin{tabular}{c} ${\color{blue}\la 45\ra\la46\ra\la47\ra\la56\ra\la57\ra\la67\ra\la68\ra\la78\ra\la89\ra}\times{\color{red} [\ast_1(P)|4\ra] [\ast_2(P)|5\ra]}$\\${\color{red} [\ast_3(P)|9\ra][\ast_4(P)|4\ra] [\ast_5(P)|5\ra] [\ast_6(P)|9\ra]}$\end{tabular}}\label{right2}
\end{equation}
\begin{equation}
\raisebox{-42pt}{\includegraphics[scale=.37]{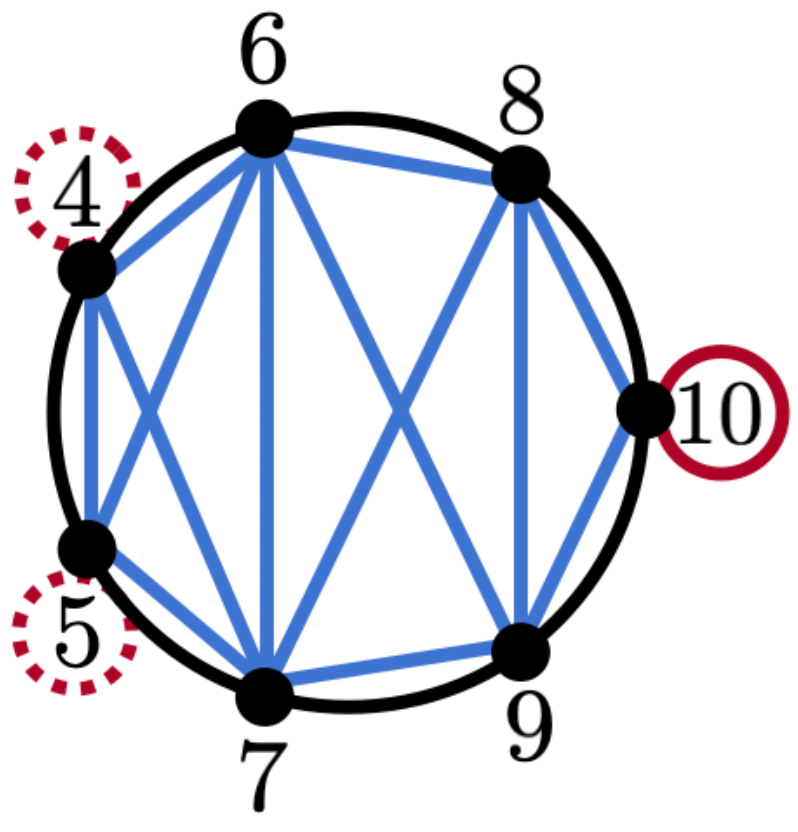}}\hspace{0.1cm} = \hspace{0.05cm} \frac{
\begin{tabular}{c}
${\color{blue}  \la\la 4|(678910)(6)(7)(678910)|5\ra\ra}$\\${\color{blue} \la\la 6|(8910)(8)(9)(8910)|7\ra\ra}\times{\color{red} [\diamond(P)(5)(4)(P)\diamond]}$\end{tabular}}{
\begin{tabular}{c}$
{\color{blue}\la 45\ra\la46\ra\la47\ra\la56\ra\la57\ra\la67\ra\la68\ra\la78\ra\la89\ra\la810\ra\la910\ra}\times{\color{red} [\ast_1(P)|4\ra]}$\\${\color{red}  [\ast_2(P)|5\ra][\ast_3(P)|10\ra][\ast_4(P)|4\ra] [\ast_5(P)|5\ra] [\ast_6(P)|10\ra]}$\end{tabular}}\label{right3}
\end{equation}
The first diagram has $(4\leftrightarrow5)$ and $(6\leftrightarrow7)$ symmetries, the sum over permutations of $4,5,6,7,8$ labels has $5!/4 = 30$ terms. The second diagram is symmetric in $(4\leftrightarrow5)$, $(6\leftrightarrow7)$, $(8\leftrightarrow9)$ and the number of permutations is $6!/8 = 90$. We clearly see the odd/even pattern emerging and are ready to formulate a general rule. 

Let us consider the right blob with labels $\{P\} = \{i,i{+}1,\dots,j{-}1\}$. One partial right blob with odd/even number of labels is
$$
\raisebox{0pt}{\includegraphics[scale=.4]{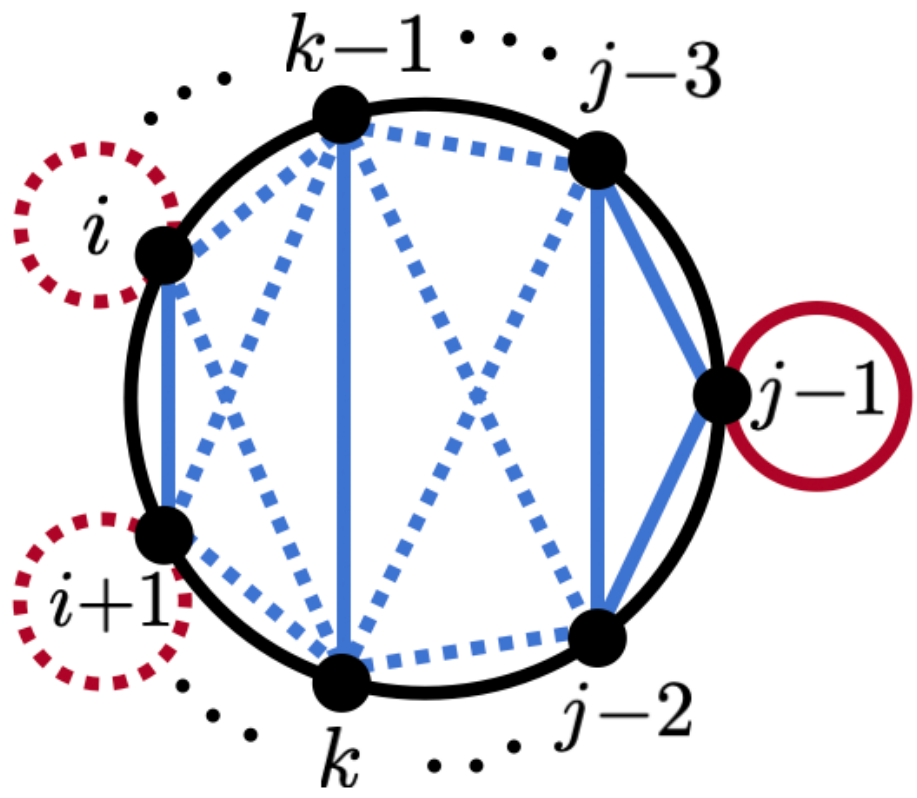}}\hspace{2cm} \raisebox{0pt}{\includegraphics[scale=.4]{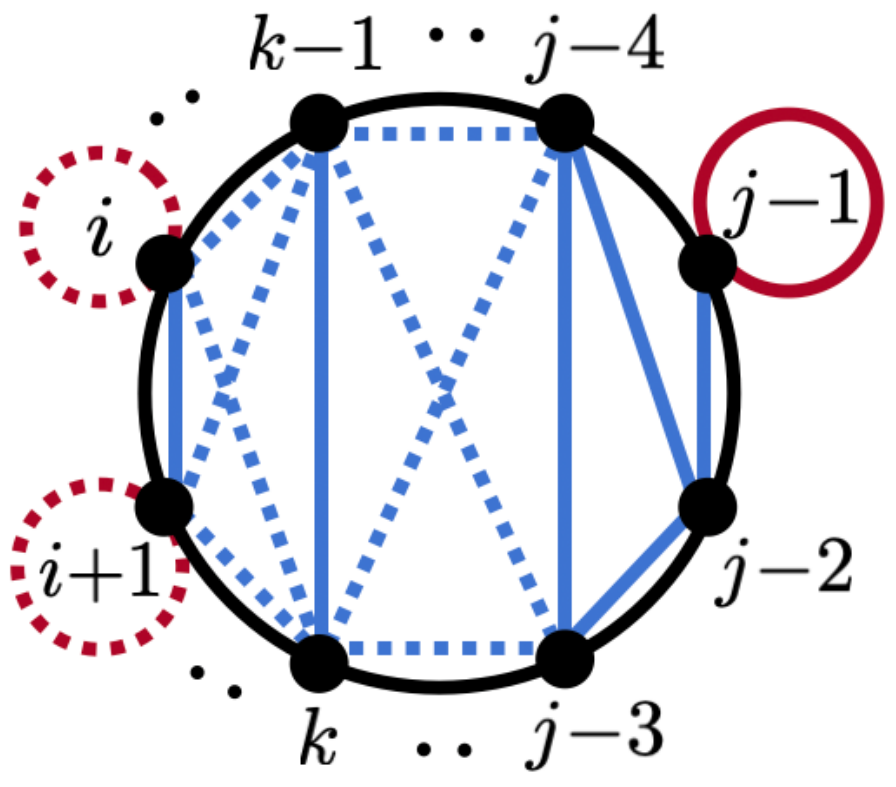}}
$$
We use the index $j{-}1$ as the origin and $i$, $i{+}1$ as two other indices which appear in the red half-poles. The diagrams are symmetric in indices along the vertical lines. The part of the right blob function $P_R$ corresponding to the red poles and red numerator is the same for any number of points,
\begin{equation}
P_R^{(1)} = {\color{red} \frac{[\diamond(P)(i{+}1)(i)(P)\diamond]}{[\ast_1(P)|i\ra][\ast_2(P)|i{+}1\ra][\ast_3(P)|j{-}1\ra][\ast_4(P)|i\ra][\ast_5(P)|i{+}1\ra][\ast_6(P)|j{-}1\ra]}}\label{PR1a}
\end{equation}
The blue poles are given by the blue lines in the figures. A generic pair of points $(k{-}1,k)$ is connected to each other and all their neighbors $k{-}3$, $k{-}2$, $k{+}1$ and $k{+}2$. 
$$
\includegraphics[scale=.35]{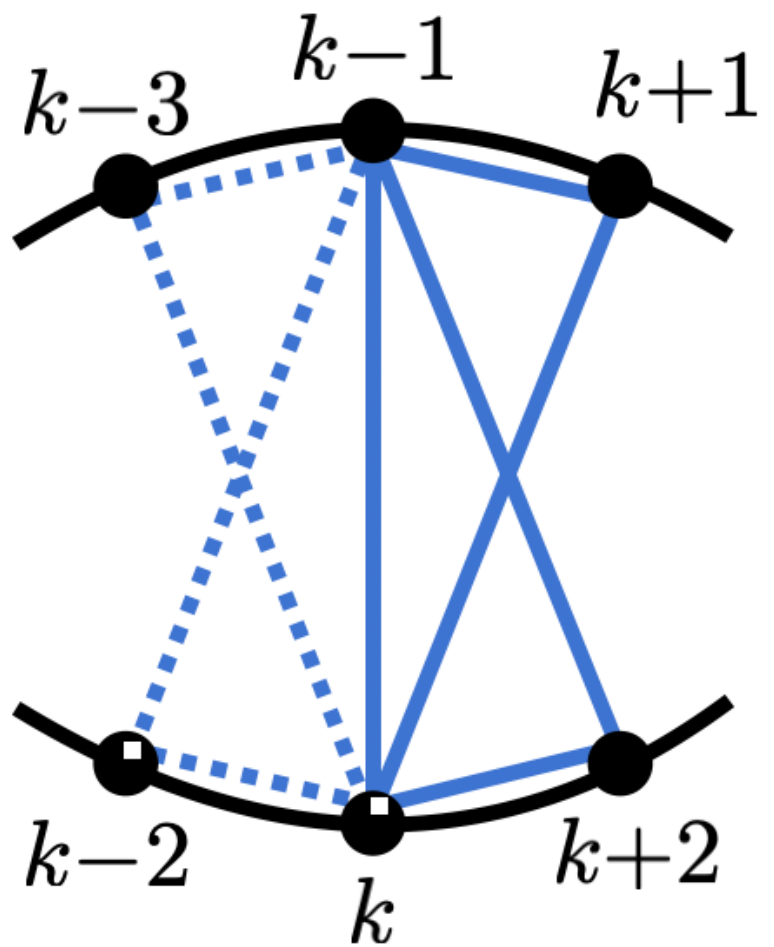}
$$
We define a building block associated with labels $(k{-}1,k)$ (correspond to thick lines in the figure above)
\begin{equation}
{\color{blue} \frac{ \la\la k{-1}|(P_k)(k{+}1)(k{+}2)(P_k)|k\ra\ra}{\la k{-}1\,k\ra\la k\,k{+}1\ra\la k{-}1\,k{+}1\ra\la k\,k{+}2\ra\la k{-}1\,k{+}2\ra}}
\end{equation}
where $P_k = p_{k{+}1}+p_{k{+}2}+\dots p_{j{-}1}$. The double-angle bracket is symmetric in $k{-}1,k$ and anti-symmetric in $k{+}1,k{+}2$. The index $k=\{i{+}1,i{+}3,\dots,j{-}4\}$ for odd number of points and $k=\{i{+}1,i{+}3,\dots,j{-}5\}$ for even number of points. There is a boundary term for $k=j{-}2$ (odd) or $k=j{-}3$ (even),
\begin{align*}
P_R^{(2)}={\color{blue}  \frac{1}{\la j{-}3\,j{-}2\ra\la j{-}2\,j{-}1\ra\la j{-}3\,j{-}1\ra}} &\qquad \mbox{for odd points}\\
P_R^{(2)}={\color{blue} \frac{[j{-}2\,j{-}1]}{\la j{-}4\,j{-}3\ra\la j{-}4\,j{-}2\ra\la j{-}3\,j{-}2\ra\la j{-}2\,j{-}1\ra}} &\qquad \mbox{for even points}
\end{align*}
We can combine all pieces in $P_R$,
\begin{align}
P_R&=P_R^{(1)}\times P_R^{(2)}\times \prod_{k} {\color{blue} \frac{ \la\la k{-1}|(P_k)(k{+}1)(k{+}2)(P_k)|k\ra\ra}{\la k{-}1\,k\ra\la k\,k{+}1\ra\la k{-}1\,k{+}1\ra\la k\,k{+}2\ra\la k{-}1\,k{+}2\ra}}\label{right4}
\end{align}
This is a rule for one partial right blob. To get the complete right blob function we have to sum over all permutations of labels $i,i{+}1,\dots, j{-}2$ modulo symmetries $(i\leftrightarrow i{+}1)$, $(i{+}2\leftrightarrow i{+}3)$, etc. For the odd number of points, we get
$$
\raisebox{-42pt}{\includegraphics[scale=.4]{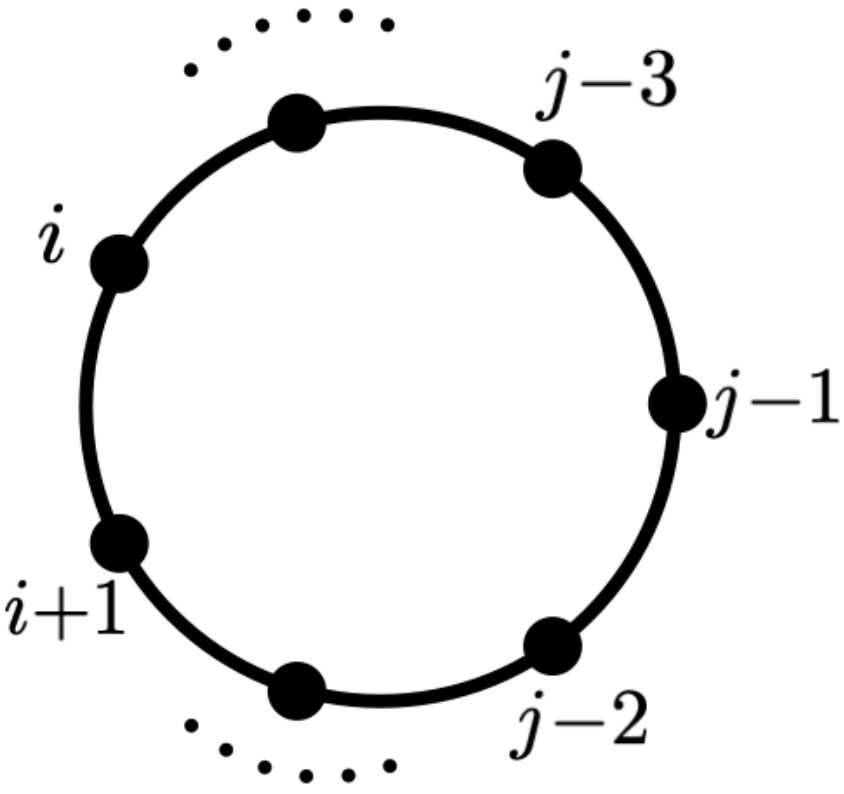}}\hspace{0.3cm} =   \sum_{{\cal P}'({i..j{-}2})} \hspace{0.3cm} \raisebox{-42pt}{\includegraphics[scale=.4]{GenRight1.pdf}}
$$
and similarly for even number of points
$$
\raisebox{-42pt}{\includegraphics[scale=.4]{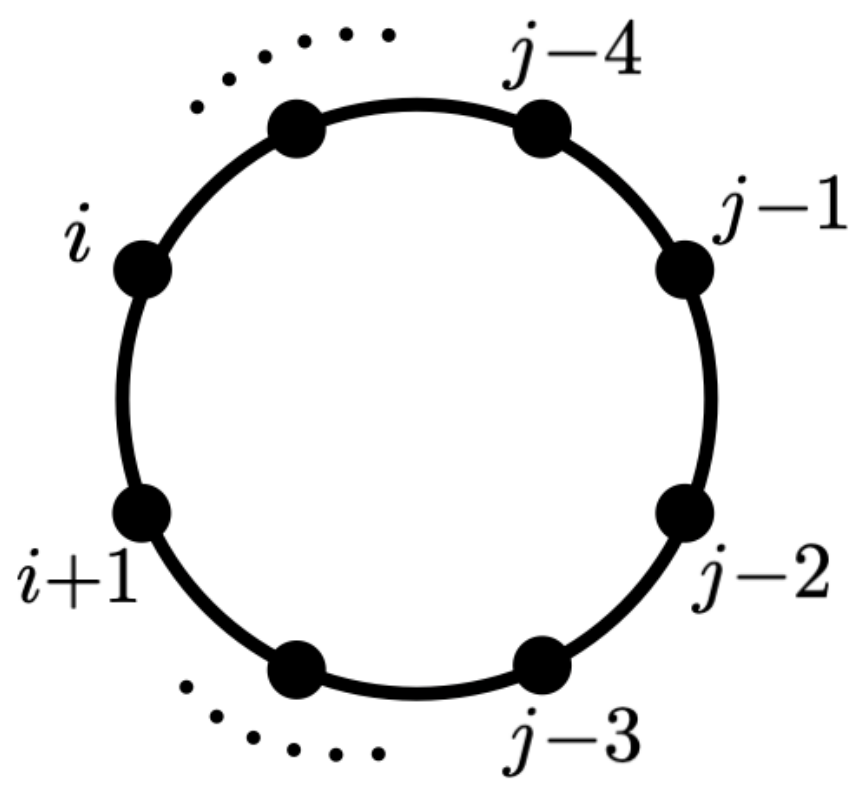}}\hspace{0.3cm} =   \sum_{{\cal P}'({i..j{-}2})} \hspace{0.3cm} \raisebox{-42pt}{\includegraphics[scale=.4]{GenRight4.pdf}}
$$
Note that the full blob is completely symmetric in all labels $(i,i{+}1,\dots,j{-}1)$ but we only sum over permutations of $(i,i{+}1,\dots j{-}2)$ (modulo symmetries).

\subsection{General formula for ${\cal G}$-invariant}

We have all ingredients needed to write the formula for the arbitrary ${\cal G}$-invariant using (\ref{Ggen}) and expressions for $P_R$ and $P_L$.
$$
{\cal G}(1,{\bf 2},3,\{Q_1\},\{P\},\{Q_2\}) =   \sum_{{\cal P}(Q_1)}\sum_{{\cal P}(Q_2)} \sum_{{\cal P}'(P)}\raisebox{-42pt}{\includegraphics[scale=.4]{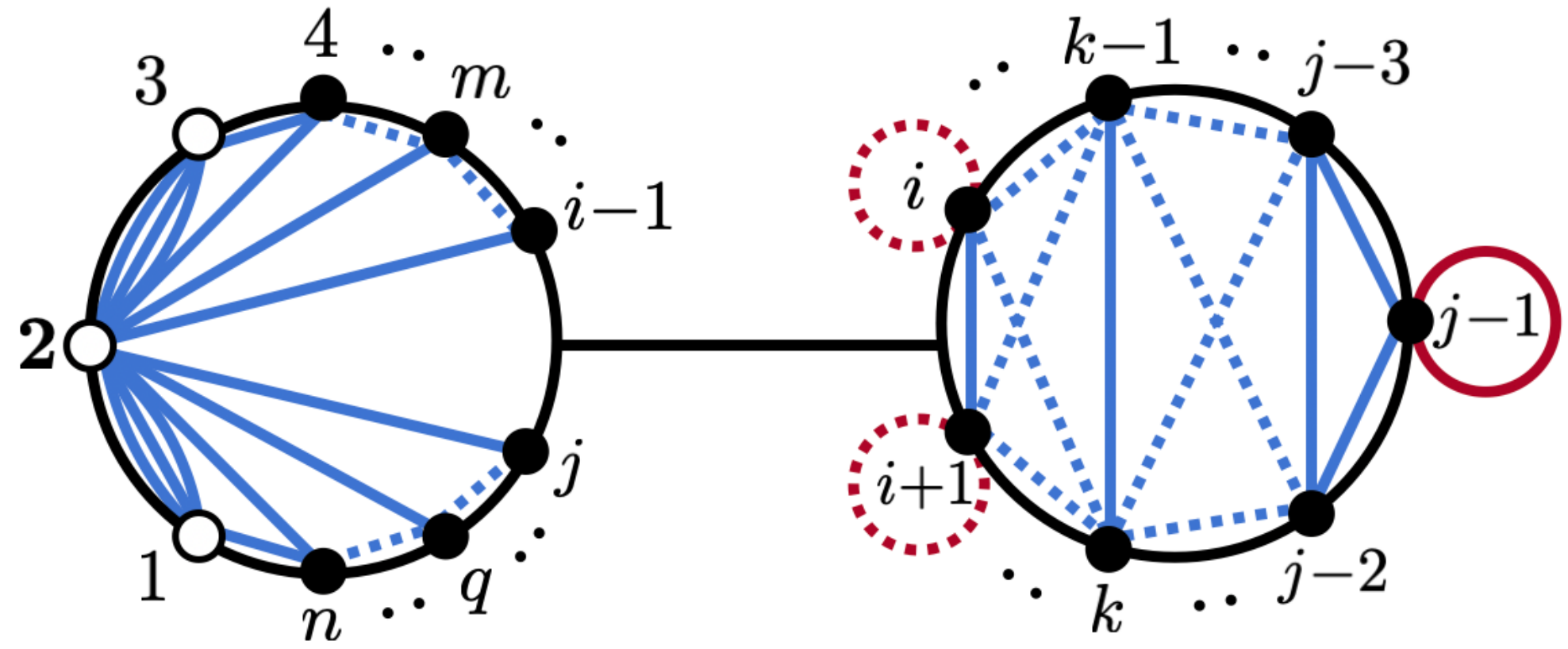}}
$$
where $Q_1=\{4,\dots,i{-}1\}$, $P=\{i,\dots,j\}$, $Q_2=\{j{+}1,\dots,n\}$. The partial factorization diagram in the figure produces the following formula
\begin{equation}
\hspace{-0.5cm}\left[\frac{\begin{tabular}{c}$\la12\ra^5\la23\ra^5 s_P^6\cdot {\color{red} \la\la2|(\widetilde{Q}_1)(P)(i{+}1)(i)(P)(\widetilde{Q})_2)|2\ra\ra}$\\
$\times\color{blue}{\left(\prod_k \la\la k{-}1|(P_k)(k{+}1)(k{+}2)(P_k)|k\ra\ra\right)}\cdot{\color{blue} \left(\prod_m \la 2|Q_{1m}|m] \right)}\cdot {\color{blue} \left(\prod_q \la 2|Q_{2q}|q]\right)}$\end{tabular}}{\begin{tabular}{c} ${\color{red} \la 2|(\widetilde{Q}_1)(P)|i\ra\la 2|(\widetilde{Q}_1)(P)|i{+}1\ra\la 2|(\widetilde{Q}_1)(P)|j{-}1\ra\la 2|(\widetilde{Q}_2)(P)|i\ra\la 2|(\widetilde{Q}_2)(P)|i{+}1\ra\la 2|(\widetilde{Q}_2)(P)|j{-}1\ra}$\\
${\color{OliveGreen} \la 2|(\widetilde{Q}_1)(P)|j\ra\la 2|(\widetilde{Q}_2)(P)|i{-}1\ra}\cdot {\color{blue} \left(\prod_m \la2 m\ra\la m{-}1\,m\ra\right)}\cdot {\color{blue} \left(\prod_q \la 2q\ra\la q\,q{+}1\ra\right)}\cdot {\color{blue}  \la i\,i{+}1\ra}$\\
${\color{blue}\la j{-}3\,j{-}2\ra\la j{-}2\,j{-}1\ra\la j{-}3\,j{-}1\ra \cdot \left(\prod_k \la k{-}1\,k\ra\la k\,k{+}1\ra\la k{-}1\,k{+}1\ra\la k\,k{+}2\ra\la k{-}1\,k{+}2\ra\right)}$\end{tabular}}\right] \label{FullOdd}
\end{equation}
where the index $k = \{i{+}1,i{+}3,\dots,j{-}4\}$ (jumps by two), index $m=\{4,\dots,i{-}1\}$ and index $q=\{j,\dots,n\}$, and we used the same notation as before,
\begin{align*}
& P = p_i + \dots + p_{j{-}1}\qquad\quad\,\,\,\widetilde{Q}_1 = p_3 + \dots p_{i{-}1}\qquad\,\,\,\, \widetilde{Q}_2 = p_j+\dots+p_n+p_1\\
& P_k = p_{k{+}1} + \dots + p_{j{-}1}\qquad Q_{1m} = p_3+ \dots p_{m{-}1}\qquad Q_{2q} = p_{q{-}1}+\dots+p_n+p_1
\end{align*}
As discussed earlier, the right blob function differs slightly if the number of points in the blob is odd or even. The difference is only in the boundary blue poles and numerator. We gave the formula for the odd number above, for the even number we have 
$$
{\cal G}(1,{\bf 2},3,\{Q_1\},\{P\},\{Q_2\}) =   \sum_{{\cal P}(Q_1)}\sum_{{\cal P}(Q_2)} \sum_{{\cal P}'(P)}\raisebox{-42pt}{\includegraphics[scale=.4]{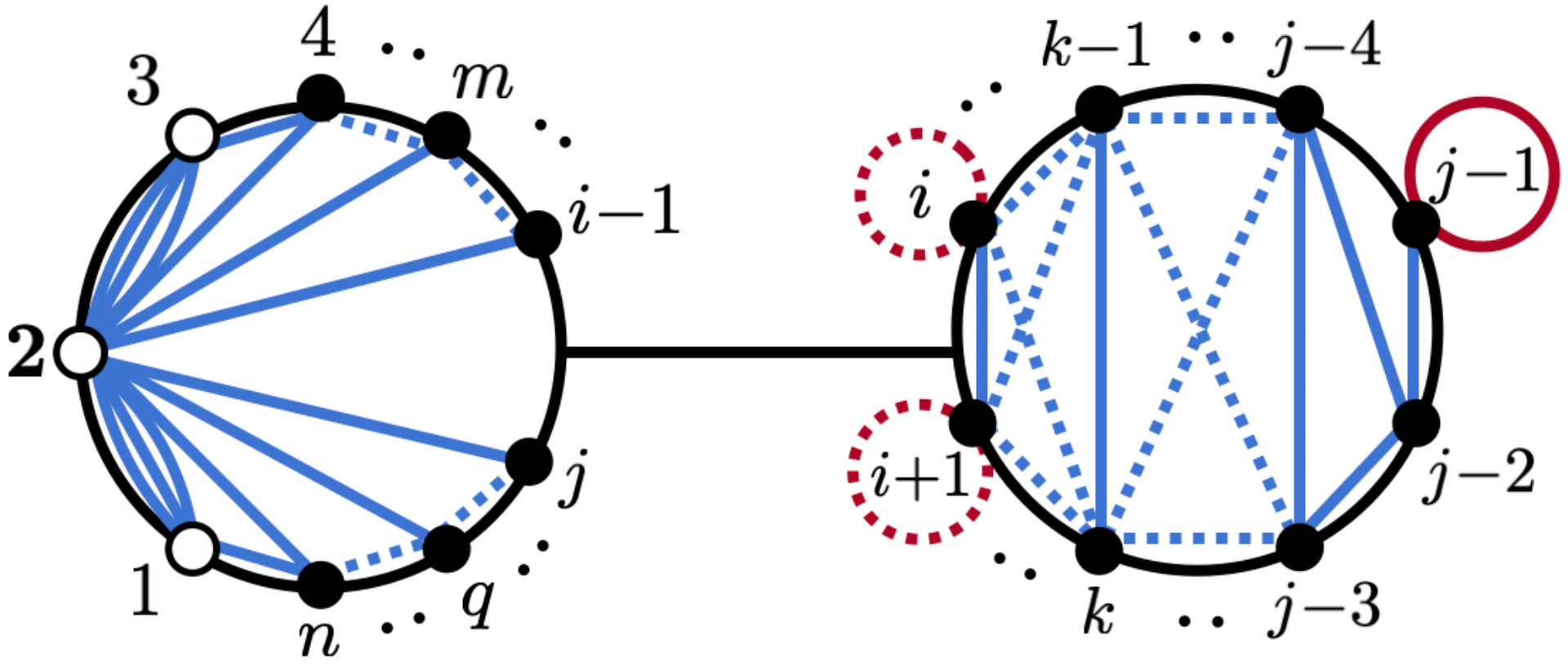}}
$$
The formula for partial factorization diagram is then 
\begin{equation}
\hspace{-0.5cm}\left[\frac{\begin{tabular}{c}$\la12\ra^5\la23\ra^5 s_P^6\cdot {\color{red} \la\la2|(\widetilde{Q}_1)(P)(i{+}1)(i)(P)(\widetilde{Q})_2)|2\ra\ra}$\\
$\times\color{blue}{\left(\prod_k \la\la k{-}1|(P_k)(k{+}1)(k{+}2)(P_k)|k\ra\ra\right)}\cdot{\color{blue} [j{-}2\,j{-}1]}\cdot{\color{blue} \left(\prod_m \la 2|Q_{1m}|m] \right)}\cdot {\color{blue} \left(\prod_q \la 2|Q_{2q}|q]\right)}$\end{tabular}}{\begin{tabular}{c} ${\color{red} \la 2|(\widetilde{Q}_1)(P)|i\ra\la 2|(\widetilde{Q}_1)(P)|i{+}1\ra\la 2|(\widetilde{Q}_1)(P)|j{-}1\ra\la 2|(\widetilde{Q}_2)(P)|i\ra\la 2|(\widetilde{Q}_2)(P)|i{+}1\ra\la 2|(\widetilde{Q}_2)(P)|j{-}1\ra}$\\
${\color{OliveGreen} \la 2|(\widetilde{Q}_1)(P)|j\ra\la 2|(\widetilde{Q}_2)(P)|i{-}1\ra}\cdot {\color{blue} \left(\prod_m \la2 m\ra\la m{-}1\,m\ra\right)}\cdot {\color{blue} \left(\prod_q \la 2q\ra\la q\,q{+}1\ra\right)}\cdot {\color{blue}  \la i\,i{+}1\ra}$\\
${\color{blue}\la j{-}4\,j{-}3\ra\la j{-}4\,j{-}2\ra\la j{-}3\,j{-}2\ra\la j{-}2\,j{-}1\ra \cdot \left(\prod_k \la k{-}1\,k\ra\la k\,k{+}1\ra\la k{-}1\,k{+}1\ra\la k\,k{+}2\ra\la k{-}1\,k{+}2\ra\right)}$\end{tabular}}\right] \label{FullEven}
\end{equation}
where the index $k$ now runs over $k = \{i{+}1,i{+}3,\dots,j{-}5\}$. Note that in both expressions ${\color{blue} \la q\,q{+}1\ra}$ in the denominator is ${\color{blue} \la 1n\ra}$ for $q=n$.

\medskip

We apply our formula and evaluate the 8pt NMHV amplitude ${\cal M}_8(1^-2^-3^-4^+5^+6^+7^+8^+)$. The expansion in terms of ${\cal G}$-invariants is
\begin{align}
{\cal M}_8 &= \sum_{{\cal P}'(1,2,3)}\sum_{{\cal P}'(4,5,6,7,8)}
\left[\begin{tabular}{c} ${\cal G}(1,{\bf 2},3,\{4,5,6\},\{7,8\},\{\}) + {\cal G}(1,{\bf 2},3,\{4,5\},\{6,7\},\{8\})$\\
$+ {\cal G}(1,{\bf 2},3,\{4,5\},\{6,7,8\},\{\})+ {\cal G}(1,{\bf 2},3,\{4\},\{5,6,7\},\{8\})$\\
$+ {\cal G}(1,{\bf 2},3,\{4\},\{5,6,7,8\},\{\})+ {\cal G}(1,{\bf 2},3,\{\},\{4,5,6,7,8\},\{\})$
\end{tabular}\right] \label{our8}
\end{align}
where we symmetrize all terms in labels $1,2,3$ and $4,5,6,7,8$ modulo symmetries of individual ${\cal G}$-invariants. Using the general formulas above we get for all six ${\cal G}$-invariants,
\begin{align}
{\cal G}(1,{\bf 2},3,\{4,5,6\},\{7,8\},\{\}) & \\
&\hspace{-4.5cm} =\sum_{{\cal P}(4,5,6)}\frac{\la12\ra^8\la23\ra^8 s_{78}^6 \cdot {\color{blue} \la 2|3|4]\la2|34|5]\la2|345|6]}}{\begin{tabular}{c}
${\color{blue} \la12\ra^3\la23\ra^3\la34\ra\la45\ra\la56\ra\la24\ra\la25\ra\la26\ra\la78\ra^2} \cdot {\color{red} \la 2|(3456)(7)|8\ra \la 2|(3456)(8)|7\ra}$\\
${\color{red}\la2|(1)(7)|8\ra\la2|(1)(8)|7\ra}\cdot {\color{OliveGreen} \la2|(3456)(78)|1\ra\la2|(1)(78)|6\ra}$\end{tabular}}\nonumber\\
&\hspace{-4.5cm} = \sum_{{\cal P}(4,5,6)}\frac{\la23\ra^6[78]^6\cdot \la12\ra[34]\la2|34|5]\la2|345|6]}{s_{178}[17][18]\la34\ra\la45\ra\la56\ra\la24\ra\la25\ra\la26\ra\la2|18|7]\la2|17|8]\la6|78|1]}\nonumber
\end{align}
\begin{align}
{\cal G}(1,{\bf 2},3,\{4,5\},\{6,7\},\{8\}) & \\
&\hspace{-4.8cm} =\sum_{{\cal P}(4,5)}\frac{\la12\ra^8\la23\ra^8 s_{67}^6 \cdot {\color{blue} \la 2|3|4]\la2|34|5]\la2|1|8]}}{\begin{tabular}{c}
${\color{blue} \la12\ra^3\la23\ra^3\la34\ra\la45\ra\la24\ra\la25\ra\la18\ra\la28\ra\la67\ra^2}\cdot{\color{red} \la 2|(345)(6)|7\ra \la 2|(345)(7)|6\ra}$\\
${\color{red}\la2|(18)(6)|7\ra\la2|(18)(7)|6\ra}\cdot {\color{OliveGreen} \la2|(345)(67)|8\ra\la2|(18)(67)|5\ra}$\end{tabular}}\nonumber\\
&\hspace{-4.8cm} = \sum_{{\cal P}(4,5)}\frac{\la12\ra^6\la23\ra^6[67]^6\cdot [34][18]\la2|34|5]}{\la34\ra\la45\ra\la24\ra\la25\ra\la18\ra\la28\ra\la2|345|6]\la2|345|7]\la2|18|6]\la2|18|7]\la2|(345)(67)|8\ra\la2|(18)(67)|5\ra}\nonumber
\end{align}
\begin{align}
{\cal G}(1,{\bf 2},3,\{4,5\},\{6,7,8\},\{\}) & \\
&\hspace{-4.8cm} =\sum_{{\cal P}(4,5)}\frac{\la12\ra^8\la23\ra^8 s_{678}^6 \cdot {\color{blue} \la 2|3|4]\la2|34|5]}\cdot {\color{red} \la\la2|(345)(678)(7)(6)(678)(1)|2\ra\ra}}
{\begin{tabular}{c}
${\color{blue} \la12\ra^3\la23\ra^3\la34\ra\la45\ra\la24\ra\la25\ra\la67\ra\la68\ra\la78\ra}\cdot {\color{red} \la 2|(345)(67)|8\ra \la 2|(345)(68)|7\ra\la 2|(345)(78)|6\ra}$\\
${\color{red}\la2|(1)(67)|8\ra\la2|(1)(68)|7\ra\la2|(1)(78)|6\ra}\cdot {\color{OliveGreen} \la2|(345)(678)|1\ra\la2|(1)(678)|5\ra}$\end{tabular}}\nonumber\\
&\hspace{-4.8cm} = -\sum_{{\cal P}(4,5)}\frac{\la23\ra^6s_{678}^6\cdot [34]\la2|34|5]\cdot\la\la2|(345)(678)(7)(6)(678)(1)|2\ra\ra}{\begin{tabular}{c} $s_{2345}\la34\ra\la45\ra\la24\ra\la25\ra\la67\ra\la68\ra\la78\ra\la 2|(345)(67)|8\ra \la 2|(345)(68)|7\ra$\\$\la 2|(345)(78)|6\ra\la8|(67)|1]\la7|(68)|1]\la6|78|1]\la5|678|1]$\end{tabular}}\nonumber
\end{align}
\begin{align}
{\cal G}(1,{\bf 2},3,\{4\},\{5,6,7\},\{8\}) & \\
&\hspace{-4.2cm} =\frac{\la12\ra^8\la23\ra^8 s_{567}^6 \cdot {\color{blue} \la 2|3|4]\la2|1|8]}\cdot {\color{red} \la\la2|(34)(567)(6)(5)(567)(81)|2\ra\ra}}
{\begin{tabular}{c}
${\color{blue} \la12\ra^3\la23\ra^3\la34\ra\la24\ra\la18\ra\la28\ra\la56\ra\la57\ra\la67\ra}\cdot {\color{red} \la 2|(34)(56)|7\ra \la 2|(34)(57)|6\ra\la 2|(34)(67)|5\ra}$\\
${\color{red}\la2|(18)(56)|7\ra\la2|(18)(57)|6\ra\la2|(18)(67)|5\ra}\cdot {\color{OliveGreen} \la2|(34)(567)|8\ra\la2|(18)(567)|4\ra}$\end{tabular}}\nonumber\\
&\hspace{-4.2cm} = -\frac{\la12\ra^6\la23\ra^6s_{678}^6\cdot [34][18]\cdot\la\la2|(34)(567)(6)(5)(567)(81)|2\ra\ra}{\begin{tabular}{c} $\la34\ra\la24\ra\la18\ra\la28\ra\la56\ra\la57\ra\la67\ra\la 2|(34)(56)|7\ra \la 2|(34)(57)|6\ra\la 2|(34)(67)|5\ra$\\ $\la2|(18)(56)|7\ra\la2|(18)(57)|6\ra\la2|(18)(67)|5\ra \la2|(34)(567)|8\ra\la2|(18)(567)|4\ra$ \end{tabular}}\nonumber
\end{align}
\begin{align}
{\cal G}(1,{\bf 2},3,\{4\},\{5,6,7,8\},\{\}) & \\
&\hspace{-4.8cm} = \hspace{-0.2cm} \sum_{{\cal P}'(5,6,7)}\hspace{-0.1cm}\frac{\la12\ra^8\la23\ra^8 s_{5678}^6 \cdot {\color{blue} \la 2|3|4]}\cdot {\color{red} \la\la2|(34)(5678)(6)(5)(5678)(1)|2\ra\ra}\cdot{\color{blue} [78]}}
{\begin{tabular}{c}
${\color{blue} \la12\ra^3\la23\ra^3\la34\ra\la24\ra\la56\ra\la57\ra\la67\ra\la78\ra}\cdot {\color{red} \la 2|(34)(567)|8\ra \la 2|(34)(678)|5\ra\la 2|(34)(578)|6\ra}$\\
${\color{red}\la2|(1)(567)|8\ra\la2|(1)(678)|5\ra\la2|(1)(578)|6\ra}\cdot {\color{OliveGreen} \la2|(34)(5678)|1\ra\la2|(1)(5678)|4\ra}$\end{tabular}}\nonumber\\
&\hspace{-4.8cm} =-\sum_{{\cal P}'(5,6,7)}\hspace{-0.1cm} \frac{\la23\ra^6s_{5678}^6\cdot [34][78]\cdot\la\la2|(34)(5678)(6)(5)(5678)(1)|2\ra\ra}{\begin{tabular}{c} $s_{234}\la34\ra\la24\ra\la56\ra\la57\ra\la67\ra\la78\ra\la8|567|1]\la5|678|1]\la6|578|1]\la4|5678|1]$\\ $\la 2|(34)(567)|8\ra \la 2|(34)(678)|5\ra\la 2|(34)(578)|6\ra$ \end{tabular}}\nonumber\\
&\hspace{-4.8cm}\mbox{where ${\cal P}'(5,6,7) = \{(5,6,7),(5,7,6),(6,7,5)\}$}\nonumber
\end{align}
\begin{align}
{\cal G}(1,{\bf 2},3,\{\},\{4,5,6,7,8\},\{\}) & \\
&\hspace{-5.6cm} = \hspace{-0.3cm} \sum_{{\cal P}'(4,5,6,7)}\hspace{-0.2cm}\frac{\la12\ra^8\la23\ra^8 s_{45678}^6 \cdot {\color{red} \la\la2|(3)(45678)(5)(4)(45678)(1)|2\ra\ra}\cdot{\color{blue} \la\la4|(678)(6)(7)(678)|5\ra\ra}}
{\begin{tabular}{c}
${\color{blue} \la12\ra^3\la23\ra^3\la45\ra\la46\ra\la47\ra\la56\ra\la57\ra\la67\ra\la68\ra\la78\ra}\cdot {\color{red} \la 2|(3)(4567)|8\ra \la 2|(3)(5678)|4\ra\la 2|(3)(4678)|5\ra}$\\
${\color{red}\la2|(1)(4567)|8\ra\la2|(1)(5678)|4\ra\la2|(1)(4678)|5\ra}\cdot {\color{OliveGreen} \la2|(3)(45678)|1\ra\la2|(1)(45678)|3\ra}$\end{tabular}}\nonumber\\
&\hspace{-5.6cm} = \hspace{-0.2cm} \sum_{{\cal P}'(4,5,6,7)}\hspace{-0.0cm} \frac{s_{45678}^6\cdot \la\la2|(3)(45678)(5)(4)(45678)(1)|2\ra\ra\cdot \la\la4|(678)(6)(7)(678)|5\ra\ra}{[12][23]\la45\ra\la46\ra\la47\ra\la56\ra\la57\ra\la67\ra\la68\ra\la78\ra\la8|12|3]\la4|12|3]\la5|12|3]\la8|23|1]\la4|23|1]\la5|23|1]}\nonumber\\
&\hspace{-4.8cm}\mbox{where ${\cal P}'(4,5,6,7) = \{(4,5,6,7),(4,6,5,7),(4,7,5,6),(5,6,4,7),(5,7,4,6),(6,7,4,5)\}$}\nonumber
\end{align}
We checked numerically that (\ref{our8}) agrees with the BCFW calculation.

\section{Discussion of the result}

In this section, we discuss some special properties of our representation of the NMHV amplitudes, such as the spurious poles cancelation, large-$z$ behavior, relation to momentum twistors and how it compares to other (BCFW) formulas in the literature. We will also elaborate on some curious features which suggest geometric interpretation.

\subsection{Factorizations}

The first crucial question is how we can be confident that our general formula gives the correct gravity amplitude. There is no actual derivation and the framework of ${\cal G}$-invariants is just an educated guess based on some similarities with Yang-Mills and there is no direct connection to BCFW, KLT or other method. There are two main consistency checks which uniquely determine if the proposed expression (\ref{GRamp}) reproduces correctly the amplitude: factorizations and spurious poles cancelation. 

The tree-level unitarity dictates that on any physical pole the $n$-pt N$^{k-2}$MHV amplitude ${\cal M}_n^{(k)}$ must factorize into a product of two lower-point amplitudes,
\begin{equation}
{\cal M}_n^{(k)} \xrightarrow[]{s_P\rightarrow0} \sum_{k_1+k_2=k{+}1}{\cal M}^{(k_1)}_L \frac{1}{s_P} {\cal M}^{(k_2)}_R
\end{equation}  
The sum turns into a sum over the helicities of the internal leg $P$. For 4D kinematics there are two types of two-particle factorizations. In the context of NMHV amplitude ${\cal M}_n$ the multi-particle factorization turns into the product of two MHV amplitudes,
$$
\includegraphics[scale=.46]{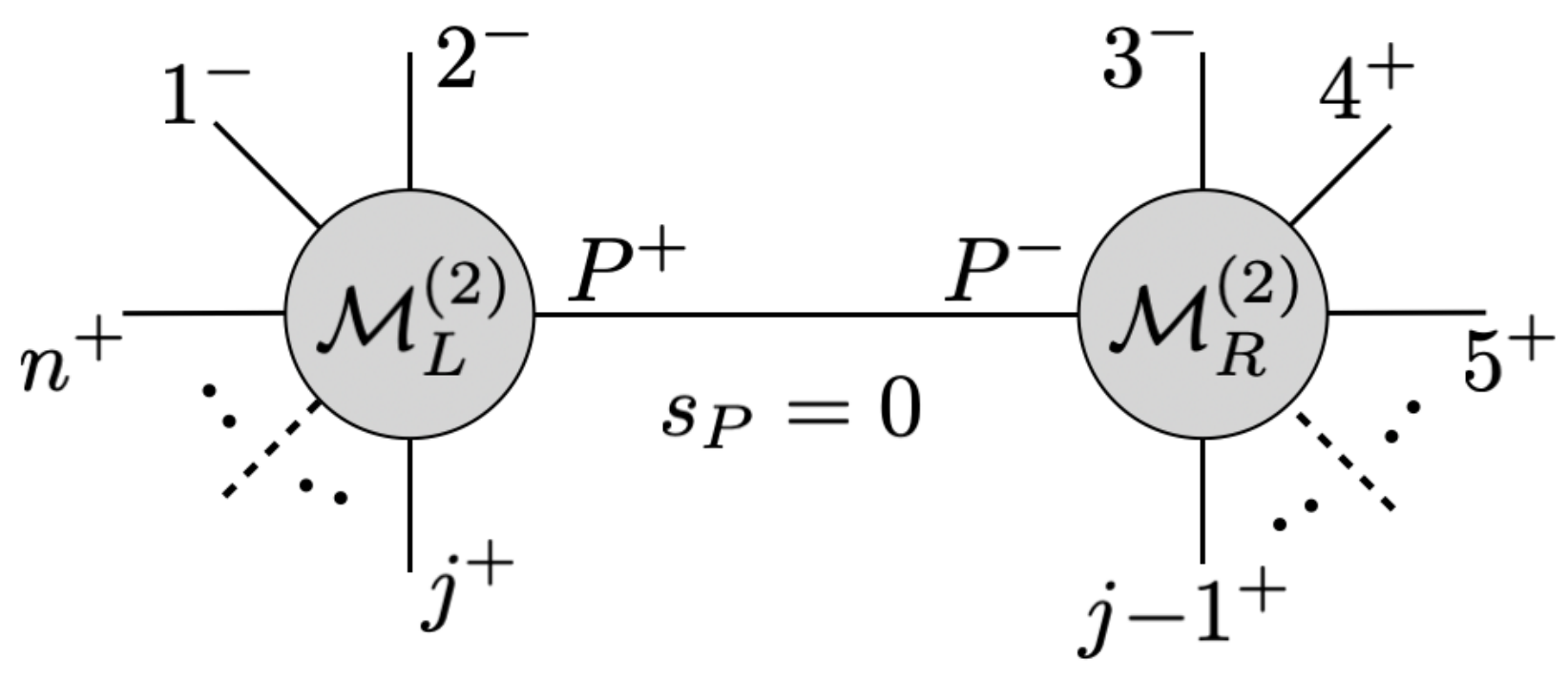}
$$
where we used the canonical ordering in the diagram but the same applies for any labeling of external legs. Importantly, two of the negative helicity gravitons are on one side of the channel and one on the other side, which is also correlated with the helicities of the internal leg $P^\pm$. The two particle factorization channels are of two types due to special 4D kinematics,
\begin{equation}
s_{ij} = \la ij\ra [ij] = 0 \rightarrow\Bigg\{ \begin{tabular}{cc} $\la ij\ra = 0$\,\,\,&\mbox{holomorphic channel}\\ $[ij]=0$\,\,\,&\mbox{anti-holomorphic channel} \end{tabular}
\end{equation}
There are two types of holomorphic factorizations (up to relabeling of legs),
$$
\raisebox{-0pt}{\includegraphics[scale=.43]{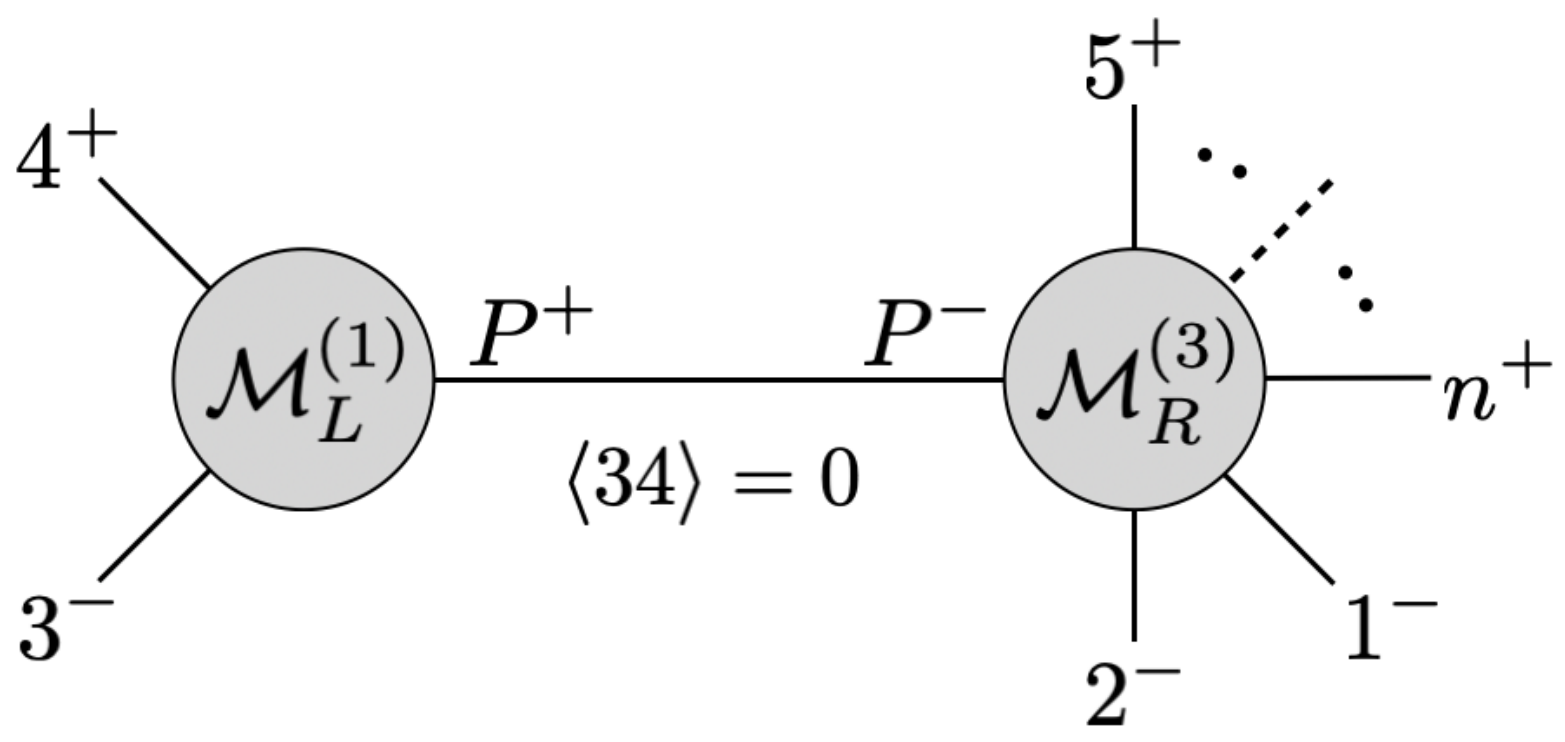}}\qquad \raisebox{-3pt}{\includegraphics[scale=.43]{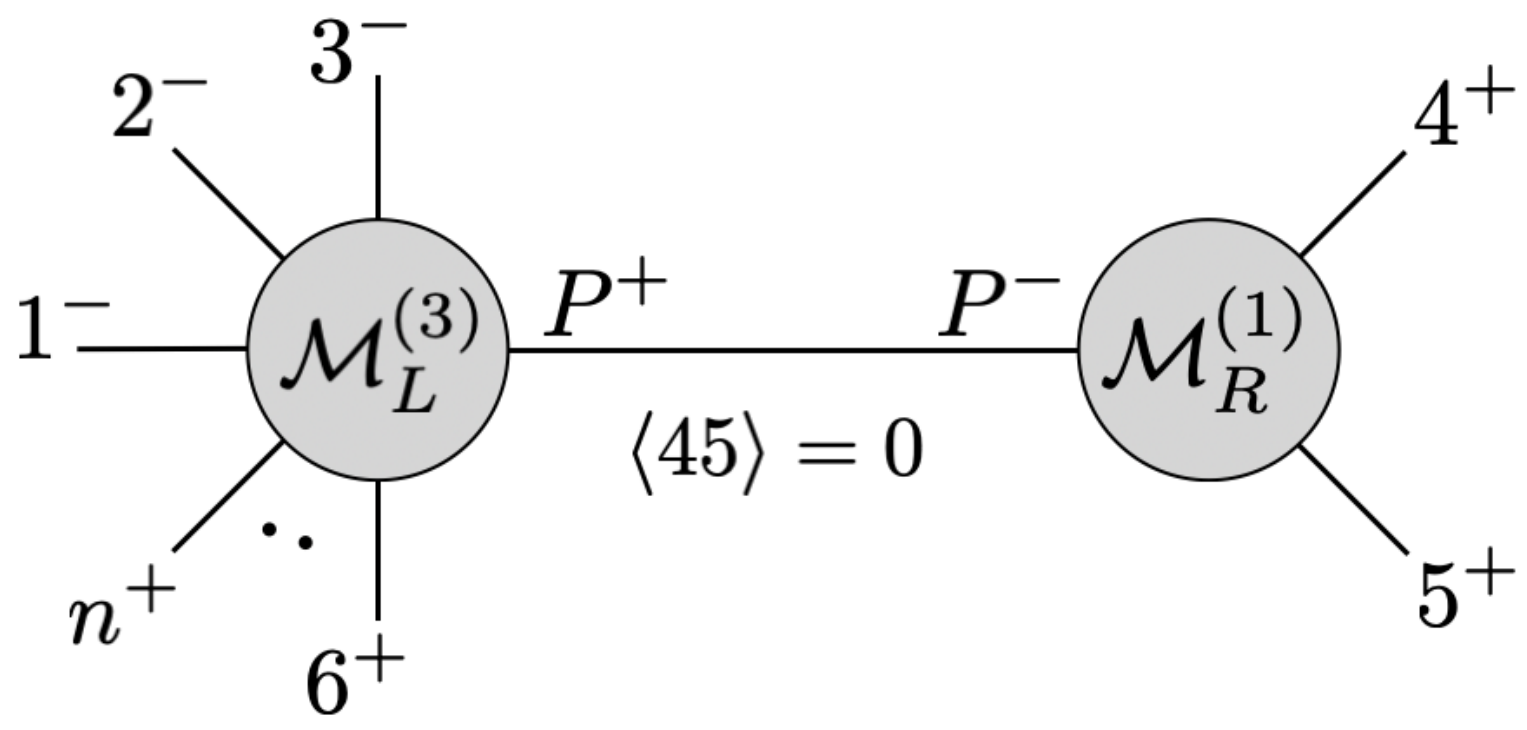}}
$$
and two kinds of anti-holomorphic factorizations,
$$
\raisebox{-0pt}{\includegraphics[scale=.43]{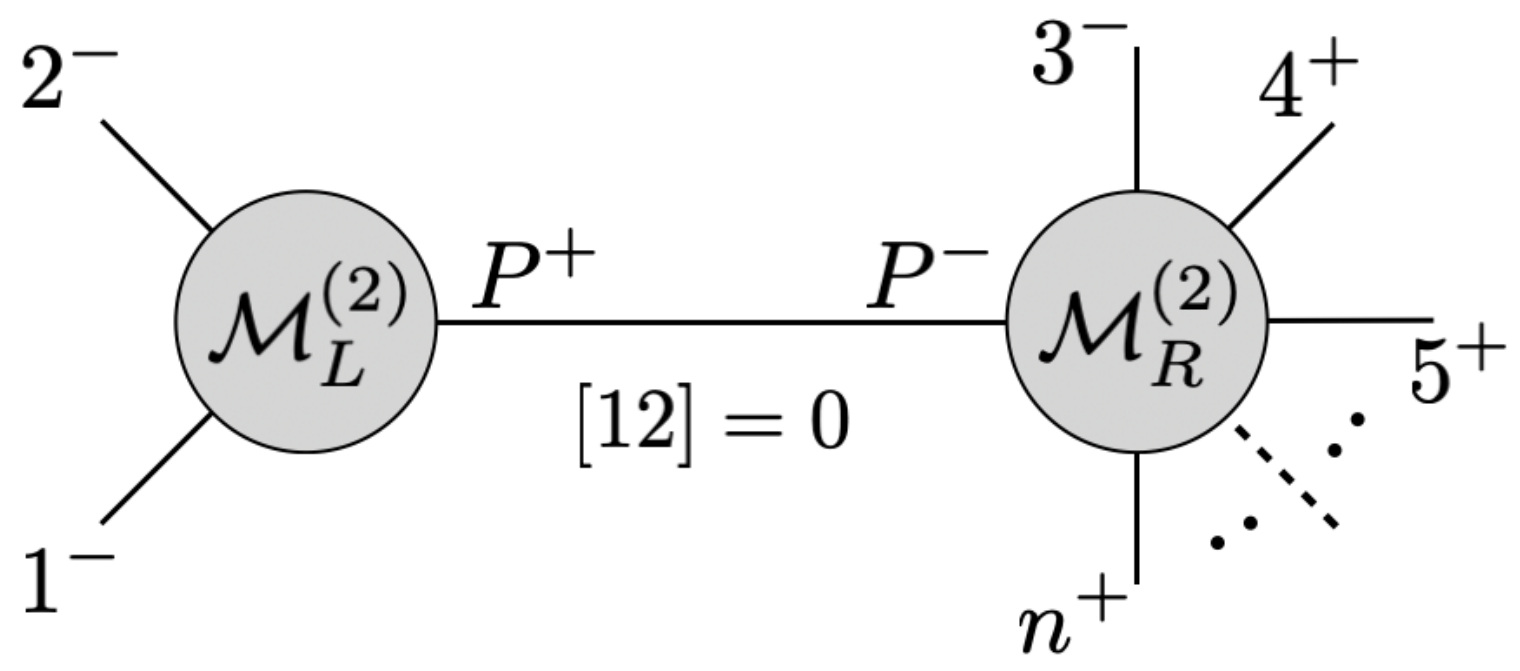}}\qquad \raisebox{-3pt}{\includegraphics[scale=.43]{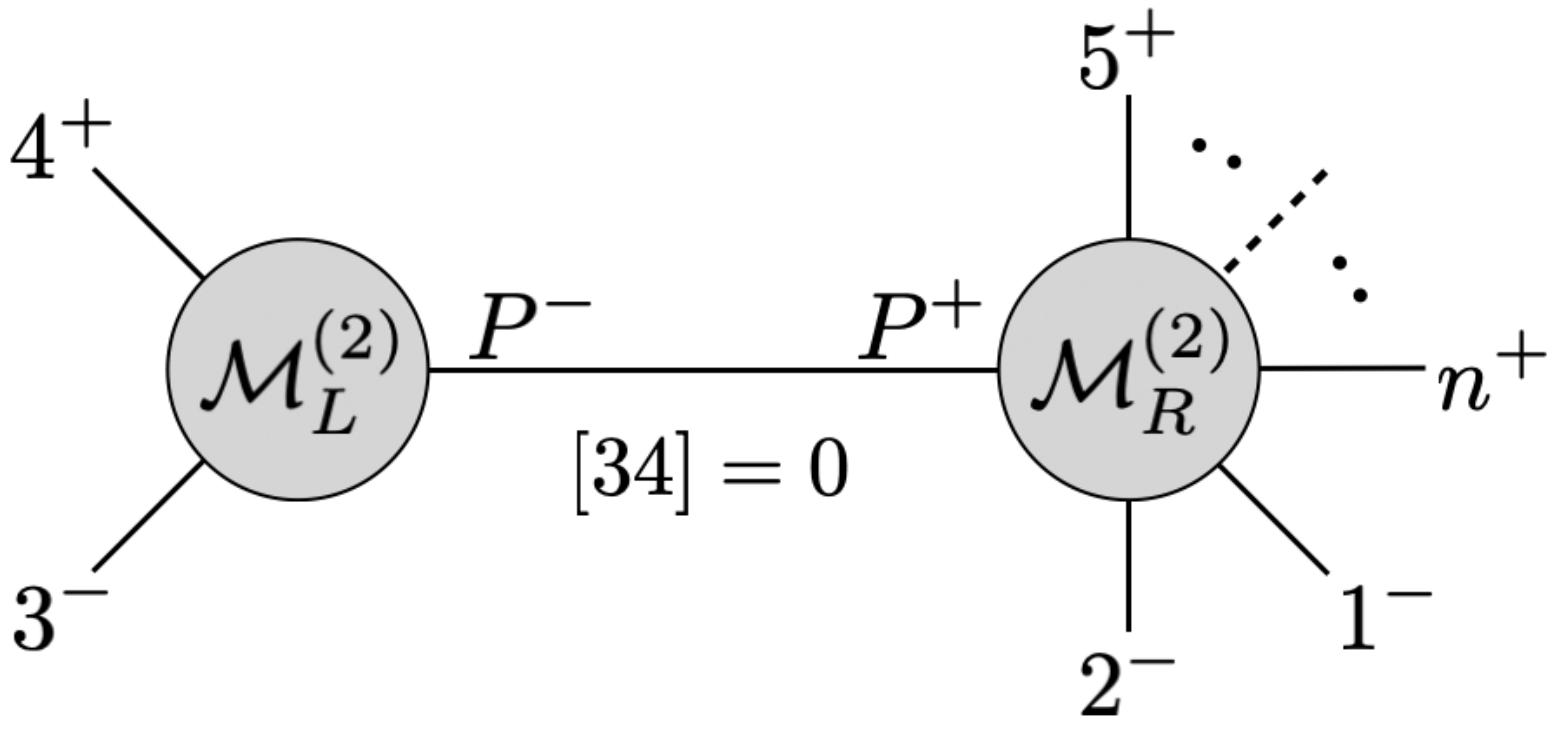}}
$$
There are no allowed factorization channels of the type $\la12\ra=0$ or $[45]=0$. We checked up to 10pt numerically that our proposal (\ref{GRamp}) correctly factorizes on all factorization channels. This itself is not enough to prove that (\ref{GRamp}) correctly reproduces the gravity amplitudes, we also need to check the spurious poles cancelations.

It is interesting to compare the factorization diagrams with the actual factorization channels. Formally, the formulas (\ref{split2}) for Yang-Mills and (\ref{GRamp}) for gravity looks like a sum over certain set of factorization channels,
$$
\raisebox{-44pt}{\includegraphics[scale=.34]{factor.pdf}}\,\,\rightarrow\,\, \raisebox{-44pt}{\includegraphics[scale=.43]{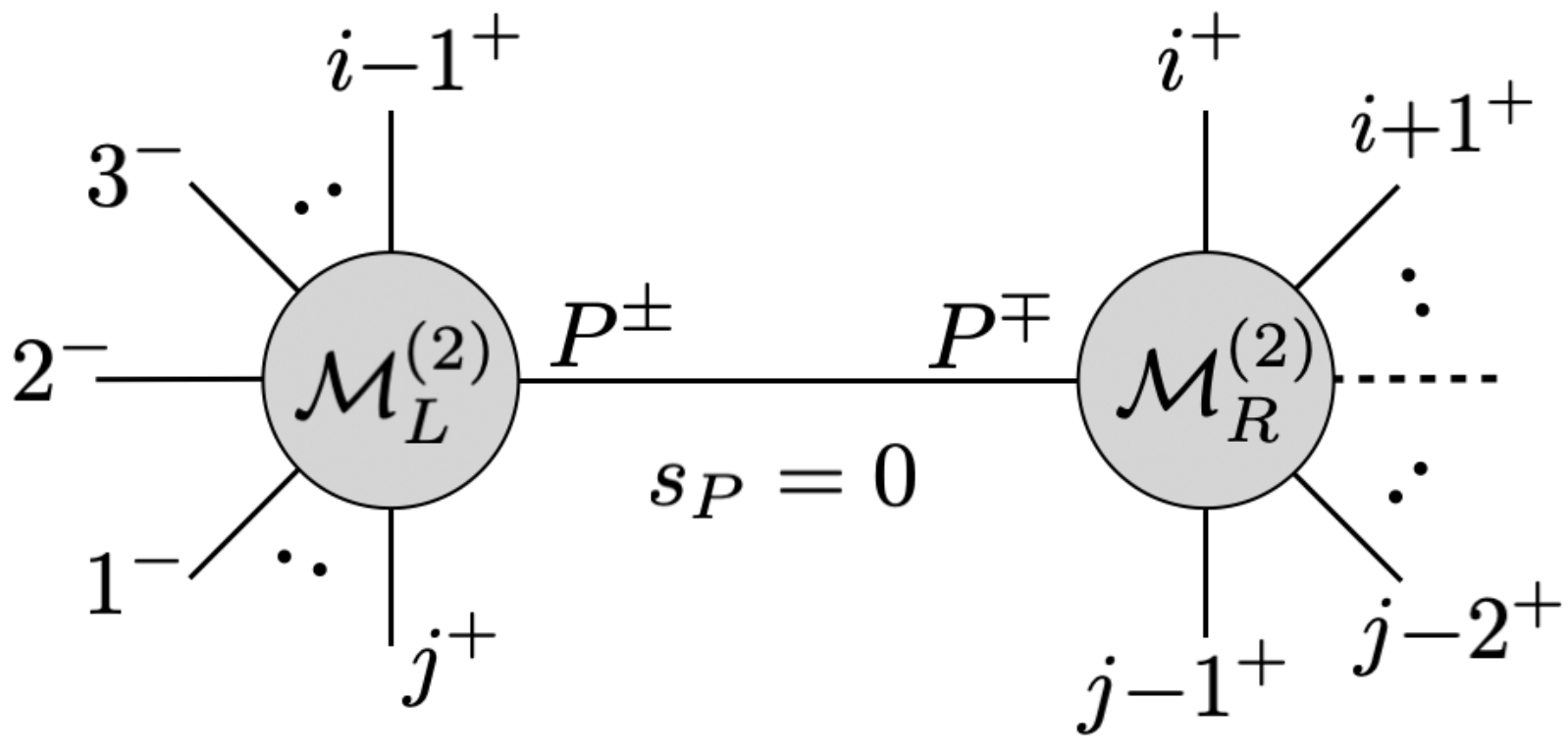}}
$$
corresponding to multi-particle and anti-holomorphic two-particle factorization channels with legs $1,2,3$ in ${\cal M}_L$ (in addition to any number of positive helicity legs). None of these poles are present in the amplitudes ${\cal A}_n$ and ${\cal M}_n$ and these factorization channels are forbidden because of wrong helicity labels. This can be also seen from the formulas for ${\cal R}$- and ${\cal G}$-invariants, as the helicity factor ${\cal H}$ always cancels $s_P$ pole in the denominator. Therefore, (\ref{split2}) and (\ref{GRamp}) are not ``sums over factorization channels".

\subsection{Spurious poles cancelation}

The individual ${\cal G}$-invariants in our representation of the NMHV gravity amplitude (\ref{GRamp}) have spurious poles, which is similar to BCFW formulas. As mentioned in the last section, matching all factorization channels is not enough to determine if (\ref{GRamp}) correctly reproduces the amplitude. We also need to check that all spurious poles cancel. While we do not have an analytic understanding of this fact, we checked numerically up to 10pt that this indeed happens and the formula (\ref{GRamp}) is spurious poles free. This concludes the proof that (\ref{GRamp}) gives the correct NMHV gravity amplitude (up to 10pt), but it is important to have an analytic proof for any $n$. We leave that for future work.

Let us quickly comment on why factorizations plus absence of spurious poles are enough to specify the amplitude uniquely. Let us combine all pieces in (\ref{GRamp}) under common denominator and denote ${\cal M}_n^{prop}$ the proposed expression for the amplitude. The difference between ${\cal M}_n^{prop}$ and the correct amplitude ${\cal M}_n$ must be a polynomial $P$ in $\lambda_i$, $\widetilde{\lambda}_i$ without any poles (otherwise, it would spoil the factorizations),
\begin{equation}
{\cal M}_n^{prop} - {\cal M}_n = P(\lambda_i,\widetilde{\lambda}_i)
\end{equation}
The mass dimension is $[P]\sim m^2$ where $\la ij\ra\sim [ij]\sim m$, and at the same time it must have the right little group weights of the amplitude under $\lambda_i\rightarrow t\lambda_i$, $\widetilde{\lambda}_i\rightarrow \frac{1}{t}\widetilde{\lambda}_i$. The dimensionality and little group weights forces $P$ to have poles which is in conflict with $P$ being the polynomial, therefore $P=0$. This also follows more generally from the $D$-dimensional analysis using polarization vectors \cite{Arkani-Hamed:2016rak,Rodina:2016mbk,Rodina:2016jyz}.

Now, we want to dive more in the spurious poles cancelation between individual ${\cal G}$-invariants. This was crucial in the geometric constructions for Yang-Mills amplitudes \cite{Hodges:2009hk} and eventually lead to the discovery of Amplituhedron \cite{Arkani-Hamed:2013jha}. We focus only on spurious poles in the six-point amplitude ${\cal M}_6$ and compare it to the six-point Yang-Mills amplitude ${\cal A}_6$. There are no spurious poles in the formula for five-point amplitudes because of special kinematics. We wrote the six-point gluon amplitude in a suggestive form (\ref{A6YM}) to show the analogy with the gravity amplitude ${\cal M}_6$ in terms of ${\cal G}$-invariants,
$$
\includegraphics[scale=.54]{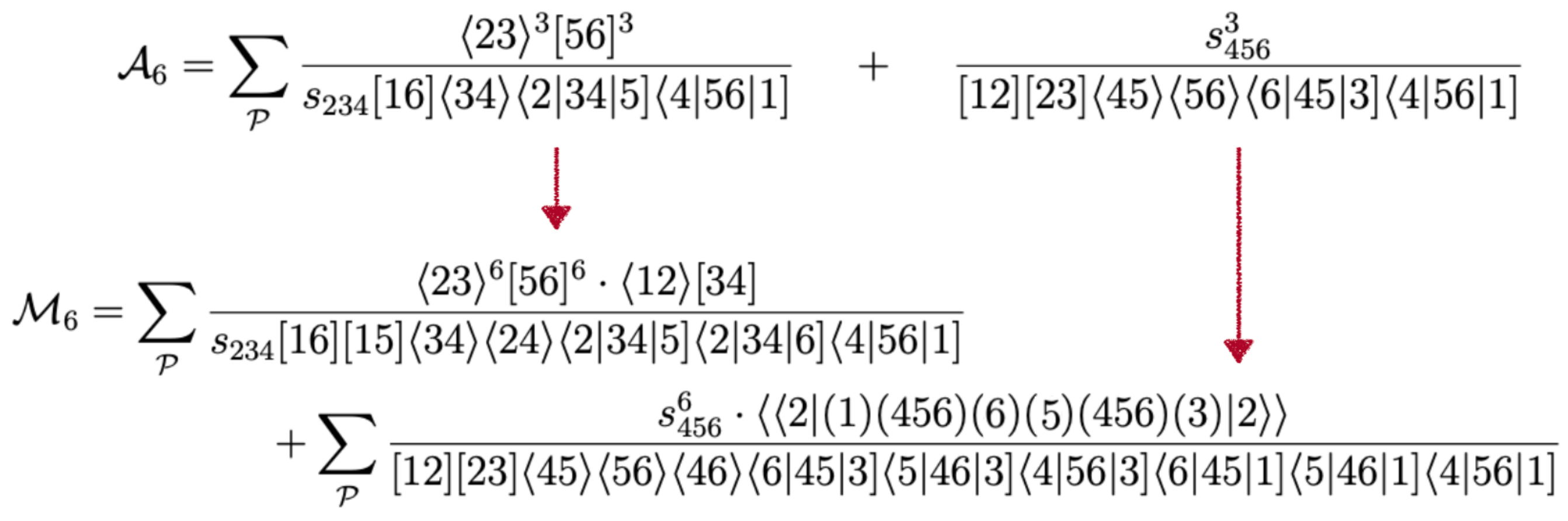}
$$
The first term in both formulas corresponds to a factorization diagram with $P=\{5,6\}$ while the factorization diagram for the second term has $P=\{4,5,6\}$ in the right blob. Here ${\cal P}$ denotes schematically the sum over permutations of a given term. The permutation sum for the first term in ${\cal A}_6$ only contains two terms, 
\begin{equation}
{\cal P}:(1,2,3,4,5,6) \rightarrow \{(1,2,3,4,5,6), (3,2,1,6,5,4)\} \label{YMperm}
\end{equation}
while the sums in ${\cal M}_6$ are over permutations of $1,2,3$ and $4,5,6$ modulo symmetries of factorization diagrams. The similarities in the gluon and graviton expressions come from the same procedure how to generate red and green poles, but there are differences in the number of these poles as well as the presence of non-trivial numerators in the graviton case. We can see that not all poles in individual terms of ${\cal A}_6$ and ${\cal M}_6$ are physical. In fact, all poles of the form $\la a|P|b]$, $\la a|(P_1)(P_2)|b\ra$ are spurious and must cancel in the sum. In the gluon case, this is a pairwise cancelation and it is an important feature of the Amplituhedron picture. For NMHV amplitudes this geometry reduces to polytopes in $\mathbb{P}^3$ and each term in ${\cal A}_6$ corresponds to a simplex. At six-point the three terms in ${\cal A}_6$ represent three simplices which are glued in the following way
$$
\includegraphics[scale=.2]{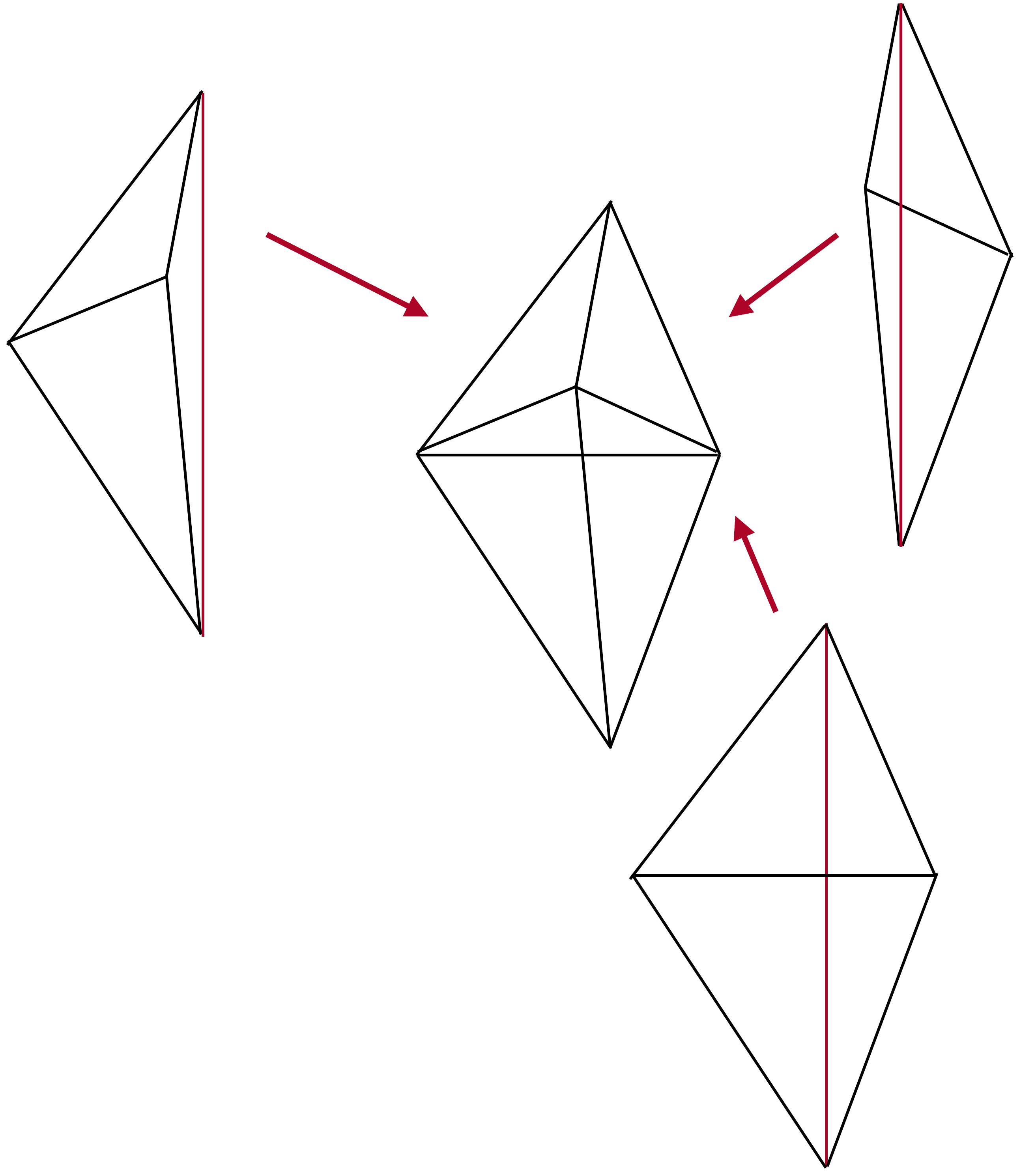}
$$
The early indication for this geometric picture was the mechanism of cancelation of spurious poles. Each spurious pole correspond to a ``spurious face" which cancels once we glue two pieces together. For this particular triangulation, there are free spurious faces reflecting three spurious poles in ${\cal A}_6$,
$$
\includegraphics[scale=.33]{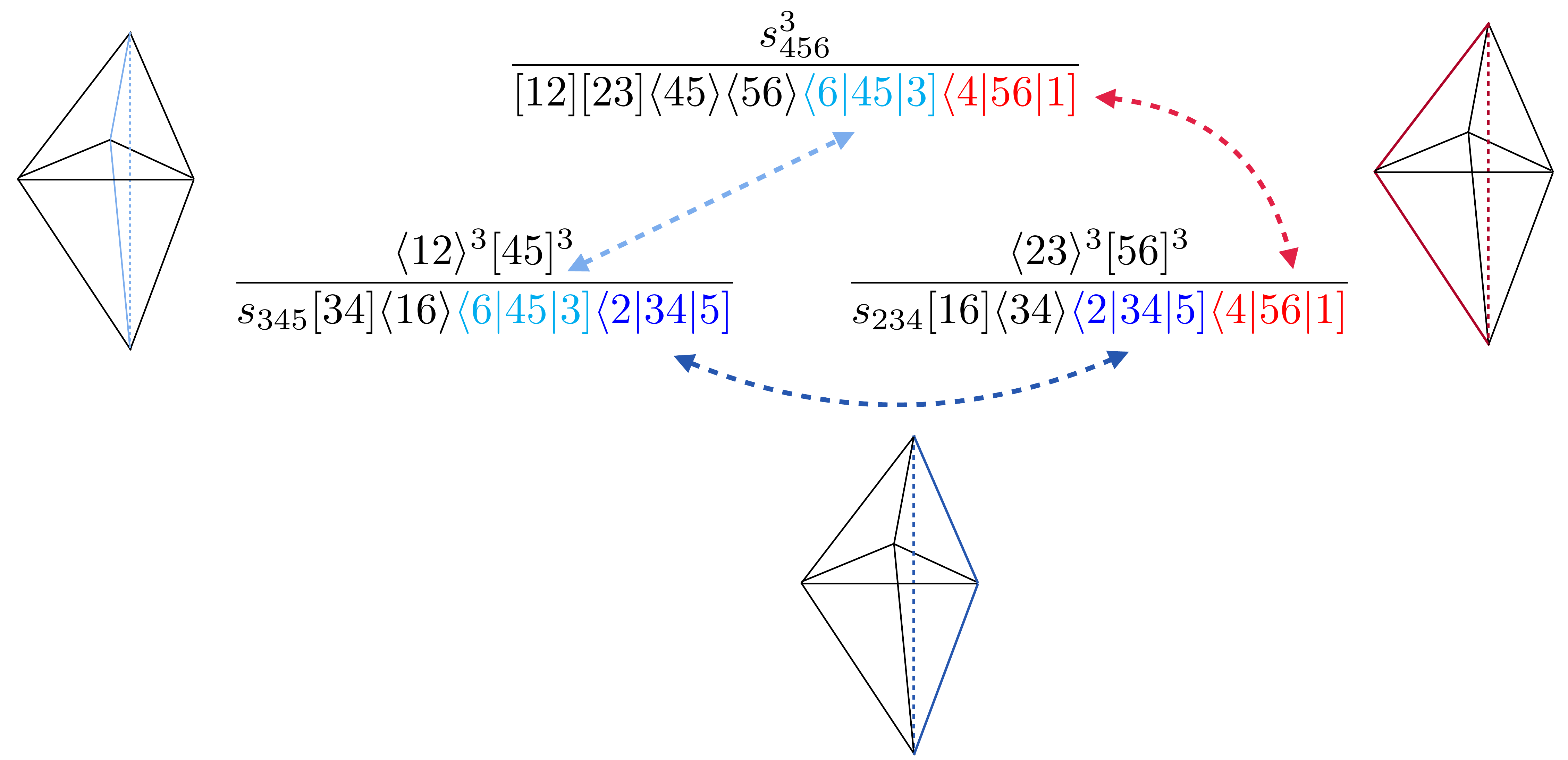}
$$
Interestingly, the pairwise cancelation of spurious poles is also a feature of our representation of the graviton amplitude ${\cal M}_6$. The web of spurious poles cancelations splits into three separate sets of ${\cal G}$-invariants with the same origin. The set of ${\cal G}$-invariants with fixed origin is spurious pole free. For example, if we choose ${\bf 2}$ as origin and keep all ${\cal G}$-invariants from the amplitude ${\cal M}_6$ with ${\cal G}(\cdot,{\bf 2},\cdot,\{\cdots\},\{\cdots\},\{\cdots\})$ we get the sum of seven terms which are spurious pole free,
\begin{align}
{\cal M}_6^{({\bf 2})} &= \sum_{{\cal P}(4,5,6)} \frac{\la23\ra^6[56]^6\cdot \la12\ra[34]}{s_{234}[16][15]\la34\ra\la24\ra\la2|34|5]\la2|34|6]\la4|56|1]}\label{GR6part}\\
& \hspace{2.5cm}+  \frac{s_{456}^6\cdot \la\la2|(1)(456)(5)(6)(456)(3)|2\ra\ra}{[12][23]\la45\ra\la56\ra\la46\ra\la6|45|3]\la5|46|3]\la4|56|3]\la6|45|1]\la5|46|1]\la4|56|1]}\nonumber
\end{align}
Six terms from the first line (\ref{GR6part}) have pairwise cancelations of dark blue and light blue poles, while each term has one red pole which cancels agains the term on the second line of (\ref{GR6part}). The web of cancelations is 
$$
\includegraphics[scale=.57]{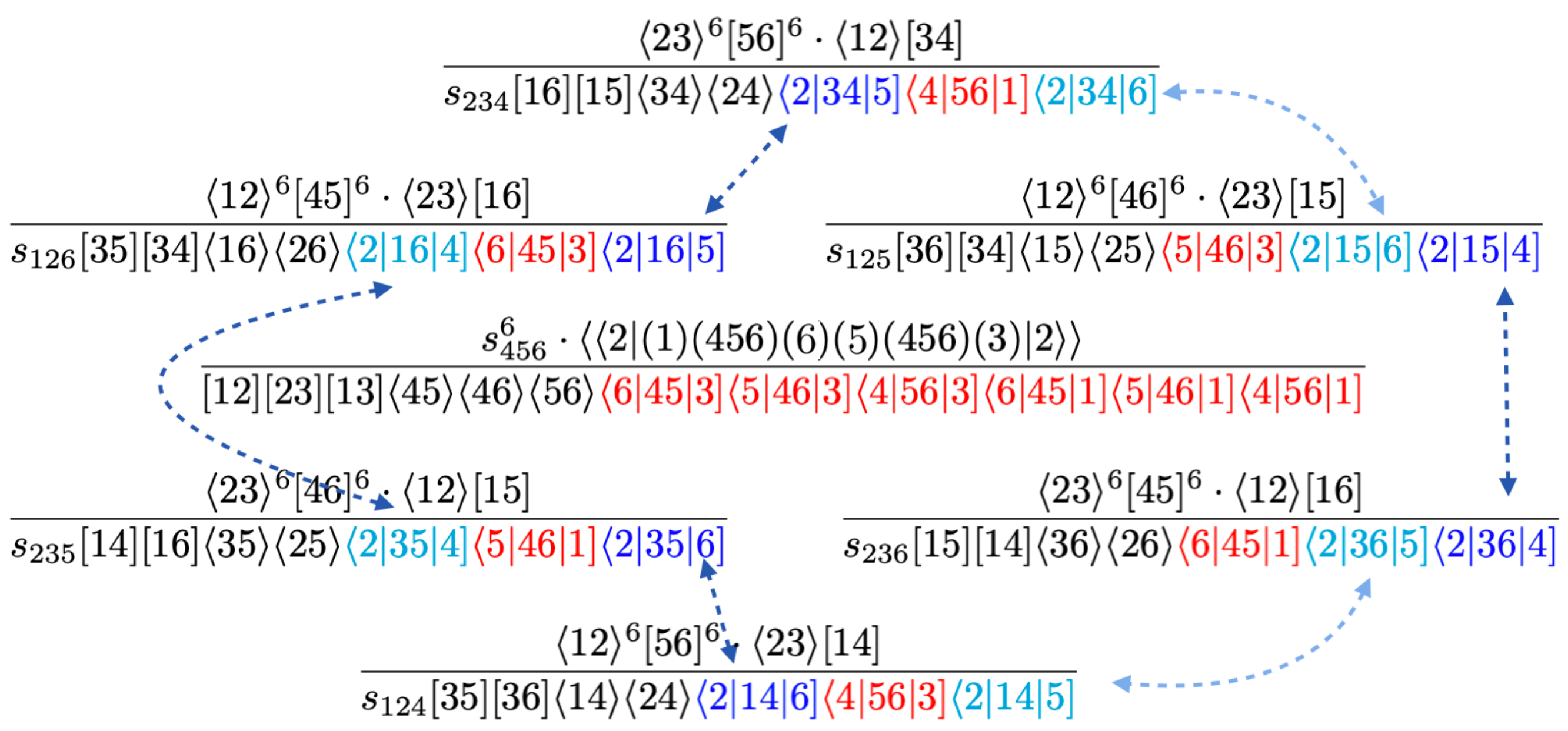}
$$
where the pair of spurious poles are always identical due to momentum conservation, ie. $\la2|16|4]=-\la2|35|4]$. We did not draw arrows for red spurious poles to keep the figure simple but it is easy to see which red poles are connected. 

This is a generalization of the Yang-Mills web which had only two terms of the first type instead of six. Note that the colors in the picture are unrelated with our labeling of red, green and blue poles. Understanding these cancelations geometrically would be an important first step in the Gravituhedron construction for ${\cal M}_6$. 

Note that spurious poles cancelation happens between terms in ${\cal M}_6^{({\bf 2})}$. We can then divide the amplitude into three pieces based on the origin in the left blob,
\begin{equation}
{\cal M}_6 = {\cal M}_6^{({\bf 1})} + {\cal M}_6^{({\bf 2})} + {\cal M}_6^{({\bf 3})}
\end{equation}
each of them is free of spurious poles. We conjecture that the same is true for any number of points and we can write a general $n$-pt amplitude ${\cal M}_n$ as the sum of three {\it partial amplitudes},
\begin{equation}
{\cal M}_n = {\cal M}_n^{({\bf 1})} + {\cal M}_n^{({\bf 2})} + {\cal M}_n^{({\bf 3})} \label{part2}
\end{equation}
where each partial amplitude
\begin{equation}
  {\cal M}_n^{({\bf 2})} =  \sum_{{\cal P}(1,3)}\sum_{P,Q_1,Q_2}  {\cal G}(1,{\bf 2},3,\{Q_1\},\{P\},\{Q_2\}) \label{part3}
\end{equation}
has no spurious poles. The sum is over all distribution of labels $4,5,\dots,n$ and symmetrized over $1,3$, modulo identical terms. We checked this conjecture numerically up to 10pt. The absence of spurious poles in (\ref{part3}) is very unusual and it would be interesting to explore if this object has some independent meaning. It is something new from the Yang-Mills case where the only combination of ${\cal R}$-invariants with no spurious poles is the scattering amplitude. Here we get three different objects with no spurious poles. Of course, we still have to sum them to get correct factorization on all physical poles. It would be interesting to understand if it is somehow related to three coordinate patches (in momentum twistor space) discussed in \cite{Armstrong:2020ljm} which helped to understand the web of cancelations for ${\cal N}=7$ BCFW terms.

\subsection{Large-$z$ behavior}

Our result is not derived from BCFW recursion relations so there is no a priori reason to expect that the individual ${\cal G}$-invariants would enjoy any special large-$z$ behavior under BCFW shifts. In Yang-Mills theory, the $n$-pt amplitude scales  for $z\rightarrow\infty$ as 
\begin{equation}
{\cal A}_n = {\cal O}\left(\frac{1}{z}\right)\quad \mbox{for shift $\{a,b\}=\{+,+\},\{-,-\},\{-,+\}$} \label{BCFWscale1}
\end{equation}
where $\pm$ denotes any positive, resp. negative helicity label. This is true for adjacent shifts where $a,b$ are adjacent labels. For non-adjacent shifts the vanishing at infinity is even stronger ${\cal O}(1/z^2)$. These are also referred to as ``good shifts" in the literature. The individual BCFW terms, aka ${\cal R}$-invariants, also scale as ${\cal O}(1/z)$ for all good shifts, making this behavior manifest term-wise. This is closely related to the manifest dual conformal symmetry, so it is natural to study it in details also for gravity. The $n$-pt gravity amplitude scales as
\begin{equation}
{\cal M}_n = {\cal O}\left(\frac{1}{z^2}\right)\quad \mbox{for shift $\{a,b\}=\{+,+\},\{-,-\},\{-,+\}$} \label{BCFWscale2}
\end{equation}
so same good shifts as in Yang-Mills. Note there is no notion of adjacent/non-adjacent shifts here. The ${\cal N}=8$ recursion does not exhibit this behavior term-by-term, and the individual terms only scale as ${\cal O}(1/z)$. It was shown in \cite{Hodges:2011wm} that the ${\cal N}=7$ recursion does make the improved ${\cal O}(1/z^2)$ scaling manifest term-by-term for certain shifts. In particular, if we use the $\{3,4\}$ shift to construct the six-point amplitude ${\cal M}_6(1^-,2^-,3^-,4^+,5^+,6^+)$, then the individual BCFW terms scale as ${\cal O}(1/z^2)$ for
\begin{equation}
\{a,b\} = \{3,+\}\,\,\,\mbox{and}\,\,\, \{a,b\} = \{-,4\}\label{N7scale}
\end{equation}
where again $\pm$ denotes any positive, resp. negative helicity label. This is a subset of all good shifts (\ref{BCFWscale2}). Numerical checks show that ${\cal G}$-invariants scale as ${\cal O}(1/z^2)$ for
\begin{equation}
\{a,b\} = \{-,+\}\label{Gscale}
\end{equation}
which does not cover all cases in (\ref{BCFWscale2}) but it is a larger set than (\ref{N7scale}). In fact, (\ref{N7scale}) is a subset of (\ref{Gscale}), and from this perspective ${\cal G}$-invariants are superior to the ${\cal N}=7$ (and also ${\cal N}=8$) BCFW terms based on the large-$z$ behavior analysis. For example, at six-point all ${\cal G}$-invariants manifest ${\cal O}(1/z^2)$ behavior under 9 shifts (\ref{Gscale}) while the ${\cal N}=7$ recursion only under 6 shifts (\ref{N7scale}). 

It has been known for long time that the manifest ${\cal O}(1/z^2)$ behavior in gravity is closely related to the permutational symmetry. And indeed, that is what seems to be happening here as well. In order to get ${\cal O}(1/z^2)$ behavior in $\{-,-\}$ or $\{+,+\}$ shifts we need to have symmetric expressions in $1,2,3$, resp. in $4,5,\dots,n$. Let us take the example of the seven-point amplitude (\ref{GR7}). The last term is already symmetric in $4,5,6,7$ and it has manifest ${\cal O}(1/z^2)$ behavior in $\{+,+\}$ shifts. If we now symmetrize in $1,2,3$ (only three terms) we get ${\cal O}(1/z^2)$ in $\{-,-\}$, and
\begin{equation}
\sum_{{\cal P}'(1,2,3)}{\cal G}(1,{\bf 2},3,\{\},\{4,5,6,7\},\{\})\quad \mbox{has}\,\, {\cal O}\left(\frac{1}{z^2}\right) \,\mbox{under all shifts}
\end{equation}
Similarly, the next-to-last term in (\ref{GR7}) must be symmetrized in $1,2,3$ to restore ${\cal O}(1/z^2)$ in $\{-,-\}$ and in $4,5,6,7$ for $\{+,+\}$,
\begin{equation}
\sum_{{\cal P}(1,2,3)}\sum_{{\cal P}'(4,5,6,7)}{\cal G}(1,{\bf 2},3,\{4\},\{5,6,7\},\{\})\quad \mbox{has}\,\, {\cal O}\left(\frac{1}{z^2}\right) \,\mbox{under all shifts}
\end{equation}
Finally, the first two terms in (\ref{GR7}) must be combined and symmetrized to reproduce ${\cal O}(1/z^2)$. This would also suggest that it is natural to consider a combination of ${\cal G}$-invariants with fixed origin and the same right blob. For example for $\{P\}=\{6,7\}$ and origin {\bf 2},
\begin{align}
{\cal G}_{67}^{({\bf 2})} &= {\cal G}(1,{\bf 2},3,\{4,5\},\{6,7\},\{\}) + {\cal G}(1,{\bf 2},3,\{\},\{6,7\},\{4,5\})\nonumber\\
&\hspace{0.3cm}+ {\cal G}(1,{\bf 2},3,\{4\},\{6,7\},\{5\})+ {\cal G}(1,{\bf 2},3,\{5\},\{6,7\},\{4\}) \label{G67}
\end{align}
If we now symmetrize this expression over $1,2,3$ we get ${\cal O}(1/z^2)$ for $\{-,-\}$ shifts, while for only one of the terms in (\ref{G67}) the symmetrization would not produce ${\cal O}(1/z^2)$. It is an open question if the fundamental building blocks are really individual ${\cal G}$-invariants which do have the enhanced ${\cal O}(1/z^2)$ behavior under $\{-,+\}$ shifts or if we should form the combinations such as (\ref{G67}) which share the same right blob. 

\subsection{Numerators and momentum twistors}

The use of momentum twistors for graviton amplitudes was proposed in \cite{Hodges:2011wm} and then discussed in great details in \cite{Armstrong:2020ljm} with some non-trivial results. The obvious problem is that momentum twistors require an ordering of external legs, hence they are very suitable for color-ordered gluon amplitudes. The external legs in gravity amplitudes are not ordered and naively there is no point of introducing them here. It was argued in \cite{Armstrong:2020ljm} that momentum twistors expose certain spurious poles cancelations in the BCFW formula for six-point NMHV amplitude. This requires using different set of momentum twistors in order to manifest spurious poles cancelations for different poles. In other words, while ``locally" momentum twistors can be used, the global picture requires switching between various local patches.

We will use momentum twistors for a different purpose: to understand the meaning of numerator factors we introduced in our rules for left and right blob functions $P_L$ and $P_R$ when constructing the $n$-pt NMHV graviton amplitude. The numerators of the left blob have a very simple meaning in the momentum twistor language and directly translate into a single four-bracket (and angle brackets),
\begin{equation}
\la 2|3|4] = \frac{\la 2345\ra}{\la34\ra\la45\ra},\qquad \la 2|34|5] = \frac{\la 2456\ra}{\la45\ra\la56\ra},\qquad \la 2|345|6] = \frac{\la 2567\ra}{\la56\ra\la67\ra},\,\, \dots
\end{equation}
For right blobs with three and more labels there is a non-trivial red numerator which requires merging between $P_L$ and $P_R$ leading to a double-angle bracket (\ref{longB2}). At six-point, it appears in ${\cal G}(1,{\bf 2},3,\{\},\{4,5,6\},\{\})$, and we can write it in terms of standard kinematical invariants (\ref{longB2}) as
\begin{align}
   \la\la 2|(1)(456)(4)(6)(456)(3)|2\ra\ra&=\la 2|(1)(456)(4)(6)(456)(3)|2\ra + s_{456}\la2|(1)(6)(4)(3)|2\ra\nonumber\\
   & \hspace{-2cm}= \la 2|(1)(65)|4\ra[46]\la6|(54)(3)|2\ra + \la 2|1|6]\la46\ra\la2|3|4]s_{456}\label{momtw}
\end{align}
The individual terms can be translated into momentum twistor space,
\begin{align}
&\la 2|(1)(65)|4\ra = \frac{\la 2164\ra}{\la16\ra},\qquad \la6|(54)(3)|2\ra = \frac{\la 6432\ra}{\la43\ra}, \qquad \la 2|1|6]=\frac{\la2165\ra}{\la16\ra\la65\ra},\nonumber\\
&\hspace{3cm}\la2|3|4] = \frac{\la 2345\ra}{\la34\ra\la45\ra},\qquad s_{456} = \frac{\la 3461\ra}{\la34\ra\la61\ra}
\end{align}
The square-bracket $[46]$ contains non-adjacent labels and the translation contains four-bracket with infinity twistor $I$,
\begin{equation}
[46] = \frac{\la I (345){\cap}(561)\ra}{\la 34\ra\la45\ra\la56\ra\la61\ra}
\end{equation}
Plugging everything into (\ref{momtw}) we get (using anti-symmetry in $4\leftrightarrow 6$),
\begin{equation}
 \la\la 2|(1)(456)(4)(6)(456)(3)|2\ra\ra =
\frac{\la 1246\ra\la2346\ra\la I(345){\cap}(561)\ra+ \la1256\ra\la1346\ra\la2345\ra\la I46\ra}{\la34\ra^2\la45\ra\la56\ra\la61\ra^2}\label{momtw2}
\end{equation}
where we rewrote $\la46\ra$ as a four-bracket $\la46\ra=\la I46\ra$ in the numerator. There are many different ways how to write the numerator (\ref{momtw2}) using Schouten identities. The invariant question is why exactly terms in (\ref{momtw2}) or originally in (\ref{momtw}) appear in this particular combination? This seems like a very complicated question in the momentum space (\ref{momtw}) but it has a nice geometric interpretation in momentum twistor space. In the framework of momentum twistor geometry \cite{ArkaniHamed:2010kv,ArkaniHamed:2010gh,Bourjaily:2013mma,Arkani-Hamed:2014dca,Bourjaily:2015jna,Bourjaily:2017wjl} we can think about the infinity twistor $I$ as a line in $\mathbb{P}^3$ and ask for special configurations of $I$ when the numerator of (\ref{momtw2}) vanishes. 

This question has multiple solutions: for $I=45$ or $I=56$ both expressions in the numerator cancel independently. But there is also (at least) one solution which requires a cancelation between both terms,
\begin{equation}
I = (234)\cap(612) \label{momtw3}
\end{equation}
This is a very surprising result: the momentum space numerator (\ref{momtw}) did not have a fixed ordering of points, in particular it was permutational invariant in $4,5,6$. However, when using canonical ordering and translating in momentum twistor space we find this intriguing geometry (\ref{momtw3}) with planes made of consecutive points.

The other example is the double-angle bracket for the right blob which appears at 8pt and higher, for example $\la\la 4|(567)(7)(5)(567)|8\ra\ra$. In momentum twistor space it takes the form,
\begin{equation}
\la \la 4|(567)(7)(5)(567)|8\ra\ra = \frac{\la4568\ra\la4678\ra\la I57\ra+\la4578\ra\la I(456){\cap}(678)\ra}{\la45\ra\la56\ra\la67\ra\la78\ra}\label{momtw4}
\end{equation}
This is a simpler expression than (\ref{momtw2}) and the numerator vanishes for $I=45$, $I=48$, $I=56$, $I=67$, $I=78$. All of these vanishings are manifest except $I=48$ which requires cancelation between both terms. We can use Shouten identities to rewrite (\ref{momtw4}) in a different form which makes cancelation for $I=48$ manifest but obscures others, and there seems to be no representation which would make all these conditions manifest at the same time.

We do not claim to have any concise story about the role of momentum twistors here, but these observations produce another evidence of potential important role of momentum twistor variables for graviton amplitudes, or in general for writing non-cyclic-ordered kinematical expressions. 
 
\subsection{Triangulation of the right blob}

Finally, let us discuss one technical but potentially important observation about the right blob function $P_R$. In section 4.4. we discussed how to obtain $P_R$ by expanding it in terms of partial right blob functions with the origin $j{-}1$. The result had to be symmetrized over all labels in the right blob except $j{-}1$. In fact, we do not have to choose any of the labels as the origin and rather use an arbitrary reference $x$. The partial right blob functions with five and six points (for canonical labels starting with $4$) are then
\begin{equation}
\raisebox{-42pt}{\includegraphics[scale=.35]{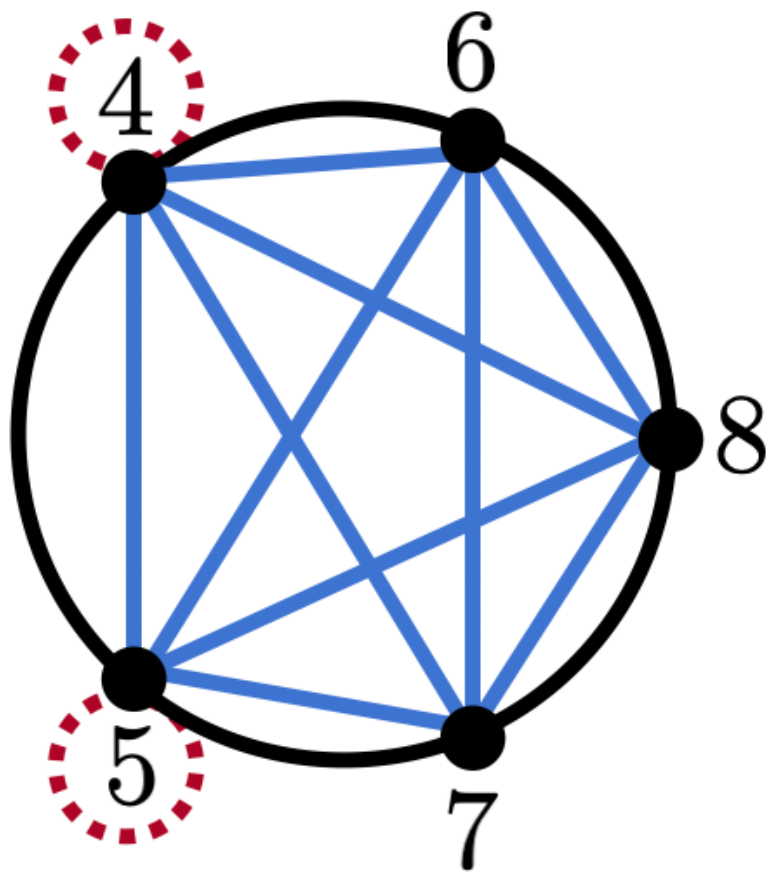}}\hspace{0.1cm} = \hspace{0.05cm} \frac{{\color{blue}  \la\la 4|(678)(6)(7)(678)|5\ra\ra\cdot \la4x\ra\la5x\ra}\times{\color{red}[\diamond (P)(4)(5)(P)\diamond]}}
{\begin{tabular}{c}
${\color{blue}\la 45\ra\la46\ra\la47\ra\la56\ra\la57\ra\la58\ra\la67\ra\la68\ra\la78\ra}\times
{\color{red} [\ast_1(P)|4\ra] }$\\${\color{red} [\ast_2(P)|5\ra][\ast_3(P)|x\ra][\ast_4(P)|4\ra] [\ast_5(P)|5\ra] [\ast_6(P)|x\ra]}$\end{tabular}}
\end{equation}
\begin{equation}
\raisebox{-42pt}{\includegraphics[scale=.35]{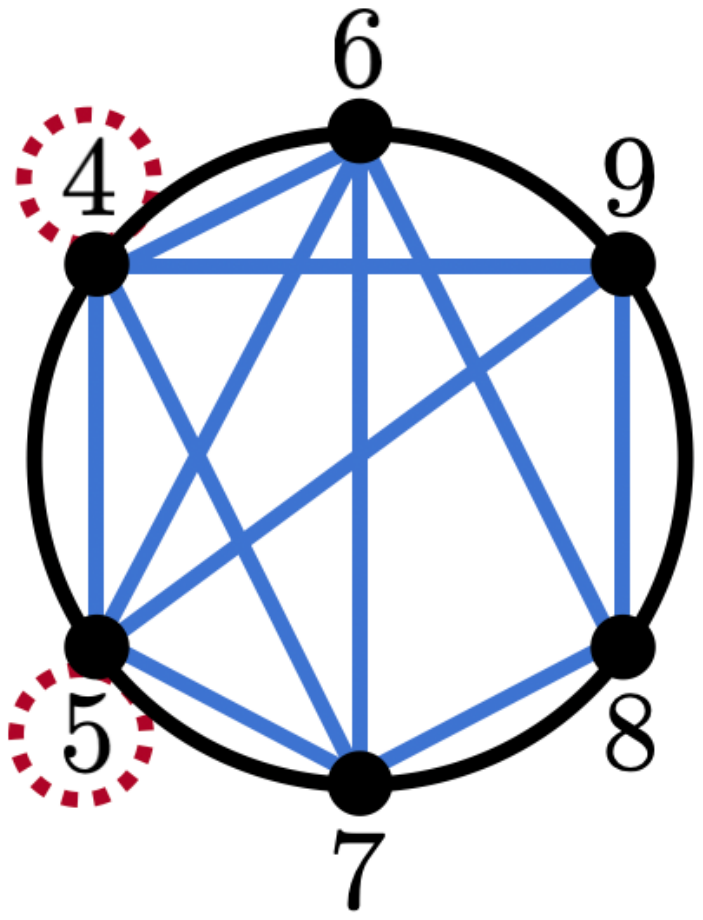}}\hspace{0.1cm} = \hspace{0.05cm} \frac{{\color{blue}  \la\la 4|(6789)(6)(7)(6789)|5\ra\ra\cdot [89]\cdot \la4x\ra\la5x\ra}\times{\color{red}[\diamond (P)(4)(5)(P)\diamond]}}
{\begin{tabular}{c}
${\color{blue}\la 45\ra\la46\ra\la47\ra\la49\ra\la56\ra\la57\ra\la59\ra\la67\ra\la68\ra\la78\ra\la89\ra}\times
{\color{red} [\ast_1(P)|4\ra] }$\\${\color{red} [\ast_2(P)|5\ra][\ast_3(P)|x\ra][\ast_4(P)|4\ra] [\ast_5(P)|5\ra] [\ast_6(P)|x\ra]}$\end{tabular}}
\end{equation}
and we have to sum over all permutations (modulo symmetries). Note that the point $j{-}1$ is no longer encircled, and the red half pole is now ${\color{red} [\ast(P)|x\ra]}$ for arbitrary reference $\lambda_x$. We get additional blue poles in the denominator and $\la4x\ra\la5x\ra$ factors in the numerator. We can easily see that by choosing $x=8$ in the first term and $x=9$ in the second term we reconstruct expressions (\ref{right1}), (\ref{right2}). The general pattern is clear, for arbitrary number of points and canonical ordering of labels we get
$$
\raisebox{0pt}{\includegraphics[scale=.38]{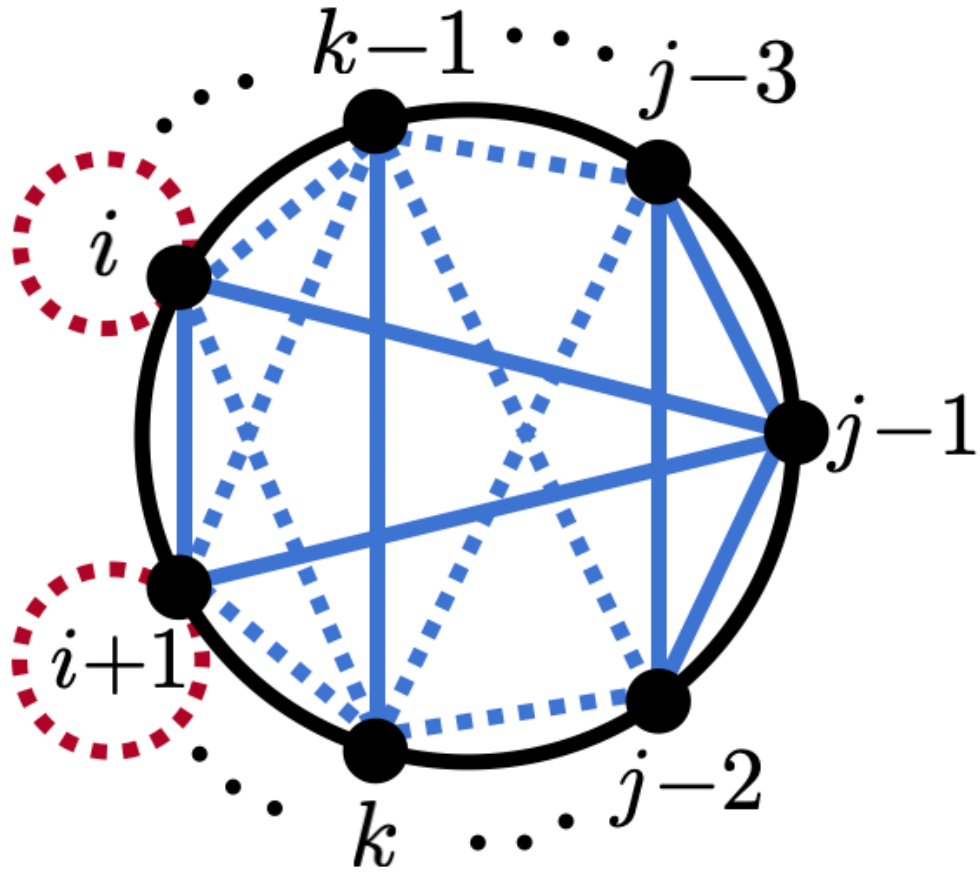}}\hspace{2cm} \raisebox{0pt}{\includegraphics[scale=.39]{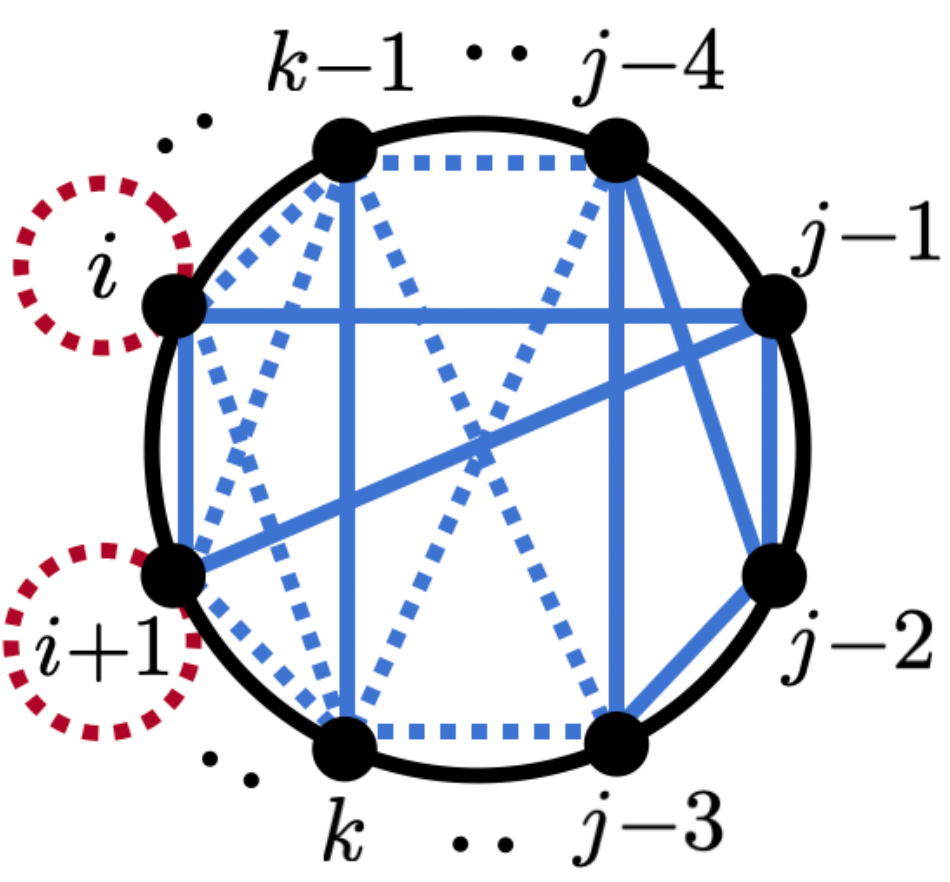}}
$$
where in the expressions (\ref{right4}) we make the replacement
\begin{equation}
\frac{1}{{\color{red} [\ast_3(P)|j{-}1\ra][\ast_6(P)|j{-}1\ra]}}\rightarrow \frac{{\color{blue} \la i\,x\ra\la i{+}1\,x\ra}}{{\color{blue} \la i\,j{-}1\ra\la i{+}1\,j{-}1\ra}\times {\color{red} [\ast_3(P)|x\ra][\ast_6(P)|x\ra]}}
\end{equation}
Introducing the reference spinor changes the permutational sum over $(i,\dots j,{-}2)$ to $(i,\dots j{-}1)$, ie. over all labels in the right blob,
\begin{equation}
\raisebox{-47pt}{\includegraphics[scale=.42]{GenRight2.pdf}}\hspace{0.3cm} =   \sum_{{\cal P}'({i..j{-}1})} \hspace{0.3cm} \raisebox{-46pt}{\includegraphics[scale=.38]{GRref3.pdf}} \label{refformula}
\end{equation}
for odd number of points and analogously for even number of points. Note that we still have to mod out by symmetries of the diagrams, $(i\leftrightarrow i{+}1)$, $(i{+}2\leftrightarrow i{+}3)$, etc. The sum (\ref{refformula}) is then independent on $\lambda_x$.

Using one point to divide a large piece into smaller pieces is very familiar from the triangulation of geometric spaces. A simple example is a polygon in the plane and the triangulation using one of the vertices or some arbitrary point,
$$
\includegraphics[scale=.27]{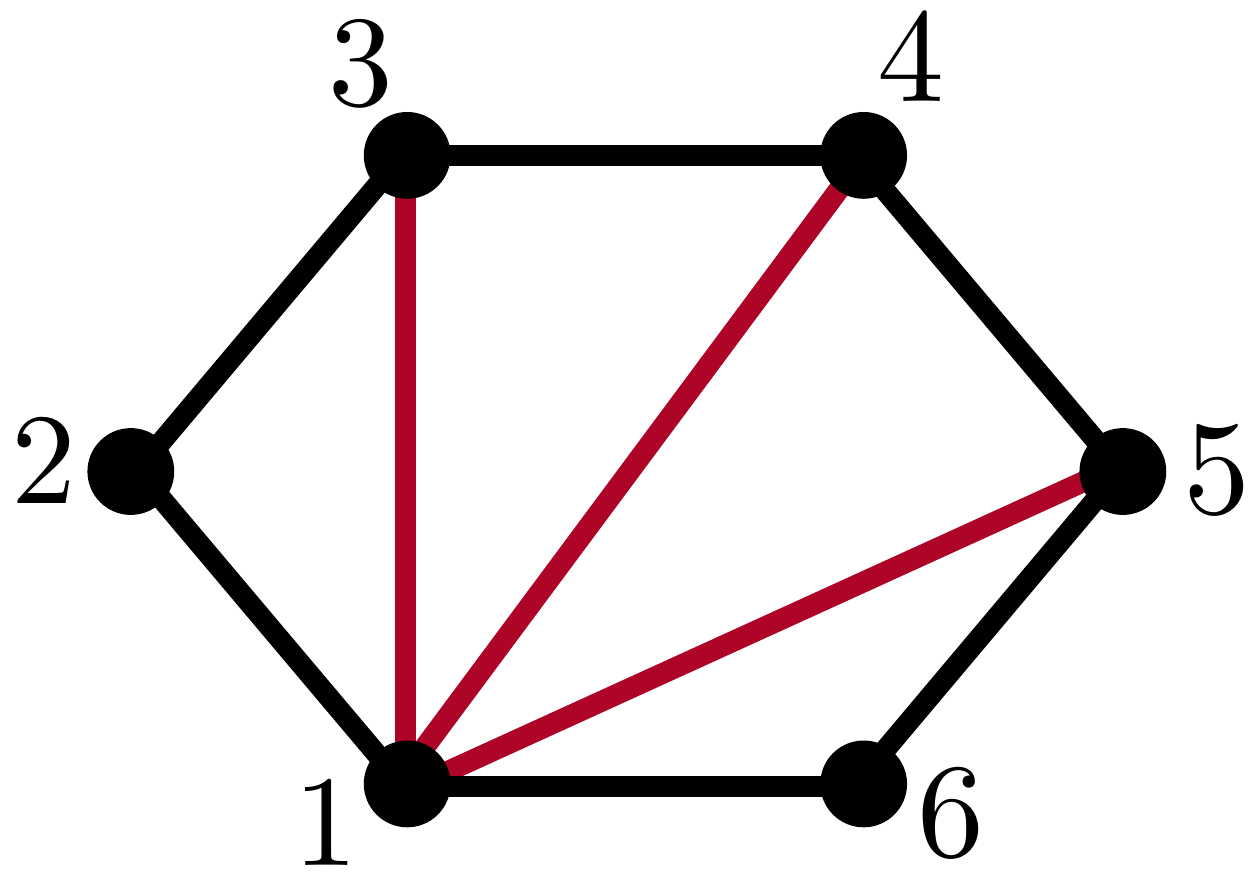}\hspace{2cm}\includegraphics[scale=.27]{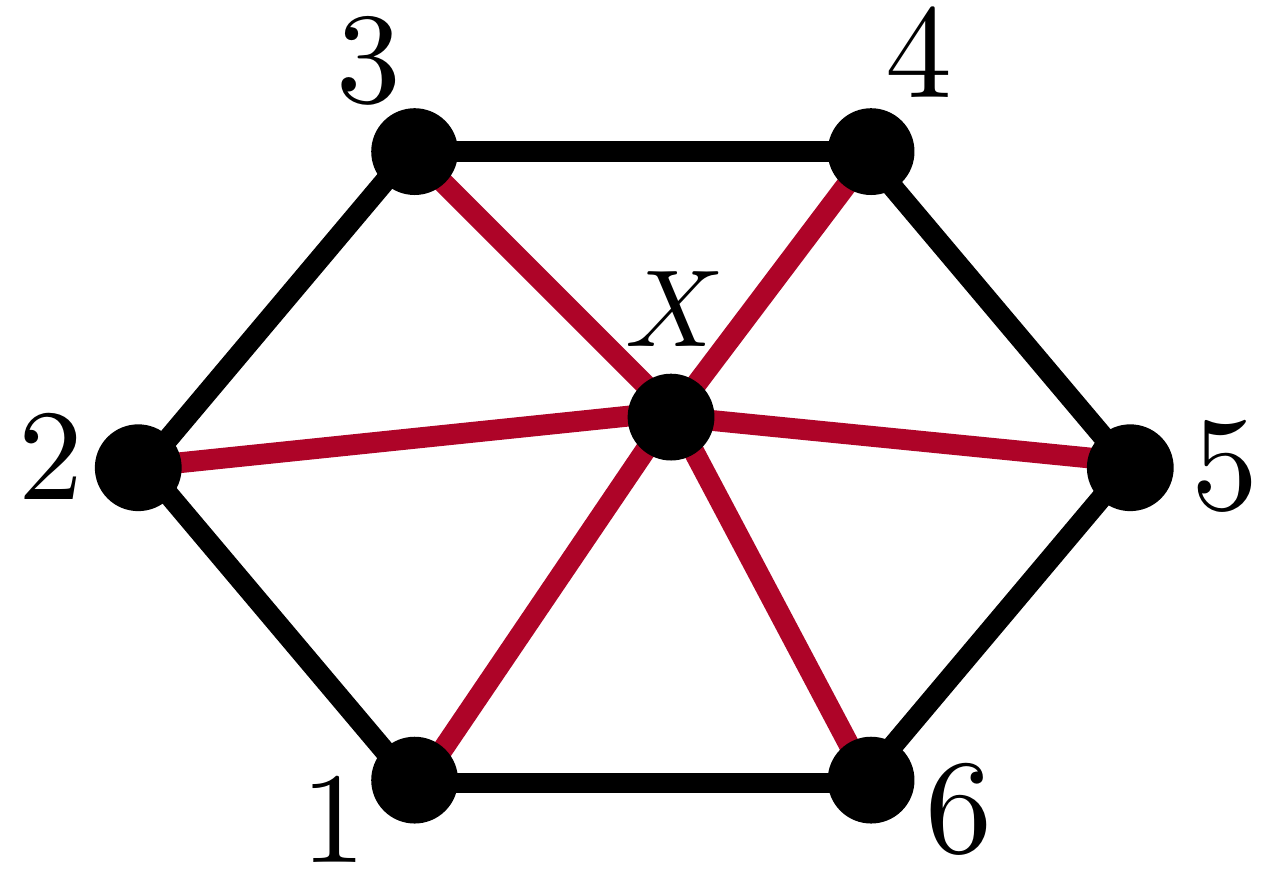}
$$
Therefore, it is suggestive that this expansion of the right blob in terms of smaller pieces is also geometric. As with this triangulation, our result does not depend on the choice of $x$ and we can use any of the labels in the right blob or just leave it generic.

\section{Supersymmetric six-point amplitude}

The important question is how to generalize our formulas for graviton helicity amplitude ${\cal M}_n(1^-2^-3^-4^+\dots n
^+)$ to $n$-pt NMHV super-amplitudes $M_n$. While we do not know the general story yet, we do have the result for the six-point super-amplitude $M_6$. In the previous sections, we argued that the ${\cal G}$-invariant expansion for gravity amplitudes is a natural counterpart of the ${\cal R}$-invariant expansion for gluons. The basis of comparison are the factorization diagrams which are identical in both cases.

In the gluon case the helicity ${\cal R}$-invariants are obtained from the supersymmetric objects $\widetilde{R}$-invariants (\ref{bosR}). We can try to ``upgrade" the ${\cal G}$-invariants in the same way and find supersymmetric graviton $G$-invariants. The natural guess is to upgrade the supersymmetric delta function from ${\cal N}=4$ to ${\cal N}=8$,
\begin{equation}
\delta^{(8)}(Q)\,\delta^{(4)}(\Xi_{2;ij}) \rightarrow \delta^{(16)}(Q)\,\delta^{(8)}(\Xi_{2;ij})
\end{equation}
where now $\eta_I$, $I=1,\dots,8$. Then for the upgraded ${\cal G}$-invariant (\ref{G56b}) we get
\begin{align}
G(1,{\bf 2},3,\{4\},\{5,6\},\{\}) &=\label{su6a}\\
  &\hspace{-2cm}\frac{\delta^{(16)}(Q)\, \delta^{(8)}(\Xi_{2;51})\cdot \la12\ra[34]}{\la12\ra^8\la56\ra^8\cdot s_{234}[56]^2[15][16]\la23\ra^2\la24\ra\la34\ra\la4|56|1]\la2|34|5]\la2|34|6]}\nonumber
\end{align}
with the same $\Xi_{2;51}$ as in the Yang-Mills case.
\begin{equation}
\Xi_{2;51}= -\la12\ra \la56\ra\cdot \Big\{[56] \eta_1 + [61] \eta_5 +[15]\eta_6\Big\}
\end{equation}
Plugging back we cancel the eight order poles in the denominator and get a proposal for the supersymmetric expression,
\begin{equation}
G(1,{\bf 2},3,\{4\},\{5,6\},\{\}) =\frac{\delta^{(16)}(Q)\, \delta^{(8)}([56] \eta_1 + [61] \eta_5 +[15]\eta_6)\cdot \la12\ra[34]}{s_{234}[56]^2[15][16]\la23\ra^2\la24\ra\la34\ra\la4|56|1]\la2|34|5]\la2|34|6]}\label{Gsusy1}
\end{equation}
We can now integrate over $\eta_1$, $\eta_2$, $\eta_3$ and extract from delta functions,
\begin{equation}
\int d^8\eta_1 d^8\eta_2 d^8\eta_3\left[\delta^{(16)}(q)\, \delta^{(8)}([56] \eta_1 + [61] \eta_5 +[15]\eta_6)\right] = \la23\ra^8[56]^8
\end{equation}
and we reconstruct the correct expression for ${\cal G}(1,{\bf 2},3,\{4\},\{5,6\},\{\})$. Similarly we can upgrade the second ${\cal G}$-invariant (\ref{G456b}) to
\begin{align}
G(1,{\bf 2},3,\{\},\{4,5,6\},\{\}) &= \label{su6b}\\
 &\hspace{-2cm}\frac{\delta^{(16)}(Q)\,\delta^{(8)}([23]\eta_1+[31]\eta_2 + [12]\eta_3)\cdot \la\la 2|(1)(456)(5)(6)(456)(3)|2\ra\ra}{s_{456}^2[12][23]\la45\ra\la46\ra\la56\ra\la6|45|3]\la5|46|3]\la4|56|3]\la6|45|1]\la5|46|1]\la4|56|1]}\nonumber
\end{align}
The naive conjecture is that the six-point NMHV ${\cal N}=8$ SUGRA amplitude is 
\begin{equation}
M_6^{\rm naive} = \hspace{-0.2cm}\sum_{{\cal P}(1,2,3)}\sum_{{\cal P}'(4,5,6)}G(1,{\bf 2},3,\{4\},\{5,6\},\{\}) + \hspace{-0.2cm}\sum_{{\cal P}'(1,2,3)}G(1,{\bf 2},3,\{\},\{4,5,6\},\{\})\label{GRsusy1}
\end{equation}
where the permutational sums are the same as in (\ref{our6}). This expression gives by construction the correct amplitude ${\cal M}_6(1^-2^-3^-4^+5^+6^+)$, but if we calculate amplitudes for other helicity configurations, we find a discrepancy. For example, for ${\cal M}_6(1^+2^+3^+4^-5^-6^-)$ the expression (\ref{GRsusy1}) gives
\begin{equation}
{\cal M}_6^{\rm naive} = \sum_{{\cal P}}\frac{\la 4|56|1]^7\cdot \la12\ra[34]}{s_{234}[56]^2[15][16]\la23\ra^2\la24\ra\la34\ra\la2|34|5]\la2|34|6]}\label{GRsusy2}
\end{equation}
where we denoted ${\cal P}\equiv\{{\cal P}(1,2,3),{\cal P}'(4,5,6)\}$. The second term in (\ref{GRsusy1}) does not contribute because the helicity factor for this helicity configuration is zero. 

Unlike in ${\cal M}_6(1^-2^-3^-4^+5^+6^+)$ now the expression (\ref{GRsusy2}) contains double poles which should cancel in the sum. It is easy to check that they do not cancel and we have to add additional simple term to cancel these double poles and reproduce the correct answer,
\begin{equation}
{\cal M}_6= \sum_{{\cal P}}\frac{\la 4|56|1]^7\cdot \la12\ra[34]}{s_{234}[56]^2[15][16]\la23\ra^2\la24\ra\la34\ra\la2|34|5]\la2|34|6]}+\frac{s_{456}^7}{\la12\ra^2\la23\ra^2\la13\ra^2[45]^2[46]^2[56]^2}\label{GRsusy3}
\end{equation}
This additional term exhibits a complete symmetry in labels $1,2,3$ and $4,5,6$. If we calculate another helicity configuration, for example ${\cal M}_6(1^-2^-3^+4^-5^+6^+)$, similar extra term is needed,
\begin{equation}
{\cal M}_6= \sum_{{\cal P}}\frac{\la24\ra^7[56]^6\cdot \la12\ra[34]}{s_{234}[15][16]\la23\ra^2\la34\ra\la2|34|5]\la2|34|6]\la4|56|1]}+\frac{\la12\ra^6[56]^6}{s_{456}\la23\ra^2\la13\ra^2[45]^2[46]^2}\label{GRsusy3}
\end{equation}
It is easy to see the supersymmetric term we need to add to (\ref{GRsusy1}) to get the correct helicity amplitudes from the superamplitude,
\begin{align}
M_6 &= \hspace{-0.2cm}\sum_{{\cal P}(1,2,3)}\sum_{{\cal P}'(4,5,6)}G(1,{\bf 2},3,\{4\},\{5,6\},\{\}) + \hspace{-0.2cm}\sum_{{\cal P}'(1,2,3)}G(1,{\bf 2},3,\{\},\{4,5,6\},\{\})\nonumber\\
&\hspace{3.5cm}+\frac{\delta^{(16)}(Q)\cdot \delta^{(8)}(\eta_4[56]+\eta_5[64]+\eta_6[45])}{s_{456}\la12\ra^2\la23\ra^2\la13\ra^2[45]^2[46]^2[56]^2}\label{GRsusy4}
\end{align}
The last term in (\ref{GRsusy4}) does not have a factorization diagram interpretation and has no image in the Yang-Mills formula for NMHV amplitudes. It is not clear yet how this generalizes to higher-point super-amplitudes but certainly additional terms beyond factorization diagrams are necessary. However, all these terms must cancel when integrated over $\eta_1$, $\eta_2$, $\eta_3$ and extracting the amplitude ${\cal M}_n(1^-2^-3^-4^+\dots n^+)$ which we correctly obtained from factorization diagrams in the last section. In other words, these extra terms do not contribute if the special labels $(1,2,3)$ completely overlap with the negative helicity gravitons.

We can compare supersymmetric $G$-invariants to the ${\cal N}=7$ $R$-invariants discussed in \cite{Armstrong:2020ljm}. For example, one of the objects (equation (81) in \cite{Armstrong:2020ljm}) is 
\begin{equation}
\frac{\la34\ra\la56\ra[12] \delta^{(7)}([45]\eta_3 + [53]\eta_4 + [34]\eta_5)}{s_{345}\la12\ra[35][34]\la61\ra\la26\ra\la6|45|3]\la6|35|4]\la6|34|5]\la2|34|5]}
\end{equation}
While we can certainly see some similarities in the numerator factors, spurious poles and the susy delta function, there are important differences in details. Nevertheless, it would be interesting to explore if and how these objects are related.

As we noted earlier, the 6pt case is a bit special and in the Yang-Mills case there are only two inequivalent representations of the gluon amplitude despite the general expression (\ref{YM1}) would suggest there are six. For higher points these accidental identifications do not happen anymore and we indeed get $n$ formulas for different choices of $\ast = 1,2,\dots,n$. In the supersymmetric case, they are all related by cyclic shifts and there is only one distinct expression. For fixed helicity amplitudes ${\cal A}_n$ all these expressions look different because the helicity amplitude is no longer cyclic.

In the graviton case the situation is similar. The choice of $\ast$ is replaced by the choice of three special labels $(a,b,c)$ in factorization diagrams. In the gluon case, for $\ast=k$ these labels must be $(k{-}1,k,k{+}1)$ due to cyclicity but for gravitons we can choose arbitrary three labels. In the supersymmetric case all choices are equivalent due to permutational symmetry and simple relabeling, but for fixed helicity amplitude we get different looking formulas. In this paper we chose $(a,b,c)=(1,2,3)$ to completely overlap with negative helicity gluons in ${\cal M}_n(1^-2^-3^-4^+\dots n^+)$ which lead to simpler expressions. If the analogy indeed goes further and our expressions do have geometric interpretation as the triangulations of Gravituhedron geometry, the choice of labels $(a,b,c)$ should be the analogue of $Z_\ast$ in the triangulation of the Amplituhedron. 

We choose a different origin and calculate the same amplitude ${\cal M}_6(1^-2^-3^-4^+5^+6^+)$. One other symmetric choice of origin is $(a,b,c)=(4,5,6)$, for which we get
\begin{equation}
{\cal M}_6= \sum_{{\cal P}}\frac{\la 1|23|4]^7\cdot [12]\la34\ra}{s_{234}[23]^2[24][34]\la56\ra^2\la15\ra\la16\ra\la5|34|2]\la6|34|2]}+\frac{s_{123}^7}{[12]^2[23]^2[13]^2\la45\ra^2\la56\ra^2\la46\ra^2}\label{GR6other}
\end{equation}
This is the analogue of the other $R$-invariant expansion of the gluon amplitude on the first line of (\ref{NMHV6}),
\begin{equation}
{\cal A}_6 = \sum_{{\cal P}} \frac{\la 1|23|4]^3}{s_{234}[23][34]\la56\ra\la16\ra\la5|34|2]} \label{YMother}
\end{equation}
where ${\cal P}$ again stands for the sum of two terms (\ref{YMperm}). Note that the third invariant is zero for the split helicity configuration. The two gluon formulas (\ref{A6YM}) and (\ref{YMother}) are the only 6pt $R$-invariant representations for gluon amplitudes, and correspond to two different triangulations of NMHV Amplituhedron, one in terms of two tetrahedra and other in terms of three tetrahedra. This geometric relation is well-known in the literature as 2-3 Pachner move. It also plays a critical role in the Grassmannian representation as residues of Grassmannian integral. More generally, any such relation between different representations of gluon amplitudes have been always associated with Global Residue Theorems and the equivalence was guaranteed by contour deformations and Cauchy formula \cite{ArkaniHamed:2009dn,ArkaniHamed:2009dg,Arkani-Hamed:2016byb}.

We can see an analogy between the first term in (\ref{GR6other}) and (\ref{YMother}), they clearly have the same structure. The second term in (\ref{GR6other}) is new and comes from the last term in (\ref{GRsusy4}) if we choose $(4,5,6)$ as origin of labels. Because it looks quite different from other terms, no spurious poles and only double poles, we speculate that it is a ``pole at infinity" in some representation (Grassmannian or other) where the relation between (\ref{GR6other}) and (\ref{our6}) is a Global Residue Theorem. If such picture exists, it would be a major hint for deeper geometric structures, it should explain the spurious poles cancelation and the relation between different representations of the amplitude. 

Finally, there exists slightly simpler version of (\ref{GR6other}),
\begin{equation}
{\cal M}_6= \sum_{{\cal P}}\frac{\la 1|23|4]^6\cdot \la3|45|6]}{s_{234}[23]^2[24][34]\la56\ra^2\la15\ra\la16\ra\la5|34|2]}+\frac{s_{123}^7}{[12]^2[23]^2[13]^2\la45\ra^2\la56\ra^2\la46\ra^2}\label{GR6other2}
\end{equation}
where ${\cal P}$ is now full ${\cal P}(1,2,3)$ and ${\cal P}(4,5,6)$ permutational sum. The relation between (\ref{GR6other}) and (\ref{GR6other2}) can be trivially proven using Shouten identities, it is just an expansion of each term in (\ref{GR6other}) in terms of two in (\ref{GR6other2}),
\begin{align}
\frac{\la 1|23|4]^7\cdot [12]\la34\ra}{s_{234}[23]^2[24][34]\la56\ra^2\la15\ra\la16\ra\la5|34|2]\la6|34|2]} &=\\ &\hspace{-6cm} \frac{\la 1|23|4]^6\cdot \la3|45|6]}{s_{234}[23]^2[24][34]\la56\ra^2\la15\ra\la16\ra\la5|34|2]} + \frac{\la 1|23|4]^6\cdot \la3|46|5]}{s_{234}[23]^2[24][34]\la56\ra^2\la15\ra\la16\ra\la6|34|2]}\nonumber
\end{align}
In (\ref{GR6other2}) each term in the sum has only one spurious pole, and it might be simpler to understand the mechanism of spurious poles cancelation. However, it is not clear if the relation (\ref{GR6other2}) is just accidental, or if it generalizes to higher $n$. In that case, it would mean we can expand ${\cal G}$-invariants in terms of even simpler objects.

\section{Outlook}

In this paper, we presented a new formula for gravity NMHV tree-level amplitudes. We conjectured this formula based on the method of factorization diagrams which originate from the $R$-invariants expansion for Yang-Mills amplitudes. The $R$-invariants manifest hidden dual conformal symmetry (and Yangian in supersymmetric case) and correspond to pieces in the triangulation of the Amplituhedron. In the gravity case, we associated the factorization diagrams with new objects, which we called ${\cal G}$-invariants. We checked up to 10pt that ${\cal G}$-invariants reproduce correctly the NMHV gravity amplitude. The striking similarity between the gravity and Yang-Mills constructions suggests that the ${\cal G}$-invariants also triangulate certain geometric object, which we tentatively called Gravituhedron. At the moment, we have neither the definition of the Gravituhedron geometry, nor the analogue of the dlog form which would reproduce the amplitude, but the detailed study of ${\cal G}$-invariants and the supersymmetric generalizations is the best start to attack this problem.

Our formula is different from the BCFW representations presented in the literature but it shares a number of similar properties: same type of spurious poles and pair-wise cancelation of these poles, or manifest ${\cal O}(1/z^2)$ behavior under certain BCFW shifts. On the other hand, in the BCFW procedure breaks the manifest symmetry of helicity amplitudes by choosing two special momenta to be shifted. Our formula manifests the $S_3\times S_{n-3}$ permutational symmetry of a helicity amplitude ${\cal M}_n(1^-2^-3^-4^+5^+\dots n^+)$. This suggests an existence of some 3-line shift recursion relations which would reproduce our representation. While 3-line shifts in gravity were studied in details in \cite{Bianchi:2008pu}, the desired recursion (if it exists) must look different. One can also try to explore more general shifts (see e.g. \cite{Cheung:2008dn,Feng:2009ei,Cohen:2010mi,Jin:2014qya,Cheung:2015cba,Cheung:2015ota,Kampf:2012fn} for various recursion relations) for that purpose. Similarly important is to extend this formalism to N$^2$MHV amplitude where the $R$-invariants are generalized to more complicated Yangian invariants, naturally labeled as cells in the positive Grassmannian $G_+(2,n)$, though they can be expressed as products of two $R$-invariants with shifted labels. It is worth noting that the closed solution to ${\cal N}=8$ BCFW recursion \cite{Drummond:2009ge} was given for the general $n$-pt N$^k$MHV amplitude in a very compact form, making it also a great target for exploration.

Finally, the most important direction is to try to geometrize the gravity amplitudes directly, ie. trying to find a geometric definition of the known amplitudes as certain differential forms over the geometric region in the momentum space. The natural starting point is the momentum Amplituhedron \cite{Damgaard:2019ztj} which (unlike the original Amplituhedron) lives directly in the spinor helicity space, and non-dlog forms associated with positive geometries \cite{Benincasa:2020uph}. However, both the geometry and the forms must be very different so new ideas are needed to make any progress here. While our results should provide a guidance how to proceed here, the burning question is to geometrize even MHV amplitudes starting at four-points! As we argued in the introduction, there are two main approaches to gravity tree-level amplitude: $d$-dimensional double-copy construction based on the connection to Yang-Mills amplitudes, best implemented in the BCJ formalism; and the four-dimensional on-shell recursion relations for helicity amplitudes based on the improved behavior of gravity amplitudes at infinity. Our formula belongs to the second category, though we do not have an explicit recursion implementation. The putative Gravituhedron geometry should likely manifest both of these set of ideas.


\section*{Acknowledgements}

We thank Nima Arkani-Hamed, Lance Dixon, Enrico Herrmann, Andrew Hodges and Arthur Lipstein for stimulating discussions. J.T. is supported in part by U.S. Department of Energy grant DE-SC0009999 and by the funds of University of California.


\bibliographystyle{JHEP}
\phantomsection
\bibliography{amp_refs}
\clearpage

\end{document}